\documentclass[a4paper,12pt]{article}
\usepackage{graphicx,rotating,hyperref,slashed,verbatim,amsmath,xcolor,amssymb,amsfonts,expdlist,colortbl,cite,youngtab}
\usepackage{amsfonts}
\usepackage[utf8]{inputenc}
\usepackage{amsthm}
\usepackage{halloweenmath}
\usepackage{physics}
\usepackage{amsmath}
\usepackage{physics}
\usepackage{mathtools}
\usepackage{hhline}
\usepackage{comment}
\usepackage{colortbl}
\usepackage{multicol}
\usepackage{pdflscape}
\usepackage{multirow}
\usepackage{lscape}
\usepackage{placeins}
\usepackage{booktabs}    
\usepackage{calc}        
\usepackage{tikz}
\usepackage[compat=1.1.0]{tikz-feynman}
\usepackage{subcaption}
\usepackage{fancyhdr}
\usepackage{cleveref}

\newcommand{\nn}{\nonumber}

\newcommand{\g}{\gamma}

\newcommand{\ab}{\alpha\beta}

\newcommand{\genbar}[1]{\,\overline{\!#1}{}}
\newcommand{\Overrightarrow}[1]{{%
	#1
}}
\newcommand{\Overleftarrow}[1]{{%
	#1
}}

\newcommand{\cbar}{\bar{c}}

\newcommand{\cbvp}{\cbar^{v'}_+}

\newcommand{\bv}{b^v_+}

\newcommand{\Dslash}{\slashed{D}}
\newcommand{\vslash}{\slashed{v}}
\newcommand{\ccdot}{\!\cdot\!}
\newcommand{\vcD}{v \ccdot D}
\newcommand{\Qbar}{\genbar{Q}}
\newcommand{\Jbar}{\genbar{J}}
\newcommand{\mJ}{\mathcal{J}}
\newcommand{\mJbar}{\genbar{\mJ}}
\newcommand{\Hc}{H_c}
\newcommand{\Hb}{H_b}

\newcommand{\lqcd}{\Lambda_{\text{QCD}}}

\newcommand{\aS}{\alpha_s}

\newcommand{\amp}[3]{\frac{\langle #1(p') |\, #2\, | #3(p) \rangle}{\sqrt{m_{\let\overline\relax#3} m_{#1}}}}

\hypersetup{colorlinks,bookmarksopen,bookmarksnumbered,
	linkcolor=blus,pdfstartview=FitH,urlcolor=rossos,citecolor=verde}

\usepackage{tabularx}

\definecolor{gold}{RGB}{187,161,79}
\definecolor{silver}{RGB}{192,192,192}
\definecolor{darkgreen}{RGB}{0, 130, 0}

\newcommand{\sfootnote}[1]{}
\definecolor{bluc}{cmyk}{1,1,0,0.1}
\definecolor{rossoCP3}{cmyk}{0,.88,.77,.40}
\definecolor{rosso}{cmyk}{0,1,1,0.4}
\definecolor{giallo}{cmyk}{0,.33,1,0}
\definecolor{rossos}{cmyk}{0,1,1,0.55}
\definecolor{rossoc}{cmyk}{0,1,1,0.2}
\definecolor{verdes}{cmyk}{0.92,0,0.59,0.4}

\hypersetup{colorlinks, bookmarksopen, bookmarksnumbered,
	citecolor=verdes, linkcolor=bluc, pdfstartview=FitH, urlcolor=rossos}

\newcommand{\mio}[1]{}

\definecolor{Gray}{gray}{0.95}

\usepackage{color}
\definecolor{rosso}{cmyk}{0,1,1,0.4}
\definecolor{rossos}{cmyk}{0,1,1,0.55}
\definecolor{rossoc}{cmyk}{0,1,1,0.2}
\definecolor{blu}{cmyk}{1,1,0,0.3}
\definecolor{blus}{cmyk}{1,1,0,0.6}
\definecolor{bluc}{cmyk}{1,1,0,0.1}
\definecolor{verde}{cmyk}{0.92,0,0.59,0.25}
\definecolor{verdec}{cmyk}{0.92,0,0.59,0.15}
\definecolor{verdes}{cmyk}{0.92,0,0.59,0.4}

\renewcommand\&{&}

\def\circa#1{\,\raise.3ex\hbox{$#1$\kern-.75em\lower1ex\hbox{$\sim$}}\,}

\newcommand{\beq}{\begin{equation}}
	\newcommand{\eeq}{\end{equation}}

\newcommand{\bea}{\begin{eqnarray}}
	\newcommand{\eea}{\end{eqnarray}}
\newcommand{\be}{\begin{equation}}
	\newcommand{\ee}{\end{equation}}
\font\tenrsfs=rsfs10 at 12pt
\font\sevenrsfs=rsfs7 at 10 pt
\font\fiversfs=rsfs5
\newfam\rsfsfam
\textfont\rsfsfam=\tenrsfs
\scriptfont\rsfsfam=\sevenrsfs
\scriptscriptfont\rsfsfam=\fiversfs
\def\mathscr#1{{\fam\rsfsfam\relax#1}}

\makeatletter

\def\hhref#1{\href{http://arxiv.org/abs/#1}{arXiv:#1}} 

\newcommand{\doi}[1]{\href{http://dx.doi.org/#1}{[doi]}}

\def\hhref#1{\href{http://arxiv.org/abs/#1}{arXiv:#1}}

\def\art{\@ifnextchar[{\eart}{\oart}}
\def\eart[#1]#2#3#4#5#6{{\rm #2}, {\em #3 \bf #4} {\rm (#6) #5} ({\em #1})}

\def\article{\@ifnextchar[{\earticle}{\oarticle}}
\def\oarticle#1#2#3#4#5#6{{\rm #1}, {\em ``#6''}, {\rm #2 #3 (#5) #4}}
\def\earticle[#1]#2#3#4#5#6#7{{\rm #2}, {\em ``#7''}, {\rm #3 #4 (#6) #5}  [\hhref{#1}]}
\def\hepart[#1]#2{{\rm #2, \em#1}}
\def\heparticle[#1]#2#3{#2, {\em ``#3''} [\hhref{#1}]}

\newcounter{alphaequation}[equation]
\def\thealphaequation{\theequation\hbox to
	0.6em{\hfil\alph{alphaequation}\hfil}}
\def\eqnsystem#1{
	\def\@eqnnum{{\rm (\thealphaequation)}}
	\def\@@eqncr{\let\@tempa\relax \ifcase\@eqcnt \def\@tempa{& & &} \or
		\def\@tempa{& &}\or \def\@tempa{&}\fi\@tempa
		\if@eqnsw\@eqnnum\refstepcounter{alphaequation}\fi
		\global\@eqnswtrue\global\@eqcnt=0\cr}
	\refstepcounter{equation} \let\@currentlabel\theequation \def\@tempb{#1}
	\ifx\@tempb\empty\else\label{#1}\fi
	\refstepcounter{alphaequation}
	\let\@currentlabel\thealphaequation
	\global\@eqnswtrue\global\@eqcnt=0 \tabskip\@centering\let\\=\@eqncr
	$$\halign to \displaywidth\bgroup \@eqnsel\hskip\@centering
	$\displaystyle\tabskip\z@{##}$&\global\@eqcnt\@ne
	\hskip2\arraycolsep\hfil${##}$\hfil& \global\@eqcnt\tw@\hskip2\arraycolsep
	$\displaystyle\tabskip\z@{##}$\hfil
	\tabskip\@centering&\llap{##}\tabskip\z@\cr}
\def\endeqnsystem{\@@eqncr\egroup$$\global\@ignoretrue} \makeatother

\definecolor{fiorentina}{rgb}{.5,0,.5}

\usepackage{tikz-feynman}

\let\oldheadrule\headrule
\renewcommand{\headrule}{\color{white}\oldheadrule}
\usepackage{graphicx} 

\newcommand{\eps}{\varepsilon}

\newcommand{\LamCst}[1][]{ {\Lambda_{c}(#1)^{+}}}

\newcommand{\mLamB}{m_{\Lambda_b}  }
\newcommand{\mLamCst}{m_{\Lambda_c^*}}

\usepackage{orcidlink}

\begin{document}

\let\thefootnote\relax\footnote{\\$^\star$ Electronic address: \textcolor{rossoCP3}{\href{mailto:vigilante.dirisi@unina.it}{\textcolor{rossoCP3}{vigilante.dirisi@unina.it}}}\\$^\dagger$ Electronic address: \textcolor{rossoCP3}{\href{mailto:davide.iacobacci@unina.it}{\textcolor{rossoCP3}{davide.iacobacci@unina.it}}}\\$^\diamondsuit$ Electronic address: \textcolor{rossoCP3}{\href{mailto:sannino@cp3.sdu.dk}{\textcolor{rossoCP3}{sannino@qtc.sdu.dk}}}}
\vspace{0.5mm}

  \begin{center}
		\boldmath

		{\textbf{\LARGE\color{rossoCP3}
\bf \Large$\Lambda_b\to \Lambda_c^*$ at $1/m_c^2$ heavy quark order
			}}
		\unboldmath
		
		\bigskip\bigskip

	 	{Vigilante~Di~ Risi$^{1,3\,\star}$, Davide~Iacobacci$^{1,3\,\dagger}$,   Francesco~Sannino$^{1,2,3\,^\diamondsuit}$\\[5mm]}

\small{$^1$ Dipartimento di Fisica ``E. Pancini", Università di Napoli Federico II - INFN sezione di Napoli, Complesso Universitario di Monte S. Angelo Edificio 6, via Cintia, 80126 Napoli, Italy}\\
\small{$^2$ Scuola Superiore Meridionale, Largo S. Marcellino, 10, 80138 Napoli NA, Italy}\\
$^3$  Quantum  Theory Center ($\hbar$QTC), Danish-IAS, IMADA, Southern Denmark Univ., Campusvej 55, 5230 Odense M, Denmark\\

		\bigskip\bigskip
		
		\thispagestyle{empty}\large
		{\bf\color{blus} Abstract}
		\begin{quote} 
		\normalsize

\noindent
We systematically compute the $\Lambda_b(p, s_b) \to
        \LamCst[2595]$ and $\Lambda_b(p, s_b) \to
        \LamCst[2625]$ form factors within the  Heavy Quark Effective Theory (HQET) framework including  $\mathcal{O}(1/m_c^2)$.
Besides taking into account the Standard Model-like vector and axial contributions, we further determine tensor and pseudo-tensor form factors. 
Our work constitutes a step forward with respect to previous analyses allowing for a comprehensive study of the matrix element parametrisation stemming from the HQET formalism. Finally, we demonstrate that the resulting form factors  agree well with lattice Quantum Chromodynamics (LQCD) determinations stressing the need and relevance of the newly derived $1/m_c^2$ corrections.  
\end{quote}
\end{center}
\newpage 
\tableofcontents

\newpage
\section{ Introduction} 

\label{sec:form_factors}
The tree-level semileptonic decays involving the $b \to c$ transition hold a fundamental role in  determining the modulus of the Cabibbo-Kobayashi-Maskawa matrix element $|V_{cb}|$ \cite{Detmold:2015aaa,Bigi:2016mdz,Bigi:2017jbd,Gambino:2019sif,Gambino:2020jvv,Simula:2021yvm,Martinelli:2021onb,Bordone:2021oof,Martinelli:2021myh,Martinelli:2021ccm,Martinelli:2022xir,Martinelli:2022ifg,Martinelli:2022fgg,Naviglio:2022obj}. They also serve as a valuable tool for investigating Lepton-Flavor-Universality Violation. In the latter scenario, the calculation of mesonic form factors is crucial for predicting the $R_{D^{(*)}}$ ratios \cite{Bernlochner:2017jka,Bernlochner:2022ywh,Bordone:2019vic}, which currently exhibit a significant $3\sigma$ tension with the current experimental average \cite{HeavyFlavorAveragingGroup:2022wzx,Murgui:2019czp,Bardhan:2016uhr}. However, the exploration of baryonic decays can offer a complementary way to probe new physics (NP) beyond the standard model (SM) in comparison to the mesonic counterparts (see Refs.~\cite{Dutta:2015ueb,Datta:2017aue,Ray:2018hrx,Penalva:2019rgt,Ferrillo:2019owd,Mu:2019bin}).

As first pointed out in \cite{Isgur:1989vq,Georgi:1990um}, the low energy dynamics of mesons and baryons containing a heavy quark enjoys a spin-flavor symmetry where symmetry-breaking effects can be taken into account in a well defined expansion in the inverse of the heavy quark masses \cite{Neubert:1993mb}.

In the context of the HQET a comprehensive computation of weak-decay form factors for the ground-state baryons $\Lambda_b\rightarrow\Lambda_c$ and mesons $B\rightarrow D^{(*)}$ processes was first derived respectively in \cite{Falk:1992ws} and \cite{Falk:1992wt}.
Instead, the focus of our study revolves around $\Lambda_b \to \Lambda_c^{*}$ decays. The computation of the form factors was initially laid out in \cite{Leibovich:1997az}, employing the Heavy Quark Expansion (HQE) to capture effects up to first order in the inverse of the heavy quark masses $\mathcal{O}(1/m_{c,b})$. Building upon this foundation, \cite{Boer:2018vpx} extended the effort, introducing a novel form factor definition and incorporating short-distance corrections up to $\mathcal{O}(\alpha_s)$.

In a recent advancement, the authors of Ref.~\cite{Papucci:2021pmj} achieved higher precision by computing the form factors up to $\mathcal{O}(1/m_{c,b},\alpha_s)$, including the contributions of new physics (NP) form factors arising from tensor and pseudo-tensor mediating currents. This computation was carried out using the definitions estabilished in \cite{Leibovich:1997az}. See also \cite{Nieves:2019kdh,Nieves:2019nol,Du:2022rbf} for applications of the HQET to  $\Lambda_b \to \Lambda_c^{(*)} $processes.

Furthermore, the full set of $\Lambda_b \to \Lambda_c^{*}$ form factors has also been established through lattice computations. Initial efforts by \cite{Meinel:2021rbm} were further refined by \cite{Meinel:2021mdj}. These computations have provided data predominantly in the near-zero recoil region $w\lesssim 1.05$, where $w=v \cdot v^{\prime}$, with $v$ and $v^{\prime}$ representing the velocities of the initial and final states.

Examining the compatibility of lattice results with HQET predictions, \cite{Papucci:2021pmj} identified significant discrepancies in the fitting process. This prompted us to go beyond the constraints the first-order approximation, incorporating terms up to $\mathcal{O}(1/m_c^2)$. At the same time, we independently cross-validated outcomes for the NP form factors, utilizing the framework established in \cite{Boer:2018vpx} as form factor basis for our analysis. One of the main achievements of this work is to show that the newly computed $1/m_c^2$ order is needed to reconcile LQCD results with HQET computations.

The work is organized in the following way. In \cref{notation}, we first revisit the notation employed for the hadronic form factors and the on-shell amplitudes and then provide a brief overview of the HQET. This section helps setting the stage for the systematic computation of hadronic matrix elements contributing to baryonic decay processes.

In \cref{sec:firstorder} we reexamine and confirm the first-order determination of the vector and axial form factors presented in \cite{Boer:2018vpx}. Furthermore we include the chromomagnetic corrections and present the results for the tensor and pseudo tensor form factors.
Moving on to \cref{secondorder}, our attention shifts to the novel determination of the $\mathcal{O}(1/m_c^2)$ contributions for the full set of form factors. The analysis takes advantage of the Residual Chiral (RC) framework as well as the Vanishing Chromomagnetic (VC) limit \cite{Bernlochner:2022ywh}. 

Finally, in \cref{fitLQCD}, we compare our analytic results to 
 LQCD data \cite{Meinel:2021mdj} and demonstrate that our predictions  align remarkably well with the lattice results.
Our findings and outlook are summarized in the conclusions presented in \cref{conclusions}. Several technical details are summarized in the appendix.

\section{Notation and theoretical framework} 
\label{notation}
This section is dedicated to setting up the notation that we will use for the hadronic form factors and the on-shell amplitudes. Furthermore, to keep the work self-contained we  provide a brief overview of the HQET.  
\subsection{$\Lambda_b\to \Lambda_c^*$  transitions }
\label{sec:helicity_form_factors}
In the following we examine a suitable parametrisation for the matrix elements stemming from the underlying currents mediating the transitions
\begin{equation}
\begin{aligned}
\label{transition}
    &\Lambda_b(p, s_b) \to
        \LamCst[2595](k, J_z \equiv s_c)                    & \text{with } J^P = 1/2^- \ ,\\
     &\Lambda_b(p, s_b) \to   \LamCst[2625](k, J_z \equiv s_c +\lambda_c)         & \text{with } J^P = 3/2^- \ .
\end{aligned}
\end{equation}
Here $p$ and $k$ are the four momenta of the initial and final state respectively. The momentum transfer is given by $q^\mu \equiv p^\mu - k^\mu$. With $J^P$ we denote the angular momentum and parity of the $\Lambda_c^{*}$ states, while $s_b$ and $J_z$ are the rest-frame helicities.
For the $\LamCst[2625]$ state, $J_z$ is the composition of the rest-frame helicity $s_c$ of a $1/2^+$ spinor
and the polarisation of the vector $\eta(\lambda_c)$ (see \cref{app:kin}).

The most general decomposition of the hadronic matrix elements for vector, axial, tensor and pseudo-tensor currents mediating the transition  $\Lambda_{b} \to \LamCst[2595]$ reads as:
\begin{equation}
\label{eq:FF-12-def}
\begin{aligned}
    \bra{\LamCst[2595](k, \eta(\lambda_c), s_c)} \genbar{c}\gamma^\mu b \ket{\Lambda^0_{b}(p, s_b)}
        & = +\genbar{u}_{\alpha}^{(1/2)}(k, \eta(\lambda_c), s_c)
            \left[\sum_{i} f_i(q^2) \Gamma^{\alpha \mu}_{V,i} \right] u(p, s_b)\,,\\
    \bra{\LamCst[2595](k, \eta(\lambda_c), s_c)} \genbar{c}\gamma^\mu \gamma_5 b \ket{\Lambda^0_{b}(p, s_b)}
        & = -\genbar{u}_{\alpha}^{(1/2)}(k, \eta(\lambda_c), s_c)
            \left[\sum_{i} g_i(q^2) \gamma_5 \Gamma^{\alpha \mu}_{A,i} \right] u(p, s_b)\,,\\
      \bra{\LamCst[2595](k, \eta(\lambda_c), s_c)} \genbar{c}\,i\sigma^{\mu\nu} q_\nu b \ket{\Lambda^0_{b}(p, s_b)} 
        & = -\genbar{u}_\alpha^{(1/2)}(k, \eta(\lambda_c), s_c)
            \left[\sum_{i} t_i(q^2)  \Gamma^{\alpha \mu}_{T,i} \right] u(p, s_b)\,, \\
       \bra{\LamCst[2595](k, \eta(\lambda_c), s_c)} \genbar{c}\, i\sigma^{\mu\nu} q_\nu\gamma_5 b \ket{\Lambda^0_{b}(p, s_b)} 
        & = -\genbar{u}_\alpha^{(1/2)}(k, \eta(\lambda_c), s_c)
            \left[\sum_{i} t^5_i(q^2) \gamma_5 \Gamma^{\alpha \mu}_{T5,i} \right] u(p, s_b)\,,      
\end{aligned}
\end{equation}
where $\genbar{u}^{(1/2)}_\alpha$ is the spin $1/2$ projection of a Rarita-Schwinger object, denoted as $u_\alpha^\text{RS}(k, \eta, s)$ (see 
 \cref{app:RSspinor}). The full set of $\Lambda_b \to \Lambda_c^*(2595)$ form factors is given by $f_i(q^2) ,g_i(q^2) ,t_i(q^2) ,t^5_i(q^2)$.

A similar parametrisation can be applied to the hadronic matrix element for the transition to
the $\LamCst[2625]$ state:
\begin{equation}
\label{eq:FF-32-def}
\begin{aligned}
    \bra{\LamCst[2625](k, \eta(\lambda_c), s_c)} \genbar{c}\gamma^\mu b \ket{\Lambda^0_{b}(p, s_b)}
        & = +\genbar{u}_\alpha^{(3/2)}(k, \eta(\lambda_c), s_c)
            \left[\sum_{i} F_i(q^2) \Gamma^{\alpha \mu}_{V,i} \right] u(p, s_b)\,,\\
    \bra{\LamCst[2625](k, \eta(\lambda_c), s_c)} \genbar{c}\gamma^\mu \gamma_5 b \ket{\Lambda^0_{b}(p, s_b)}
        & = -\genbar{u}_\alpha^{(3/2)}(k, \eta(\lambda_c), s_c)
            \left[\sum_{i} G_i(q^2) \gamma_5 \Gamma^{\alpha \mu}_{A,i} \right] u(p, s_b)\,,\\
     \bra{\LamCst[2595](k, \eta(\lambda_c), s_c)} \genbar{c}\,i\sigma^{\mu\nu} q_\nu b \ket{\Lambda^0_{b}(p, s_b)} 
        & = -\genbar{u}_\alpha^{(3/2)}(k, \eta(\lambda_c), s_c)
            \left[\sum_{i} T_i(q^2)  \Gamma^{\alpha \mu}_{T,i} \right] u(p, s_b)\,, \\
       \bra{\LamCst[2595](k, \eta(\lambda_c), s_c)} \genbar{c}\,i\sigma^{\mu\nu} q_\nu\gamma_5 b \ket{\Lambda^0_{b}(p, s_b)} 
        & = -\genbar{u}_\alpha^{(3/2)}(k, \eta(\lambda_c), s_c)
            \left[\sum_{i} T^5_i(q^2) \gamma_5 \Gamma^{\alpha \mu}_{T5,i} \right] u(p, s_b)\,.    
\end{aligned}
\end{equation}
Here $\genbar{u}^{(3/2)}_\alpha$ is the spin $3/2$ projection of a Rarita-Schwinger spinor and $F_i(q^2) ,G_i(q^2) ,T_i(q^2) ,T^5_i(q^2)$ are the $\Lambda_b \to \Lambda_c^*(2625)$ form factors.

Following \cite{Papucci:2021pmj}, we express the matrix elements of the currents via on-shell amplitudes, by contracting them with states having well-defined polarisation. The relevant Dirac structures $\Gamma_{V(A, T, T5),i}^{\alpha\mu}$ are listed in Appendix \ref{app:ff-details}.
Then, the on-shell amplitudes are defined as
\begin{equation}
    \label{hel-amp}
    \mathcal{A}_\Gamma(s_b, s_c, \lambda_c, \lambda_q)
        \equiv \bra{\Lambda_c^*(s_c, \eta(\lambda_c))}\bar{c} \  \Gamma^\mu \eps_\mu^*(\lambda_q)b\ket{\Lambda_b(s_b)}\, .
\end{equation}
where for the SM-like currents $\Gamma^{\mu}=\gamma^{\mu},\gamma^{\mu}\gamma_5$, eq. \eqref{hel-amp} coincide with helicity amplitudes and the polarisation vectors $\eps_\mu^*(\lambda_q)$ constitutes a basis for the exchange of a virtual $W$ boson with polarisation  \mbox{$\lambda_q \in \lbrace t, 0,
+1, -1\rbrace$} (see \cref{app:kin}). However, in the case where the decay is mediated by the (pseudo)-tensor current $\Bar{c}\sigma^{\mu\nu}b$ ($\Bar{c}\sigma^{\mu\nu}\gamma_5b$), it becomes necessary to establish a basis for the exchange of a fictitious particle with $J^{P}=1^-\bigoplus1^+$, momentum $q^\mu$ and mass $\sqrt{q^2}$. As outlined in \cite{Papucci:2021pmj}, a suitable representation for the polarisation is given by $q_{\nu}\varepsilon_{\mu}(\lambda_q)$.

In order to systematically calculate the full set of form factors, we match the on-shell amplitudes with fixed total angular momentum $J$ of the $\Lambda_c^*$ state in the full QCD theory to the HQET ones. More details on the computation of the on-shell amplitudes can be found in \cref{app:ff-details}. The next sections will provide an overview of the HQET and a detailed analysis of the matrix elements that are needed to express the on-shell amplitudes up to the next-to-next-to leading order within the HQE framework.

\subsection{Heavy Quark Effective Theory in a nutshell}
\label{HQETreview}
We now provide a concise overview of the fundamentals of the HQET framework that we will use throughout this work. We will employ the notation of the references \cite{Neubert:1993mb,Bernlochner:2022ywh}. 
At the heart of the HQET there is a heavy quark $Q$ with mass $m_Q$, inside a hadron almost on-shell. This permits the following decomposition in terms of the large and small component fields $Q_+$ and $Q_-$:
\begin{equation}
	\label{eqn:masssub}
	Q^v_{+}(x) = e^{im_Q v \cdot x}\, \Pi_{+} Q(x)\, , \quad Q^v_{-}(x) = e^{im_Q v \cdot x}\, \Pi_{-} Q(x)\, ,
\end{equation}
with $\Pi_{\pm} = (1 \pm \vslash)/2$  projector operators,  $v$ denoting the heavy quark velocity which is defined up to reparametrisations. In fact the effective Lagrangian is invariant under $v \to v^{\prime} = v + k/m_Q$, 
where $k$ is a residual momentum of order $ \lqcd$ \cite{Neubert:1993mb,Luke:1992cs}.

The QCD Lagrangian for the heavy quark decomposes as follows:
\begin{equation}
	\label{eqn:fullL}
	\mathcal{L}_{\text{QCD}} 
	= \Qbar^v_+ i \vcD Q^v_+ + \Qbar^v_+ i \Dslash_\perp Q^v_-  
	+ \Qbar^v_- i \Dslash_\perp Q^v_+ - \Qbar^v_- (i \vcD + 2m_Q) Q^v_-\,,
\end{equation}
where $D^\mu$ is the gauge covariant derivative of QCD, 
and $D^\mu_\perp = D^\mu - (\vcD)v^\mu$ denotes the transverse derivative orthogonal to the heavy quark velocity.  $Q_+^v$ represents the nearly onshell degree of freedom while $Q_-^v$ corresponds to fluctuations suppressed by the heavy quark mass. Consequently, integrating out $Q_-^v$ 
yields an effective theory for the large component $Q_+^v$ of the heavy quark field, which reads
\begin{equation}
	\label{eqn:eftL}
	\mathcal{L}_{\text{HQET}} = \Qbar^v_+ i \vcD Q^v_+ +  \Qbar^v_+ i\Dslash_\perp \frac{1}{i \vcD + 2m_Q}  i\Dslash_\perp Q^v_+\, .
\end{equation}
This form is well-suited for an expansion in terms of the inverse of (twice) the heavy quark mass given by 
\begin{equation}
\label{HQETlag}
    \mathcal{L}_{\text{HQET}}  = \sum_{n = 0}\mathcal{L}_n/(2m_Q)^n \ .
\end{equation}
By expanding eq. \eqref{HQETlag} up to the second order, one finds \cite{Neubert:1993mb,Falk:1992ws,Bernlochner:2022ywh}
\begin{align}
\label{lagrcorr}
	\mathcal{L}_0 & = \Qbar^v_+ i \vcD Q^v_+ \,,\\
	\mathcal{L}_1 &  =  -\Qbar^v_+ \bigg[D^2 + Z\frac{g}{2} \sigma_{\ab} G^{\ab} \bigg]Q^v_+ = \mathcal{L}_{\text{kin}}+ \mathcal{L}_{\text{mag}}\,,\\
	\mathcal{L}_2 &  =  g\Qbar^v_+ \bigg[Z_1 v_\beta  D_\alpha G^{\ab}  - i Z_2 v_\alpha \sigma_{\beta\gamma} D^\gamma G^{\ab} \bigg]Q^v_+ \,.
\end{align}
Here, $ig\, G^{\ab} = [D^\alpha, D^\beta]$ is the field strength and $\sigma^{\ab} \equiv \frac{i}{2}[\g^\alpha, \g^\beta]$.
The coefficients $Z$, $Z_1$ and $Z_2$ are renormalisation factors.
Their deviation from unity becomes significant when addressing corrections at $\mathcal{O}(\aS/m_{Q})$ and beyond (see Refs.~\cite{Neubert:1993mb,Eichten:1990vp,Falk:1990pz,Falk:1992wt}). For the scope of our study, these deviations can be disregarded.

Furthermore, a source term $\Jbar Q$ for the full QCD  can now be written in terms of $Q^v_\pm$ as 
\begin{equation}
	\label{eqn:JQexp}
	\Jbar Q  \equiv \Jbar^v \mJ_{\text{HQET}} Q^v_+ = \Jbar^v \bigg[1 + \frac{1}{i \vcD + 2m_Q}i\Dslash_\perp \bigg] Q^v_+  =  \Jbar^v \bigg[1 + \Pi_-\bigg(\frac{i\Dslash}{2m_Q} - \frac{\Dslash\Dslash}{4m_Q^2} + \ldots\bigg)\bigg] Q^v_+\, ,
\end{equation}
where $J^v = e^{im_Q v\cdot x}J$.\\
Identifying the coefficients $\mJ_n$ in the current $\mJ_{\text{HQET}} = 1 + \Pi_-\sum_{n = 1}\mJ_n/(2m_Q)^n$ with the expansion above we get, up to second order \cite{Bernlochner:2022ywh}
\begin{align}
 &\mJ_1 = i \Dslash \, ,\\ 
 &\mJ_2 = -\Dslash \Dslash \ .
\end{align} 
The conjugate is defined as
\begin{equation}
	\mJbar_n \equiv \gamma^0 \overleftarrow{\mJ}_n^\dagger \gamma^0\,,
\end{equation}
with $\overleftarrow{\mJ}$ ($\overrightarrow{\mJ}$) indicating the action of the derivatives to the left and right, respectively.
 
In this study, we focus on the computation of matrix elements for hadrons undergoing  a $b \to c$ transition ($\Hb \to \Hc$). The QCD matrix elements are denoted as $\langle \Hc | \cbar \Gamma b | \Hb \rangle$, where $\Gamma$ represents a generic Dirac matrix. The states $\ket{\Hc}$ and $\ket{\Hb}$ correspond to the QCD baryon states and are normalized as:

\begin{equation}
    \langle H_{b(c)}(p')|H_{b(c)}(p)\rangle=2p^0(2\pi)^{3}\delta^{(3)}(\Vec{p}-\Vec{p'}) \ .
\end{equation}

The matching onto HQET up to second-order is \cite{Bernlochner:2022ywh}
\begin{align}
	\frac{\langle \Hc | \cbar \,\Gamma\, b | \Hb \rangle}{\sqrt{m_{\Hc} m_{\Hb}}}  
	& \simeq \big\langle \Hc^{v'} \big| \Bar{c}^{v'}_{+} \, \Gamma \, b^{v}_{+} \big| \Hb^v \big\rangle  \label{eqn:QCDmatchexp} +\\
	& + \frac{1}{2m_c} \big\langle \Hc^{v'} \big|  \big(\Bar{c}^{v'}_{+}\Overleftarrow{\mJbar}^{\prime}_1+ \mathcal{L}'_1 \circ c^{v'}_{+} \big) \Gamma \, b^{v}_{+} \big| \Hb^v \big\rangle \nn+ \\
	& + \frac{1}{2m_b} \big\langle \Hc^{v'} \big| \Bar{c}^{v'}_{+}  \, \Gamma  \big(\Overrightarrow{\mJ}_1 b^{v}_{+} + b^{v}_{+} \circ \mathcal{L}_1 \big)  \big| \Hb^v \big\rangle \nn +\\
	& + \frac{1}{4m_c^2} \big\langle \Hc^{v'} \big| \big(\Bar{c}^{v'}_{+}\Overleftarrow{\mJbar}^{\prime}_2\Pi_-' + \mathcal{L}'_2 \circ \Bar{c}^{v'}_{+}
		+ \mathcal{L}'_1 \circ \Bar{c}^{v'}_{+}\Overleftarrow{\mJbar}^{\prime}_1\Pi_-'
		+ \frac{1}{2}\,\mathcal{L}'_1 \circ \mathcal{L}'_1 \circ \Bar{c}^{v'}_{+}\big) \Gamma \, b^{v}_{+} \big| \Hb^v \big\rangle \nn +\\
	& + \frac{1}{4m_b^2} \big\langle \Hc^{v'} \big| \Bar{c}^{v'}_{+} \, \Gamma \big(\Pi_-\Overrightarrow{\mJ}_2 b^{v}_{+}  + b^{v}_{+} \circ \mathcal{L}_2
		+ \Pi_-\Overrightarrow{\mJ}_1 b^{v}_{+} \circ \mathcal{L}_1 
		+ \frac{1}{2}\,b^{v}_{+} \circ \mathcal{L}_1 \circ \mathcal{L}_1 \big) \big| \Hb^v \big\rangle \nn+ \\
	&+ \frac{1}{4m_c m_b} \big\langle \Hc^{v'} \big| \big(\Bar{c}^{v'}_{+}\Overleftarrow{\mJbar}^{\prime}_1 + \mathcal{L}'_1 \circ \Bar{c}^{v'}_{+} \big)
		\Gamma \big(\Overrightarrow{\mJ}_1 b^{v}_{+} + b^{v}_{+} \circ \mathcal{L}_1 \big) \big| \Hb^v \big\rangle\,, \nn
\end{align}
where $b_{+}^{v}$ and $c_{+}^{v'}$ are the large components of the quark fields. We adopt the convention where terms involving the $c$-quark are denoted with primes, while those involving the $b$-quark remain unprimed. The matrix elements on the right hand side are evaluated on the HQET states, which are defined as the eigenstates of $\mathcal{L}_0$ and normalized as follows:
\begin{equation}
    \langle H^{v'}(p')|H^{v}(p)\rangle=2v^0\delta_{vv^{'}}(2\pi)^3\delta^{(3)}(\Vec{p}-\Vec{p'}) \ .
\end{equation}

Non-local terms, arising from the operator product with the HQET Lagrangian, emerge from the matching procedure, because the states in the full theory differ from those in the effective theory, as highlighted in \cite{Neubert:1993mb}. Here, following \cite{Bernlochner:2022ywh} $\circ$ symbol denotes such operator product, for example when we write above the term $ \mathcal{L}'_1 \circ \Bar{c}_{+}^{v'}(z)$ this means: 
\begin{equation}
	 \mathcal{L}'_1 \circ \Bar{c}_{+}^{v'}(z)  = i\int d^4 x \,  \mathcal{L}'_1(x) \Bar{c}_{+}^{v'}(z) \nn \ .
\end{equation}

\section{Next-to-leading order corrections to  $\Lambda_b\to \Lambda_c^*$ warm-up}
\label{sec:firstorder}
In this section we re-derive the baryon matrix elements for the transition $\Lambda_b\to \Lambda_c^*$ employing  the HQE eq. \eqref{eqn:QCDmatchexp} up to $\mathcal{O}(\alpha_s,1/m_{b,c})$. 

Apart from computing the vector and axial current matrix elements that were already presented in \cite{Boer:2018vpx}, we also perform a  determination of the tensor and pseudo-tensor current matrix elements. Furthermore, we consider the effects of the chromomagnetic operator $\mathcal{L}_{\text{mag}}$, that were neglected in   \cite{Boer:2018vpx}.

Starting from eq. \eqref{eqn:QCDmatchexp},  the hadronic matrix  in the HQE  can be written at leading order (LO) as: 
\begin{equation}
\label{eq:hqet_expansion}
    \bra{\Lambda_c^{*}(k, \eta, s_c)} \bar{c}^{v'}_{+} \Gamma b^{v}_{+} \ket{\Lambda_b(p, s_b)}
    = \sqrt{4} \bar{u}_\alpha(\mLamCst v', \eta, s_c) \Gamma u(\mLamB v, s_b) \zeta^\alpha(w)\,,
\end{equation}
where $w \equiv v\cdot v'= (\mLamB^2+\mLamCst^2-q^2)/(2 \mLamB\mLamCst)$,
$v$ and $v'$ are the four-velocities of the initial and final states, respectively,
and $\Gamma$ denotes a given Dirac structure.
 Following \cite{Falk:1991nq},  the light-state transition
amplitude $\zeta^{\alpha}(w)$ can be written as
\begin{equation}
    \zeta^\alpha(w) = \zeta(w) (v - v')^\alpha\,.
\end{equation}
Therefore, at leading order in the HQE all the form factors can be expressed in terms of the single amplitude $\zeta(w)$, independently from the Dirac structure of the current.

The  local $1/m$ corrections to the current induce a shift that we can account for by operating the following substitution:
\begin{equation}
    \Gamma^{\mu} \to \Gamma^{\mu} + \varepsilon_b\Delta J^{\mu}_{\Gamma}+\varepsilon_c\Delta \genbar{J}^{\mu}_{\Gamma} \, ,
\end{equation}
where $\varepsilon_{b,c}=1/(2m_{b,c})$ are the HQE parameters and 
\begin{align}
\label{eq:m_shift1}
   \Delta J_\Gamma^{\mu} = \genbar{c}^{v'}_{+} \Gamma^{\mu} \mJ_{1}b^{v}_{+} \, ,\\
   \label{eq:m_shift2}
  \Delta \genbar{ J}_\Gamma^{\mu} =  \genbar{c}^{v'}_{+}\mJbar_{1} \Gamma^{\mu} b^{v}_{+} \ .
\end{align}
The hadronic matrix elements of $\Delta J_\Gamma^{\mu}$ and $\Delta \genbar{J}_\Gamma^{\mu}$
can be parametrised as:
\begin{align}
\label{currentcorrection}
& \bra{\Lambda_c^{*}(k, \eta, s_c)}\Delta J_\Gamma^{\mu} \ket{\Lambda_b(p, s_b)}=\sqrt{4} \genbar{u}_\alpha(\mLamCst v', \eta, s_c)\Gamma^{\mu}\gamma_{\beta}u(\mLamB v, s_b)  \zeta_{b}^{\alpha\beta}(w)\,, \\
& \bra{\Lambda_c^{*}(k, \eta, s_c)}\Delta \genbar{J}_\Gamma^{\mu} \ket{\Lambda_b(p, s_b)}=\sqrt{4} \genbar{u}_\alpha(\mLamCst v', \eta, s_c)\gamma_{\beta}\Gamma^{\mu}u(\mLamB v, s_b)  \zeta_{c}^{\alpha\beta}(w)\,,
\end{align}
where the most general decomposition of the light-state transition amplitude is
\begin{equation}
\zeta^{\alpha\beta}_{(q)}(w)=(v-v')^{\alpha}\left[\zeta_{1}^{(q)}(w)v^{\beta}+\zeta^{(q)}_{2}(w)v'^{\beta}\right]+g^{\alpha\beta}\zeta^{(q)}_3(w)\,.
\end{equation}

At this stage the $1/m$ corrections are described by six subleading Isgur-Wise (IW) functions, however they can be related via the equations of motion.
In particular we have that $v_{\beta}\zeta^{\alpha\beta}_{(b)}=0$, and
$v'_{\beta}\zeta^{\alpha\beta}_{(c)}=0$. This leads to the following relations:
\begin{align}
\zeta_1^{(b)}(w)+w\zeta_{2}^{(b)}(w)+\zeta_{3}^{(b)}(w)&=0 \,,\\
w\zeta_1^{(c)}(w)+\zeta_{2}^{(c)}(w)&=0 \,.
\end{align}

Furthermore from the modified Ward identity \cite{Boer:2018vpx,Bernlochner:2022ywh}
\begin{equation}
i\partial_\alpha[\Qbar^{v'}_+\Gamma Q^v_+]=\Qbar^{v'}_+ i \overset{\leftarrow}{D}_{\alpha}\Gamma Q^v_+ +\Qbar^{v'}_+\Gamma iD_{\alpha} Q^v_+ \, ,
\end{equation}

we recover the following relations among leading and subleading IW functions:
\begin{align}
\zeta_{1}^{(b)}(w)+\zeta_{1}^{(c)}(w)&=\bar{\Lambda}\zeta(w) \,, \\
\zeta_{2}^{(b)}(w)+\zeta_{2}^{(c)}(w)&=-\bar{\Lambda}'\zeta(w) \,, \\
\zeta_{3}^{(b)}(w)+\zeta_{3}^{(c)}(w)&=0\,,
\end{align}
where the important parameters $\Bar{\Lambda}$, $\Bar{\Lambda}^{\prime}$ represent the difference between the mass of the baryon states and that of the heavy quark at leading order in HQE:
\begin{equation}
    \begin{aligned}
        &\Bar{\Lambda}=\mLamB-m_{b} \ , \\
        &\Bar{\Lambda}^{\prime}=\mLamCst-m_{c} \ .
    \end{aligned}
\end{equation}
Since the phase factor in eq. \eqref{eqn:masssub} effectively
removes the mass of the heavy quark $m_Q$ from the states, these parameters control the space-time dependence of the HQET states, as discussed in \cite{Neubert:1993mb,Falk:1992ws}.
The relations above reduce the initial six subleading IW functions
to one independent subleading IW function. Inspired by \cite{Boer:2018vpx}, we choose to express all in terms of $\zeta$ and $\zeta_{\text{SL}}\equiv\zeta_3^{(b)} = - \zeta_3^{(c)}$:
\begin{equation}
    \begin{aligned}
    \label{systIW}
        \zeta_1^{(b)} & = -\frac{\zeta_{\text{SL}}}{1 - w^2} + \frac{w \, \zeta}{1 - w^2}\left(\bar{\Lambda}' - \bar{\Lambda} w\right)\,,  \ & \   \zeta_2^{(b)} & = +\frac{w \, \zeta_{\text{SL}}}{1 - w^2} - \frac{\zeta}{1 - w^2}\left(\bar{\Lambda}' - \bar{\Lambda} w\right)\,,\\
        \zeta_1^{(c)} & = +\frac{\zeta_{\text{SL}}}{1 - w^2} - \frac{\zeta}{1 - w^2}\left(w \bar{\Lambda}' - \bar{\Lambda}\right)\,, \  & \ \zeta_2^{(c)} & = -\frac{w\, \zeta_{\text{SL}}}{1 - w^2} + \frac{w \zeta}{1 - w^2}\left(w \bar{\Lambda}' - \bar{\Lambda}\right)\,.
    \end{aligned}
\end{equation}

The relations among the IW functions, obtained via the equations of motions and the modified Ward identity, are singular at $w=1$. In fact, for $w=1$  we need to use, instead, the relation
\begin{equation}
    \zeta_{\text{SL}}+\zeta(\bar{\Lambda}-\bar{\Lambda}')=0 \ .
\end{equation}
This equation permits to verify that the HQE form factors obey, up to second order, the endpoint relations reported in \cite{Hiller:2013cza,Hiller:2021zth}. 

Before going into the details of the matrix element calculations, it is worth mentioning that we have  cross checked the outcomes presented in reference \cite{Boer:2018vpx} pertaining to both the axial and vector form factors. Furthermore, our findings coincide with the results presented in \cite{Papucci:2021pmj}, including the tensor and pseudo-tensor form factors up to the next-to leading order in the HQE as well as the chromomagnetic contributions.
We note that in \cite{Papucci:2021pmj} the authors employed the form factor basis introduced in \cite{Leibovich:1997az} with  $\zeta$ and $\zeta_1^{(c)}$ chosen to be independent IW functions.

We are now ready to investigate the implications stemming from the non-local $1/m$ corrections and the hard gluon exchange.

\subsection{Vector and Axial current matrix elements}
\label{sec:axialvector}
In order to include both local $1/m$ and $\alpha_s$ corrections to the vector current matrix element, we use the approach introduced in Ref. \cite{Boer:2018vpx}, employing the following substitution:
\begin{equation}
\label{eq:hqet_vect_current}
\gamma^{\mu}\to J^{\mu}_V=C_1(w) \gamma^{\mu} + C_2(w) v^\mu + C_3(w) v'^\mu +\varepsilon_b\Delta J_V^{\mu}+\varepsilon_c\Delta \genbar{ J}_V^{\mu}
+\mathcal{O}(\alpha_s / m, 1/m^2)\, .
\end{equation}
The corrections $\Delta J_V^\mu$ are given by eqs. \eqref{eq:m_shift1} and \eqref{eq:m_shift2}, with $\Gamma^\mu=\gamma^{\mu}$ and $C_i$ are the Wilson coefficients  obtained by matching the HQET to QCD at the energy scale $\mu=\sqrt{m_c m_b}$ \cite{Papucci:2021pmj}.
The Wilson coefficients depend on  $w = v \cdot v'$, which is the recoil parameter.
The expression for the Wilson coefficients $C_i$ up to $\mathcal{O}(\alpha_s)$ can be found in \cite{Bernlochner:2017jka}. In our notation, the Wilson coefficients are related to those in Ref. \cite{Bernlochner:2017jka} by: $C_i=\hat{\alpha}_s C_{V_{i}}+\delta_{i1}$, where $\hat{\alpha}_s=\alpha_s/\pi$. 

In eq. \eqref{eq:hqet_vect_current} we consider only the local power corrections $\Delta J_V^{\mu}$ and $\Delta \genbar{J}_V^{\mu}$. Other $1/m$ local operators arise in the HQE from QCD hard corrections and then contribute at the order $\alpha_s /m$.
Since they are beyond the precision we aim for in the analysis proposed in this work, we neglect them.

From eq. \eqref{currentcorrection}  we re-obtain the following contributions already present in \cite{Boer:2018vpx}  \\
\begin{equation}
\begin{aligned}
 \bra{\Lambda_c^{*}(k, \eta, s_c)}    \Delta J^\mu_{V} \ket{\Lambda_b(p, s_b)}
         = &2\genbar{u}_\alpha(\mLamCst v', \eta, s_c)\gamma^{\mu}u(\mLamB v, s_b) v^\alpha\left(\zeta^{(b)}_{1}(w)-\zeta^{(b)}_{2}(w)\right)+ \\
        +&4\genbar{u}_\alpha(\mLamCst v', \eta, s_c) u(\mLamB v, s_b)v^{\alpha}v'^{\mu}\zeta^{(b)}_{2}(w)+ \\
         +&2\genbar{u}_{\alpha}(\mLamCst v', \eta, s_c)\gamma^\mu\gamma^\alpha u(\mLamB v, s_b)\zeta^{(b)}_3(w)\,, \\
 \bra{\Lambda_c^{*}(k, \eta, s_c)}    \Delta \genbar{ J}_V^{\mu} \ket{\Lambda_b(p, s_b)}
        = &2\genbar{u}_\alpha(\mLamCst v', \eta, s_c)\gamma^{\mu}u(\mLamB v, s_b) v^\alpha \left(\zeta^{(c)}_{2}(w)-\zeta^{(c)}_{1}(w)\right)+ \\
         +&4\genbar{u}_\alpha(\mLamCst v', \eta, s_c) u(\mLamB v, s_b)v^{\alpha}v^{\mu}\zeta^{(c)}_{1}(w) +\\
          +&2\genbar{u}_\alpha(\mLamCst v', \eta, s_c) \gamma^\alpha\gamma^\mu u(\mLamB v, s_b)\zeta^{(c)}_{3}(w)\,.
\end{aligned}
\end{equation}
The same procedure is applied to the axial-vector current yielding \cite{Boer:2018vpx}:
\begin{equation}
\label{eq:hqet_axial_current}
\gamma^{\mu}\gamma_5\to    J^{\mu}_A=C_1^{(5)}(w)\gamma^{\mu}\gamma^5+ C_2^{(5)}(w) v^\mu\gamma^5 + C_3^{(5)}(w) v'^\mu\gamma^5 +\varepsilon_b \Delta J_A^{\mu}+\varepsilon_c \Delta \genbar{J}_A^{\mu}+\mathcal{O}(\alpha_s / m, 1/m^2)\, .
\end{equation}

The Wilson coefficients $C^{5}_{i}$ are related to those reported in \cite{Bernlochner:2017jka} by $C^{5}_{i}=\hat{\alpha}_sC_{A_{i}}+\delta_{i1}$.  The $1/m$ pure current corrections  $\Delta J_A^{\mu}$ and $\Delta \genbar{J}_A^{\mu}$ can be computed from eq. \eqref{currentcorrection} with $\Gamma^\mu= \gamma^\mu \gamma_5$. We obtain:
\begin{equation}
\begin{aligned}
\bra{\Lambda_c^{*}(k, \eta, s_c)} \Delta J_{A}^{\mu} \ket{\Lambda_b(p, s_b)}=&2\bar{u}_\alpha(\mLamCst v', \eta, s_c)\gamma^{\mu}\gamma^5 u(\mLamB v, s_b)v^\alpha\left(\zeta^{(b)}_{1}(w)+\zeta^{(b)}_{2}(w)\right) -\\
 -&4\bar{u}_\alpha(\mLamCst v', \eta, s_c)\gamma^{5} u(\mLamB v, s_b)v^{\alpha}v'_{\mu}\zeta^{(b)}_{2}(w) +\\
 +&2\bar{u}_{\alpha}(\mLamCst v', \eta, s_c)\gamma^{\mu}\gamma^5\gamma^{\alpha}u(\mLamB v, s_b)\zeta^{(b)}_3(w)\,, \\
\bra{\Lambda_c^{*}(k, \eta, s_c)} \Delta \genbar{J}_{A}^{\mu} \ket{\Lambda_b(p, s_b)}=&2\bar{u}_\alpha(\mLamCst v', \eta, s_c)\gamma^{\mu}\gamma^{5}u(\mLamB v, s_b)v^\alpha\left(\zeta^{(c)}_{1}(w)+\zeta^{(c)}_{2}(w)\right)+ \\
 +&4\bar{u}_\alpha(\mLamCst v', \eta, s_c) \gamma^5 u(\mLamB v, s_b)v^{\alpha}v_{\mu}\zeta^{(c)}_{1}(w) +\\
  +&2\bar{u}_{\alpha}(\mLamCst v', \eta, s_c)\gamma^{\alpha}\gamma^{\mu}\gamma^5u(\mLamB v, s_b)\zeta^{(c)}_3(w)\,. 
\end{aligned}
\end{equation}

Besides the contributions from local operators describing the corrections to the infinite mass limit of the Heavy Quark (HQ) currents, also the effects from
non-local insertions of the HQET Lagrangian at power $1/m$ in the HQ currents need to be included. 

Non-local insertions of
the kinetic operator can be parametrised as:
\begin{align}
&\bra{\Lambda_c^{*}(k, \eta, s_c)} \mathcal{L}_\text{kin}\circ \genbar{c}^{v'}_{+}\Gamma b^{v}_{+}  \ket{\Lambda_b(p, s_b)}=\sqrt{4}\eta^{(b)}_\text{kin}(w)\,v^{\alpha}\genbar{u}_\alpha(\mLamCst v', \eta, s_c)\Gamma u(\mLamB v, s_b) \, , \\
&\bra{\Lambda_c^{*}(k, \eta, s_c)} \mathcal{L}^{\prime}_\text{kin}\circ \genbar{c}^{v'}_{+}\Gamma b^{v}_{+}  \ket{\Lambda_b(p, s_b)}=\sqrt{4}\eta^{(c)}_\text{kin}(w)\,v^{\alpha}\genbar{u}_\alpha(\mLamCst v', \eta, s_c)\Gamma u(\mLamB v, s_b) \, ,
\end{align}
with $\Gamma=\gamma^{\mu}, \gamma^{\mu}\gamma_{5}$.

Clearly, they give rise to a $w$-dependent shift $\eta^{(b,c)}_\text{kin}(w)$
to the leading-power IW function $\zeta(w)$. At order $\mathcal{O}(\alpha_s, 1/m)$ we can absorb this shift into the
definition of $\zeta$:
\begin{equation}
    \zeta(w) + \varepsilon_c \eta^{(c)}_\text{kin}(w) + \varepsilon_b \eta^{(b)}_\text{kin}(w)
    \to \zeta(w)\,.
\end{equation}

Nevertheless, as stressed in \cite{Bernlochner:2017jka,Leibovich:1997az}, the above redefinition leads to corrections of order $1/m^2$ which we need to take into account if we aim to explore the second order of the HQE.  

Non-local insertion of the chromomagnetic operator gives two contributions which are:
\begin{align}
\label{eq:eta_mag_b}
     &\bra{\Lambda_c^{*}(k, \eta, s_c)} \mathcal{L}_\text{mag}\circ \genbar{c}^{v'}_{+}\Gamma b^{v}_{+}  \ket{\Lambda_b(p, s_b)} = \eta_\text{mag}^{(b)}(w)\,g_{\mu \alpha} v'_\nu \genbar{u}^\alpha(\mLamCst v', \eta, s_c) \   \Gamma \Pi_+ \  i \sigma^{\mu\nu} u(\mLamB v, s_b)\,, \\
     \label{eq:eta_mag_c}
  & \bra{\Lambda_c^{*}(k, \eta, s_c)} \mathcal{L}^{\prime}_\text{mag}\circ \genbar{c}^{v'}_{+}\Gamma b^{v}_{+}  \ket{\Lambda_b(p, s_b)} = \eta_\text{mag}^{(c)}(w)\,g_{\mu \alpha} v_\nu\genbar{u}^\alpha(\mLamCst v', \eta, s_c) \  i \sigma^{\mu\nu} \Pi^{\prime}_+ \Gamma u(\mLamB v, s_b) \ .
\end{align}

The chromomagnetic correction to the $b$ vector current is given by:
\begin{equation}
    \varepsilon_b\eta_\text{mag}^{(b)}(w) \left[2(w-1)\genbar{u}_{\alpha}g^{\mu\alpha} u-(w-1)\genbar{u}_{\alpha}\gamma^\alpha\gamma^\mu u +2\genbar{u}_{\alpha} \gamma^\alpha v'^\mu u - 2\genbar{u}_{\alpha}v^\alpha v'^\mu  u + \genbar{u}_{J \,\alpha}v^{\alpha}\gamma^{\mu}u \right].
\end{equation}

Analogously, the chromomagnetic correction to the $c$ vector current is given by:
\begin{equation}
    \varepsilon_c \eta_\text{mag}^{(c)}(w)\left[(1-w)\genbar{u}_{\alpha}\gamma^\alpha\gamma^\mu u -2\genbar{u}_{\alpha} \gamma^\alpha v^\mu u + \genbar{u}_{J \,\alpha}v^{\alpha} \gamma^{\mu}u \right].
\end{equation}

The chromomagnetic correction to the $b$ and $c$ axial currents are respectively:
\begin{equation}
\begin{split}
    \varepsilon_b\eta_\text{mag}^{(b)}(w) \{-2(1+w)\genbar{u}_{\alpha}g^{\mu\alpha}\gamma_5 u+(1+w)\genbar{u}_{\alpha}\gamma^\alpha\gamma^\mu\gamma_5 u+\\
    +2\genbar{u}_{\alpha}\gamma^\alpha v'^\mu  \gamma_5 u + 2\Bar{u}_{J \, \alpha}v^\alpha v'^\mu  \gamma_5 u - \genbar{u}_{J \,\alpha} v^{\alpha} \gamma^{\mu}\gamma_5 u \} \ , \\
    \end{split}
\end{equation}
\begin{equation}
    \varepsilon_c \eta_\text{mag}^{(c)}(w)\left[-(w+1)\genbar{u}_{\alpha}\gamma^\alpha \gamma^\mu \gamma_5 u -2\genbar{u}_{\alpha}\gamma^\alpha v^\mu  \gamma_5 u + \genbar{u}_{J \,\alpha} v^{\alpha} \gamma^{\mu}\gamma_5 u \right].
\end{equation}

Here we went beyond the result of \cite{Boer:2018vpx} with respect to the determination of the non-local magnetic contributions which, however, in another basis were determined in \cite{Papucci:2021pmj}. 

\subsection{ Vector and axial form factors in the VC limit}
\label{sec:vector_axial_form_factor}
By virtue of the identity $v_{\alpha}(1+\vslash)\sigma^{\alpha\beta}(1+\vslash)=0$ the chromomagnetic correction is zero in the exact zero recoil limit ($w=1$) and, as argued in \cite{Neubert:1992hb,Neubert:1992wq,Neubert:1992pn}, is expected to be numerically small with respect to the local current corrections. Then, as a first approximation, the computation of the form factors can be performed by considering the Vanishing Chromomagnetic limit (VC) (as also motivated in \cite{Falk:1992wt,Leibovich:1997az,Bernlochner:2022ywh}) in which we set the field strength $G_{\alpha \beta }=0$ and the chromomagnetic correction results automatically to zero.

Furthermore, at order $\mathcal{O}(1/m)$, the correction to the leading order HQ amplitude deriving from the insertion of the kinetic lagrangian can be reabsorbed through a redefinition of the IW function $\zeta$. Since we aim at extending the HQE to order $1/m_{c}^{2}$, we must take into account the contribution to $\mathcal{O}(1/m_c)$  of the kinetic operator, whereas the $\mathcal{O}(1/m_b)$  kinetic operator contribution can still be reabsorbed into the definition of the leading IW function $\zeta$ because otherwise it appears at order $1/m_{b}^{2}$ that is neglected here.

To the order we are working here theoretical consistency requires to use directly the quark masses in the HQE for the form factors rather than approximate them by the baryon masses while still agreeing with the result of  \cite{Boer:2018vpx} valid to the next leading order. Nevertheless, we have also explicitly checked the results against the case in which the baryon masses are used also in the HQE as we shall discuss in the section about the results.

Concerning the final state $\LamCst[2595]$ for the vector current the form factors are:

\begin{align}
    f_{1/2,0}=& \frac{\zeta s_- \sqrt{s_+}}{2(\mLamB \mLamCst)^{3/2}}  \left(C_1+\frac{C_2 s_+}{2 \mLamB (\mLamB+\mLamCst)}+\frac{C_3 s_+}{2 \mLamCst (\mLamB+\mLamCst)}\right)+\nn \\+&\varepsilon_b \frac{ \sqrt{s_+} (\mLamB-\mLamCst)}{(\mLamB+\mLamCst)\sqrt{\mLamB \mLamCst}}  \left(\zeta \bar\Lambda'- 2 \zeta_{\text{SL}}-\frac{\zeta\bar\Lambda \left(\mLamB^2+\mLamCst^2-q^2\right)}{2 \mLamB \mLamCst} \right)+\nn \\+&\varepsilon_c \frac{ \sqrt{s_+} (\mLamB-\mLamCst)}{(\mLamB+\mLamCst)\sqrt{\mLamB \mLamCst}}  \left(\zeta \bar\Lambda- 2 \zeta_{\text{SL}}-\frac{\zeta\bar\Lambda' \left(\mLamB^2+\mLamCst^2-q^2\right)}{2 \mLamB \mLamCst} \right)+\varepsilon_c\frac{\eta^{(c)}_\text{kin}  s_- \sqrt{s_+} }{2 (\mLamB \mLamCst)^{3/2}} \ , \\
    f_{1/2,t}=&\frac{\zeta s_+ \sqrt{s_-}}{2(\mLamB \mLamCst)^{3/2}}  \left(C_1+\frac{C_2 s_+}{2 \mLamB (\mLamB-\mLamCst)}+\frac{C_3 s_+}{2 \mLamCst (\mLamB-\mLamCst)}\right)+\nn \\+&\varepsilon_b \frac{ \sqrt{s_-} (\mLamB+\mLamCst)}{(\mLamB-\mLamCst)\sqrt{\mLamB \mLamCst}}  \left(\zeta \bar\Lambda'- 2 \zeta_{\text{SL}}-\frac{\zeta\bar\Lambda \left(\mLamB^2+\mLamCst^2-q^2\right)}{2 \mLamB \mLamCst} \right)+\nn \\+&\varepsilon_c \frac{ \sqrt{s_-} (\mLamB+\mLamCst)}{(\mLamB-\mLamCst)\sqrt{\mLamB \mLamCst}}  \left(\zeta \bar\Lambda- 2 \zeta_{\text{SL}}-\frac{\zeta\bar\Lambda' \left(\mLamB^2+\mLamCst^2-q^2\right)}{2 \mLamB \mLamCst} \right)+\varepsilon_c\frac{\eta^{(c)}_\text{kin}  s_+ \sqrt{s_-} }{2 (\mLamB \mLamCst)^{3/2}} \ , \\
    f_{1/2,\perp}=&\frac{C_1 \zeta  s_- \sqrt{s_+}}{2 (\mLamB \mLamCst)^{3/2}}+ \varepsilon_b \frac{\sqrt{s_+}}{\sqrt{\mLamB \mLamCst}} \left(\frac{\zeta \bar\Lambda  \left(\mLamB^2+\mLamCst^2-q^2\right)}{2 \mLamB \mLamCst}-\bar\Lambda'\zeta\right)+\nn \\+&\varepsilon_c \frac{\sqrt{s_+}}{\sqrt{\mLamB \mLamCst}} \left(\bar\Lambda\zeta-2\zeta_{\text{SL}}-\frac{\zeta \bar\Lambda'  \left(\mLamB^2+\mLamCst^2-q^2\right)}{2 \mLamB \mLamCst}\right)+\varepsilon_c\frac{\eta^{(c)}_\text{kin}  s_- \sqrt{s_+} }{2 (\mLamB \mLamCst)^{3/2}} \ ,
\end{align}
while for the axial form factors we obtain:
\begin{align}
    g_{1/2,0}=& \frac{\zeta s_+ \sqrt{s_-}}{2(\mLamB \mLamCst)^{3/2}}  \left(C^5_1-\frac{C^5_2 s_-}{2 \mLamB (\mLamB-\mLamCst)}-\frac{C^5_3 s_-}{2 \mLamCst (\mLamB-\mLamCst)}\right)+\nn \\+&\varepsilon_b \frac{ \sqrt{s_-} (\mLamB+\mLamCst)}{(\mLamB-\mLamCst)\sqrt{\mLamB \mLamCst}}  \left(\zeta \bar\Lambda'+2 \zeta_{\text{SL}}-\frac{\zeta\bar\Lambda \left(\mLamB^2+\mLamCst^2-q^2\right)}{2 \mLamB \mLamCst} \right)+\nn \\+&\varepsilon_c \frac{ \sqrt{s_-} (\mLamB-\mLamCst)}{(\mLamB+\mLamCst)\sqrt{\mLamB \mLamCst}}  \left(\zeta \bar\Lambda- 2 \zeta_{\text{SL}}-\frac{\zeta\bar\Lambda' \left(\mLamB^2+\mLamCst^2-q^2\right)}{2 \mLamB \mLamCst} \right)+\varepsilon_c\frac{\eta^{(c)}_\text{kin}  s_+ \sqrt{s_-} }{2 (\mLamB \mLamCst)^{3/2}}, \\
    g_{1/2,t}=&\frac{\zeta s_- \sqrt{s_+}}{2(\mLamB \mLamCst)^{3/2}}  \left(C^5_1+\frac{C^5_2 s_+}{2 \mLamB (\mLamB+\mLamCst)}+\frac{C^5_3 s_+}{2 \mLamCst (\mLamB+\mLamCst)}\right)+\nn \\+&\varepsilon_b \frac{ \sqrt{s_+} (\mLamB+\mLamCst)}{(\mLamB-\mLamCst)\sqrt{\mLamB \mLamCst}}  \left(\zeta \bar\Lambda'+ 2 \zeta_{\text{SL}}-\frac{\zeta\bar\Lambda \left(\mLamB^2+\mLamCst^2-q^2\right)}{2 \mLamB \mLamCst} \right)+\nn \\+&\varepsilon_c \frac{ \sqrt{s_+} (\mLamB+\mLamCst)}{(\mLamB-\mLamCst)\sqrt{\mLamB \mLamCst}}  \left(\zeta \bar\Lambda- 2 \zeta_{\text{SL}}-\frac{\zeta\bar\Lambda' \left(\mLamB^2+\mLamCst^2-q^2\right)}{2 \mLamB \mLamCst} \right)+\varepsilon_c\frac{\eta^{(c)}_\text{kin}  s_- \sqrt{s_+} }{2 (\mLamB \mLamCst)^{3/2}} \ , \\
    g_{1/2,\perp}=&\frac{C^5_1 \zeta  s_+ \sqrt{s_-}}{2 (\mLamB \mLamCst)^{3/2}}+ \varepsilon_b \frac{\sqrt{s_-}}{\sqrt{\mLamB \mLamCst}} \left(\frac{\zeta \bar\Lambda  \left(\mLamB^2+\mLamCst^2-q^2\right)}{2 \mLamB \mLamCst}- \zeta \bar\Lambda'\right)+\nn \\+&\varepsilon_c \frac{\sqrt{s_-}}{\sqrt{\mLamB \mLamCst}} \left(\zeta\bar\Lambda-2\zeta_{\text{SL}}-\frac{\zeta \bar\Lambda'  \left(\mLamB^2+\mLamCst^2-q^2\right)}{2 \mLamB \mLamCst}\right)+\varepsilon_c\frac{\eta^{(c)}_\text{kin}  s_+ \sqrt{s_-} }{2 (\mLamB \mLamCst)^{3/2}} \ .
\end{align}

Considering the final state $\LamCst[2625]$, the vector form factors are:
\begin{align}
    F_{1/2,0}=&  \frac{\zeta s_- \sqrt{s_+}}{2(\mLamB \mLamCst)^{3/2}}  \left(C_1+\frac{C_2 s_+}{2 \mLamB (\mLamB+\mLamCst)}+\frac{C_3 s_+}{2 \mLamCst (\mLamB+\mLamCst)}\right)+\nn \\+&\varepsilon_b \frac{ \sqrt{s_+} (\mLamB-\mLamCst)}{(\mLamB+\mLamCst)\sqrt{\mLamB \mLamCst}}  \left(\zeta \bar\Lambda'+  \zeta_{\text{SL}}-\frac{\zeta\bar\Lambda \left(\mLamB^2+\mLamCst^2-q^2\right)}{2 \mLamB \mLamCst} \right)+\nn \\+&\varepsilon_c \frac{ \sqrt{s_+} (\mLamB-\mLamCst)}{(\mLamB+\mLamCst)\sqrt{\mLamB \mLamCst}}  \left(\zeta \bar\Lambda+ \zeta_{\text{SL}}-\frac{\zeta\bar\Lambda' \left(\mLamB^2+\mLamCst^2-q^2\right)}{2 \mLamB \mLamCst} \right)+ \frac{\sqrt{s_+}s_-}{2(\mLamB\mLamCst)^{3/2}}\varepsilon_c\eta^{(c)}_{\text{kin}}, \\
    F_{1/2,t}=& \frac{\zeta s_+ \sqrt{s_-}}{2(\mLamB \mLamCst)^{3/2}} \left( C_1+\frac{\mLamCst(\mLamB^2-\mLamCst^2+q^2)}{2(\mLamB-\mLamCst)\mLamB \mLamCst}C_2+\frac{\mLamB(\mLamB^2-\mLamCst^2-q^2)}{2(\mLamB-\mLamCst)\mLamB \mLamCst}C_3\right)+\nn \\
    &+\varepsilon_b \frac{(\mLamB+\mLamCst)\sqrt{s_-}}{(\mLamB-\mLamCst)\sqrt{\mLamB \mLamCst}}\left(\zeta \bar\Lambda'+\zeta_{\text{SL}}-\frac{\zeta \bar\Lambda(\mLamB^2+\mLamCst^2-q^2)}{2\mLamB\mLamCst}\right) +\nn \\
    &+\varepsilon_c \frac{(\mLamB+\mLamCst)\sqrt{s_-}}{(\mLamB-\mLamCst)\sqrt{\mLamB \mLamCst}}\left(\zeta\bar\Lambda+\zeta_{\text{SL}}-\frac{\zeta \bar\Lambda'(\mLamB^2+\mLamCst^2-q^2)}{2\mLamB\mLamCst}\right)+\frac{\sqrt{s_-}s_+}{2(\mLamB\mLamCst)^{3/2}}\varepsilon_c\eta^{(c)}_{\text{kin}}\,, \\   
    F_{1/2,\perp}=& \frac{\zeta\,\sqrt{s_+}s_-}{2(\mLamB \mLamCst)^{3/2}}C_1+\varepsilon_b\zeta\frac{\sqrt{s_+}}{\sqrt{\mLamB\mLamCst}}\left(\frac{\bar\Lambda(\mLamB^2+\mLamCst^2-q^2)}{2\mLamB\mLamCst}-\bar\Lambda'\right)+ \nn \\
    &+\varepsilon_c \frac{\sqrt{s_+}}{\sqrt{\mLamB\mLamCst}}\left(\zeta\bar\Lambda+\zeta_{\text{SL}}-\zeta\frac{\bar\Lambda'(\mLamB^2+\mLamCst^2-q^2)}{2\mLamB\mLamCst}\right)+\frac{\sqrt{s_+}s_-}{2(\mLamB\mLamCst)^{3/2}}\varepsilon_c\eta^{(c)}_{\text{kin}}, \\
    F_{3/2,\perp}=&-\varepsilon_b\frac{\sqrt{s_+}}{\sqrt{\mLamB\mLamCst}}\zeta_{\text{SL}},
\end{align}

while the axial form factors read as:
\begin{align}
    G_{1/2,0}=& \frac{\zeta s_+ \sqrt{s_-}}{2(\mLamB \mLamCst)^{3/2}}  \left(C^5_1-\frac{C^5_2 s_-}{2 \mLamB (\mLamB-\mLamCst)}-\frac{C^5_3 s_-}{2 \mLamCst (\mLamB-\mLamCst)}\right)+\nn \\+&\varepsilon_b \frac{ \sqrt{s_-} (\mLamB+\mLamCst)}{(\mLamB-\mLamCst)\sqrt{\mLamB \mLamCst}}  \left(\zeta \bar\Lambda'- \zeta_{\text{SL}}-\frac{\zeta\bar\Lambda \left(\mLamB^2+\mLamCst^2-q^2\right)}{2 \mLamB \mLamCst} \right)+\nn \\+&\varepsilon_c \frac{ \sqrt{s_-} (\mLamB+\mLamCst)}{(\mLamB-\mLamCst)\sqrt{\mLamB \mLamCst}}  \left(\zeta \bar\Lambda+ \zeta_{\text{SL}}-\frac{\zeta\bar\Lambda' \left(\mLamB^2+\mLamCst^2-q^2\right)}{2 \mLamB \mLamCst} \right)+\varepsilon_c \frac{\sqrt{s_-}s_+}{2(\mLamB\mLamCst)^{3/2}}\eta^{(c)}_{\text{kin}} ,\\
    G_{1/2,t}=&\frac{\zeta s_- \sqrt{s_+}}{2(\mLamB \mLamCst)^{3/2}} \left( C^5_1-\frac{(\mLamB^2-\mLamCst^2+q^2)}{2(\mLamB+\mLamCst)\mLamB }C^5_2-\frac{(\mLamB^2-\mLamCst^2-q^2)}{2(\mLamB+\mLamCst) \mLamCst}C^5_3\right)+\nn \\
    &+\varepsilon_b \frac{(\mLamB-\mLamCst)\sqrt{s_+}}{(\mLamB+\mLamCst)\sqrt{\mLamB \mLamCst}}\left(\zeta \bar\Lambda'-\zeta_{\text{SL}}-\frac{\zeta \bar\Lambda(\mLamB^2+\mLamCst^2-q^2)}{2\mLamB\mLamCst}\right) +\nn \\
    &+\varepsilon_c \frac{(\mLamB-\mLamCst)\sqrt{s_+}}{(\mLamB+\mLamCst)\sqrt{\mLamB \mLamCst}}\left(\zeta\bar\Lambda'+\zeta_{\text{SL}}-\frac{\zeta \bar\Lambda(\mLamB^2+\mLamCst^2-q^2)}{2\mLamB\mLamCst}\right)+\varepsilon_c\frac{\sqrt{s_+}s_-}{2(\mLamB\mLamCst)^{3/2}}\eta^{(c)}_{\text{kin}},\\
    G_{1/2,\perp}=&\frac{\zeta\,\sqrt{s_-}s_+}{2(\mLamB \mLamCst)^{3/2}}C^5_1+\varepsilon_b\zeta\frac{\sqrt{s_-}}{\sqrt{\mLamB\mLamCst}}\left(\frac{\bar\Lambda(\mLamB^2+\mLamCst^2-q^2)}{2\mLamB\mLamCst}-\bar\Lambda'\right)+ \nn \\
    &+\varepsilon_c \frac{\sqrt{s_-}}{\sqrt{\mLamB\mLamCst}}\left(\zeta\bar\Lambda+\zeta_{\text{SL}}-\zeta\frac{\bar\Lambda'(\mLamB^2+\mLamCst^2-q^2)}{2\mLamB\mLamCst}\right)+ \varepsilon_c\frac{\sqrt{s_-}s_+}{2(\mLamB\mLamCst)^{3/2}}\eta^{(c)}_{\text{kin}},\\
    G_{3/2,\perp}=&-\varepsilon_b\frac{\sqrt{s_-}}{\sqrt{\mLamB\mLamCst}}\zeta_{\text{SL}}.
\end{align}

\subsection{The  chromomagnetic corrections for vector and axial form factors}
\label{sec:vector_axial_form_factor_chromomagnetic}

The contributions of the chromomagnetic operators result in the following shifts to the form factors reported in \cref{sec:vector_axial_form_factor} :
\begin{align}
f^{mag}_{1/2,0}=&\frac{s_-\sqrt{s_+}}{2(\mLamB\mLamCst)^{3/2}}(\varepsilon_b \eta^{(b)}_{\text{mag}}-\varepsilon_c\eta^{(c)}_{\text{mag}}) \,,\\
 f^{mag}_{1/2,t}=&\frac{s_+\sqrt{s_-}}{2(\mLamB\mLamCst)^{3/2}}(\varepsilon_b \eta^{(b)}_{\text{mag}}-\varepsilon_c\eta^{(c)}_{\text{mag}})\,,\\
f^{mag}_{1/2,\perp}=&-\frac{s_-\sqrt{s_+}}{2(\mLamB\mLamCst)^{3/2}}\varepsilon_c\eta^{(c)}_{\text{mag}} \,,
\end{align}
for the vector form factors, while for the axial-vector form factors we obtain:
\begin{align}
g^{mag}_{1/2,0}=& \frac{s_+\sqrt{s_-}}{2(\mLamB\mLamCst)^{3/2}}(-\varepsilon_b \eta^{(b)}_{\text{mag}}-\varepsilon_c\eta^{(c)}_{\text{mag}}) \,,\\
g^{mag}_{1/2,t}=&\frac{s_-\sqrt{s_+}}{2(\mLamB\mLamCst)^{3/2}}(-\varepsilon_b \eta^{(b)}_{\text{mag}}-\varepsilon_c\eta^{(c)}_{\text{mag}})\,,\\
g^{mag}_{1/2,\perp}=&-\frac{s_+\sqrt{s_-}}{2(\mLamB\mLamCst)^{3/2}}\varepsilon_c\eta^{(c)}_{\text{mag}}\,.
\end{align}
 Concerning the $\LamCst[2625]$ final state, the contributions of the chromomagnetic operator to the form factors are:
\begin{align}
 F^{mag}_{1/2,0} = &\frac{ s_- \sqrt{ s_+}}{4(\mLamB\mLamCst)^{3/2}}(-\varepsilon_b\eta^{(b)}_\text{mag}+\varepsilon_c\eta^{(c)}_\text{mag})\,, \\
    F^{mag}_{1/2,t} = &\frac{ s_+ \sqrt{ s_-}}{4(\mLamB\mLamCst)^{3/2}}(-\varepsilon_b\eta^{(b)}_\text{mag}+\varepsilon_c\eta^{(c)}_\text{mag})\, , \\
    F^{mag}_{1/2,\perp} =& \frac{ s_- \sqrt{ s_+}}{4(\mLamB\mLamCst)^{3/2}}\varepsilon_c\,\eta^{(c)}_\text{mag}\, , \\
     F^{mag}_{3/2,\perp} =&-\frac{ s_- \sqrt{ s_+}}{4(\mLamB\mLamCst)^{3/2}}\varepsilon_b\,\eta^{(b)}_\text{mag} \ ,
\end{align}
while for the axial-vector form factor we obtain:
\begin{align}
     G^{mag}_{1/2,0} =& \frac{ s_+ \sqrt{ s_-}}{4(\mLamB\mLamCst)^{3/2}}(\varepsilon_b\eta^{(b)}_\text{mag}+\varepsilon_c\eta^{(c)}_\text{mag})\, , \\
       G^{mag}_{1/2,t} =&\frac{ s_- \sqrt{ s_+}}{4(\mLamB\mLamCst)^{3/2}}(\varepsilon_b\eta^{(b)}_\text{mag}+\varepsilon_c\eta^{(c)}_\text{mag})\, , \\
     G^{mag}_{1/2,\perp} =&\frac{ s_+ \sqrt{ s_-}}{4(\mLamB\mLamCst)^{3/2}}\varepsilon_c\eta^{(c)}_\text{mag} \, , \\
 G^{mag}_{3/2,\perp} =&-\frac{ s_+ \sqrt{ s_-}}{4(\mLamB\mLamCst)^{3/2}}\varepsilon_b\eta^{(b)}_\text{mag}  \,.
\end{align}

\subsection{Tensor and pseudo-tensor current matrix elements}
We now obtain the tensor and pseudo-tensor current matrix elements by replacing $\Gamma$ with $i\sigma_{\mu \nu}q^{\nu},i\sigma_{\mu \nu}q^{\nu}\gamma_5$ in eq. \eqref{eq:hqet_expansion}:
\begin{align}
 & \sqrt{4} \zeta(w)\{-(\mLamB+\mLamCst)\genbar{u}_\alpha \gamma^\mu v^\alpha u +\mLamB\genbar{u}_\alpha v^\mu v^\alpha u+\mLamCst)\bar{u}_\alpha v'^\mu v^\alpha u\}\,, \\
&\sqrt{4} \zeta(w)\{(\mLamB-\mLamCst)\genbar{u}_\alpha \gamma^\mu v^\alpha \gamma_5 u +\mLamB\genbar{u}_\alpha v^\mu v^\alpha\gamma_5 u+\mLamCst)\genbar{u}_\alpha v'^\mu v^\alpha \gamma_5 u\}\, .
\end{align}

In order to include $1/m$ and $\alpha_s$ corrections to the tensor current matrix element, we consider the operator shift:
\begin{equation}
\begin{split}
    i\sigma^{\mu\nu}q_{\nu} \to C_{T_1}(w)i\sigma^{\mu\nu}q_{\nu}+C_{T_2}(w)(v\cdot q \gamma^{\mu}-v^{\mu}\slashed{q}) 
    +C_{T_3}(w)(v'\cdot q \gamma^{\mu}-v'^{\mu} \slashed{q}) \\
    +C_{T_4}(w)(v'^{\mu}v\cdot q - v^{\mu} v'\cdot q)
    +\varepsilon_b \Delta J_{T}^{\mu}+\varepsilon_c\Delta \genbar{J}_{T}^{\mu}
+\mathcal{O}(\alpha_s / m, 1/m^2)\, .
\end{split}
\end{equation}
Our definition of the $C_{T_i}$ coefficients is related to the one given in \cite{Bernlochner:2017jka} by the following substitutions: $\hat{\alpha}_sC_{T_{i}}+\delta_{i1} \to C_{T_i}$ and $i \hat{\alpha}_s C_{T_4} \rightarrow C_{T_4}$.
From eq. \eqref{eq:hqet_expansion} we find the leading order hadronic matrix element including the short distance QCD corrections:
\begin{equation}
\begin{split}
     2 \zeta(w)\Bar{u}_{\alpha}v^{\alpha}\gamma^{\mu}u\bigg[-C_{T_1}(\mLamB+\mLamCst)+C_{T_2}(\mLamB-\mLamCst w)+ C_{T_3}(w\mLamB-\mLamCst)\bigg] +\\
+2 \zeta(w)\Bar{u}_{\alpha}v^{\alpha} v^{\mu}u\bigg[C_{T_1}\mLamB +C_{T_2}(\mLamCst-\mLamB )+ C_{T_4}(\mLamCst-w\mLamB)\bigg] +\\
+2 \zeta(w)\Bar{u}_{\alpha}v^{\alpha} v'^{\mu}u\bigg[C_{T_1}\mLamCst +C_{T_3}(\mLamCst-\mLamB )+ C_{T_4}(\mLamB-w \mLamCst)\bigg] \ .
\end{split}
\end{equation}
We consider only the local power corrections $\Delta J_{T}^{\mu}$ and $\Delta \genbar{J}_{T}^{\mu}$ derived from eq. \eqref{currentcorrection} with  $\Gamma^\mu=i\sigma^{\mu\nu}q_{\nu}$. The other local operators of the tensor current contribute at order $\alpha_s/m$, which goes beyond the precision we are aiming to achieve here.  The hadronic matrix elements of $\Delta J_{T}^{\mu} $ and $\Delta \genbar{J}_{T}^{\mu}$ can be written as follows:

\begin{align}
    \bra{\Lambda_c^{*}(k, \eta, s_c)}\Delta J^{\mu}_{T}& \ket{\Lambda_b(p, s_b)} = 4\genbar{u}_{\alpha}g^{\mu\alpha}u(\mLamB-\mLamCst)\zeta^{(b)}_{3} + 2\genbar{u}_{\alpha}\gamma^{\alpha}\gamma^{\mu}u(\mLamCst-\mLamB)\zeta^{(b)}_{3}\notag +\\
    &+  2\genbar{u}_{\alpha}\gamma^{\alpha}v^{\mu}u \, \mLamB\zeta_{3}^{(b)}+2\genbar{u}_{\alpha}\gamma^{\alpha}v'^{\mu}u \, \mLamCst \zeta_{3}^{(b)} \notag+\\   &+\genbar{u}_{\alpha}v^{\alpha}\gamma^{\mu}u\bigg[-4\mLamB \zeta_{3}^{(b)}-2(\mLamB+\mLamCst)\zeta_{1}^{(b)}+2\bigg(\frac{-\mLamB^2-\mLamB\mLamCst+q^2}{\mLamCst}\zeta_{2}^{(b)}\bigg)\bigg] \notag+ \\
&+2\genbar{u}_{\alpha}v^{\alpha}v^{\mu}u\,\mLamB(\zeta_{1}^{(b)}+\zeta_{2}^{(b)}) + \genbar{u}_{\alpha}v^{\alpha}v'^{\mu}u\big[2\mLamCst\zeta_{1}^{(b)}+(4\mLamB-2\mLamCst)\zeta_{2}^{(b)}]\notag\, , \\ 
\end{align}
\begin{equation}
\begin{split}
    \bra{\Lambda_c^{*}(k, \eta, s_c)}\Delta \genbar{J}^{\mu}_{T} \ket{\Lambda_b(p, s_b)} = 2\genbar{u}_{\alpha}\gamma^{\alpha}\gamma^{\mu}u(\mLamCst-\mLamB)\zeta_{3}^{(c)} + 2\genbar{u}_{\alpha}\gamma^{\alpha}v^{\mu}u\, \mLamB\zeta_{3}^{(c)}+ \\
    +2\genbar{u}_{\alpha}\gamma^{\alpha}v'^{\mu}u\, \mLamCst\zeta_{3}^{(c)} + \genbar{u}_{\alpha}v^{\alpha}\gamma^{\mu}u\bigg[ -2(\mLamB+\mLamCst)\zeta_{2}^{(c)}+2\bigg(\frac{-\mLamCst^2-\mLamB\,\mLamCst+q^2}{\mLamB}\bigg)\zeta_{1}^{(c)}\bigg]+ \\
   + \genbar{u}_{\alpha}v^{\alpha}v^{\mu}u\bigg[(4\mLamCst-2\mLamB)\zeta_{1}^{(c)}+2\mLamB\zeta_{2}^{(c)}\bigg]+ 2\genbar{u}_{\alpha}v^{\alpha}v'^{\mu}u\,\mLamCst(\zeta_{1}^{(c)}+\zeta_{2}^{(c)}) \ .
\end{split}
\end{equation}
The short distance corrections to the pseudo-tensor current are related to those of the tensor current via the identity:
\begin{equation}
    \sigma_{\mu \nu } \gamma_5 = \frac{i}{2} \varepsilon_{\mu \nu \rho \sigma} \sigma^{\rho \sigma} \ .
\end{equation}
We get:
\begin{equation}
\begin{aligned}
    i\sigma^{\mu\nu}q_{\nu}\gamma_5 \rightarrow \, &
 C_{T_1}(w) i\sigma^{\mu\nu}q_{\nu}\gamma_5 -\frac{1}{2} \varepsilon_{\mu \nu \rho \sigma} q^{\nu} i C_{T_2}(w)(v^{\rho} \gamma^{\sigma}- v^{\sigma} \gamma^{\rho}) -\\
 -&\frac{1}{2} \varepsilon_{\mu \nu \rho \sigma} q^{\nu} i C_{T_3}(w)(v'^{\rho} \gamma^{\sigma}- v'^{\sigma} \gamma^{\rho}) 
 -\frac{1}{2} \varepsilon_{\mu \nu \rho \sigma} q^{\nu} C_{T_4}(w)(v'^{\rho} v^{\sigma}- v'^{\sigma} v^{\rho}) +\\+& \varepsilon_b \Delta J_{T5}^{\mu}+\varepsilon_c\Delta \genbar{J}_{T5}^{\mu}
+\mathcal{O}(\alpha_s / m, 1/m^2)\,\ .
\end{aligned}
\end{equation}
The hadronic matrix element of the short distance corrections become:
\begin{align}
     2 \zeta(w) C_{T_1}&\big[(\mLamB-\mLamCst) \, \Bar{u}_{\alpha}v^{\alpha}\gamma^{\mu}\gamma^{5} u +\mLamB  \, \Bar{u}_{\alpha}v^{\alpha} v^{\mu}\gamma^{5} u 
+ \mLamCst\,\Bar{u}_{\alpha}v^{\alpha} v'^{\mu}\gamma^{5} u \big] \notag +\\
&+ 2 i \zeta (w) C_{T_2} \Bar{u}_{\alpha}v^{\alpha} \varepsilon_{\mu \nu \rho \sigma} k^{\nu}v^{\rho} \gamma^{\sigma} u + 2 i \zeta(w) C_{T_3}  \Bar{u}_{\alpha}v^{\alpha}\varepsilon_{\mu \nu \rho \sigma} p^{\nu}v'^{\sigma} \gamma^{\rho}\ u \,.
\end{align}

From eq. \eqref{currentcorrection} with $\Gamma^\mu =i\sigma^{\mu\nu}q_{\nu}\gamma_5$ we obtain the hadronic matrix elements of $\Delta J^{\mu}_{T5}$ and $\Delta \genbar{J}^{\mu}_{T5}$:
\begin{align}
    \bra{\Lambda_c^{*}(k, \eta, s_c)}&\Delta J^{\mu}_{T5} \ket{\Lambda_b(p, s_b)} = 4\Bar{u}_{\alpha}g^{\mu\alpha}\gamma^{5} u(\mLamB+\mLamCst)\zeta^{(b)}_{3} -2\Bar{u}_{\alpha}\gamma^{\alpha}\gamma^{\mu}\gamma^5 u(\mLamCst+\mLamB)\zeta^{(b)}_{3}\notag -\\
    &-  2\Bar{u}_{\alpha}\gamma^{\alpha}\gamma^{5}v^{\mu}u \, \mLamB\zeta_{3}^{(b)}-2\Bar{u}_{\alpha}\gamma^{\alpha}\gamma^{5}v'^{\mu}u \, \mLamCst \zeta_{3}^{(b)}\notag+ \\  &+ \Bar{u}_{\alpha}v^{\alpha}\gamma^{\mu}\gamma^{5}u\bigg[4\mLamB \zeta_{3}^{(b)}+2(\mLamB-\mLamCst)\zeta_{1}^{(b)}+2\bigg(\frac{\mLamB^2-\mLamB\mLamCst-q^2}{\mLamCst}\zeta_{2}^{(b)}\bigg)\bigg]\notag +\\
&+2\Bar{u}_{\alpha}v^{\alpha}v^{\mu}\gamma^{5} u\,\mLamB(\zeta_{1}^{(b)}-\zeta_{2}^{(b)}) + \Bar{u}_{\alpha}v^{\alpha}v'^{\mu} \gamma^{5} u\big[2\mLamCst\zeta_{1}^{(b)}+(4\mLamB+2\mLamCst)\zeta_{2}^{(b)}] \ ,\notag \\
\end{align}
\begin{equation}
\begin{split}
    \bra{\Lambda_c^{*}(k, \eta, s_c)}\Delta \genbar{J}^{\mu}_{T5} \ket{\Lambda_b(p, s_b)} = 2\Bar{u}_{\alpha}\gamma^{\alpha}\gamma^{\mu}\gamma^{5} u(\mLamCst+\mLamB)\zeta_{3}^{(c)} + 2\Bar{u}_{\alpha}\gamma^{\alpha}v^{\mu}\gamma^{5} u\, \mLamB\zeta_{3}^{(c)}+ \\
    +2\Bar{u}_{\alpha}\gamma^{\alpha}v'^{\mu}\gamma^{5} u\, \mLamCst\zeta_{3}^{(c)} + \Bar{u}_{\alpha}v^{\alpha}\gamma^{\mu} \gamma^{5} u\bigg[ 2(\mLamB-\mLamCst)\zeta_{2}^{(c)}+2\bigg(\frac{-\mLamCst^2+\mLamB\,\mLamCst+q^2}{\mLamB}\bigg)\zeta_{1}^{(c)}\bigg]+ \\
  +  \Bar{u}_{\alpha}v^{\alpha}v^{\mu}\gamma^{5} u\bigg[(4\mLamCst+2\mLamB)\zeta_{1}^{(c)}+2\mLamB\zeta_{2}^{(c)}\bigg]+ 2\Bar{u}_{\alpha}v^{\alpha}v'^{\mu}\gamma^{5} u\,\mLamCst(\zeta_{2}^{(c)}-\zeta_{1}^{(c)}) \ .
\end{split}
\end{equation}

Non-local insertions of
the kinetic operator can be parametrised as:

\begin{align}
&\bra{\Lambda_c^{*}(k, \eta, s_c)} \mathcal{L}_\text{kin}\circ \Bar{c}^{v'}_{+}\Gamma b  ^{v}_{+}\ket{\Lambda_b(p, s_b)}=\sqrt{4}\eta^{(b)}_\text{kin}(w)\,v^{\alpha}\bar{u}_\alpha(\mLamCst v', \eta, s_c)\Gamma u(\mLamB v, s_b) \, , \\
&\bra{\Lambda_c^{*}(k, \eta, s_c)} \mathcal{L}^{\prime}_\text{kin}\circ \Bar{c}^{v'}_{+}\Gamma b^{v}_{+}  \ket{\Lambda_b(p, s_b)}=\sqrt{4}\eta^{(c)}_\text{kin}(w)\,v^{\alpha}\bar{u}_\alpha(\mLamCst v', \eta, s_c)\Gamma u(\mLamB v, s_b) \, ,
\end{align}
with $\Gamma=i\sigma^{\mu\nu}q_{\nu}, i\sigma^{\mu\nu}q_{\nu}\gamma_{5}$. The above corresponds to a $w$-dependent shift $\eta_\text{kin}(w)$ with the same Dirac structure of  the tree-level leading order IW function $\zeta(w)$. At order $\mathcal{O}(1/m)$ we can again absorb it by redefining the leading order IW function $\zeta(w)$, yet at $\mathcal{O}(1/m^2)$ we must retain it .

The chromomagnetic correction to the $b$ tensor current can be parametrised as in eq. \eqref{eq:eta_mag_b} with $\Gamma=i \sigma^{\rho \sigma}q_{\sigma}$ and yields:
\begin{align}
    \varepsilon_b\eta^{(b)}_\text{mag}(w)&\{2(1-w)(\mLamB+\mLamCst)\,\genbar{u}_{\alpha}g^{\alpha \rho}u+ 
    \,(1-w)(\mLamB+\mLamCst)\,\genbar{u}_{\alpha}\gamma^{\alpha}\gamma^{\rho}u
    \notag -\\& - 
    (\mLamB+\mLamCst)\,\genbar{u}_{\alpha}v^{\alpha}\gamma^{\rho}u+
    \mLamB(w+1)\,\genbar{u}_{\alpha}\gamma^{\alpha}v^{\rho}u +\\&+ 
     [\mLamCst(w-1)-2\mLamB]\,\genbar{u}_{\alpha}\gamma^{\alpha}v'^{\rho}u \notag -
   \mLamB \, \genbar{u}_{\alpha}v^{\alpha}v^{\rho}u +(2\mLamB+
   \mLamCst )\, \genbar{u}_{\alpha}v^{\alpha}v'^{\rho}u\,\} \, . 
\end{align}

Analogously from eq. \eqref{eq:eta_mag_c} with $\Gamma=i \sigma^{\rho \sigma}q_{\sigma}$ we obtain the chromomagnetic correction to the $c$ tensor current:
\begin{equation}
\begin{split}
    &\varepsilon_c\eta^{(c)}_\text{mag}(w)\{\,(w-1)(\mLamB+\mLamCst)\,\genbar{u}_{\alpha}\gamma^{\alpha}\gamma^{\rho}u- \\&
   - (\mLamB+\mLamCst)\,\genbar{u}_{\alpha}v^{\alpha}\gamma^{\rho}u+
    [\mLamB(-w+1)+2\mLamCst]\,\genbar{u}_{\alpha}\gamma^{\alpha}v^{\rho}u - \\&-
     \mLamCst(w+1)\,\genbar{u}_{\alpha}\gamma^{\alpha}v'^{\rho}u +
   \mLamB \, \genbar{u}_{\alpha}v^{\alpha}v^{\rho}u +
   \mLamCst \, \genbar{u}_{\alpha}v^{\alpha}v'^{\rho}u\,\} \ . \\
   \end{split} 
\end{equation}

By imposing $\Gamma=i \sigma^{\rho \sigma}\gamma_5 q_{\sigma}$ in eqs. \eqref{eq:eta_mag_b} and \eqref{eq:eta_mag_c}, we obtain that the chromomagnetic correction to the $b$ and $c$ pseudo-tensor currents is respectively given by:
\begin{align}
    &\varepsilon_b\eta^{(b)}_\text{mag}(w)\{-2(w+1)(\mLamB-\mLamCst)\,\genbar{u}_{\alpha}g^{\alpha \rho}\gamma_5 u+ 
    \,(w+1)(\mLamB-\mLamCst),\genbar{u}_{\alpha}\gamma^{\alpha}\gamma^{\rho}\gamma_5 u \notag +\\&+
    (-\mLamB+\mLamCst)\,\genbar{u}_{\alpha}v^{\alpha}\gamma^{\rho}\gamma_5 u+
    \mLamB(-w+1)\,\genbar{u}_{\alpha}\gamma^{\alpha}v^{\rho}\gamma_5 u \notag +
     [-\mLamCst(w+1)+2\mLamB]\,\genbar{u}_{\alpha}\gamma^{\alpha}v'^{\rho}\gamma_5 u \notag +\\ &+
   \mLamB \, \genbar{u}_{\alpha}v^{\alpha}v^{\rho}\gamma_5 u +(2\mLamB-
   \mLamCst )\, \genbar{u}_{\alpha}v^{\alpha}v'^{\rho} \gamma_5 u\,\}  \, ,
   \end{align}
   \begin{equation}
\begin{split}
   & \varepsilon_c\eta^{(c)}_\text{mag}(w)\{\,-(w+1)(\mLamB-\mLamCst)\,\genbar{u}_{\alpha}\gamma^{\alpha}\gamma^{\rho}\gamma_5 u+ \\&+
    (\mLamB-\mLamCst)\,\genbar{u}_{\alpha}v^{\alpha}\gamma^{\rho}\gamma_5 u+
    [-\mLamB(w+1)+2\mLamCst]\,\genbar{u}_{\alpha}\gamma^{\alpha}v^{\rho} \gamma_5 u - \\&-
     \mLamCst(w-1)\,\genbar{u}_{\alpha}\gamma^{\alpha}v'^{\rho} \gamma_5 u +
   \mLamB \, \genbar{u}_{\alpha}v^{\alpha}v^{\rho} \gamma_5 u +
   \mLamCst \, \genbar{u}_{\alpha}v^{\alpha}v'^{\rho} \gamma_5 u\,\} \ .
   \end{split}
\end{equation}

As expected, at zero recoil the chromomagnetic corrections vanish for each current.
\subsection{Tensor and pseudo-Tensor form factors in the VC limit}
\label{sec:tensor_pseudotensor_form_factors_VC_limit}
In the following we list the expressions of the QCD form factors obtained from the matching on the HQE on-shell amplitudes in the VC limit. As we have already done for vector and axial current form factors, we explicitly report the kinetic operator contribution $\mathcal{O}(1/m_{c})$ .

Concerning the $\LamCst[2595]$ final
state we find for the tensor form factors:
\begin{align}
t_{1/2,0}=&\frac{\zeta  \mLamCst s_- \sqrt{s_+}}{(\mLamB \mLamCst)^{3/2}} \left(C_{T_1} -C_{T_2}+ C_{T_3}-\frac{C_{T_4} s_+}{2 \mLamB \mLamCst}\right)+\nn \\+&\varepsilon_b\frac{\mLamCst \sqrt{s_+}}{\sqrt{\mLamB \mLamCst}} \left(\frac{\zeta \bar\Lambda \left(\mLamB^2+\mLamCst^2-q^2\right)}{\mLamB \mLamCst}- 2 \zeta \bar\Lambda' + 4 \zeta_{\text{SL}} \right)+\nn \\+& \varepsilon_c\frac{\mLamCst \sqrt{s_+}}{\sqrt{\mLamB \mLamCst}} \left(2 \zeta \bar\Lambda - 4 \zeta_{\text{SL}}-\frac{\zeta \bar\Lambda' \left(\mLamB^2+\mLamCst^2-q^2\right)}{\mLamB \mLamCst} \right)+\varepsilon_c\frac{\eta^{(c)}_\text{kin}  \mLamCst s_- \sqrt{s_+} }{(\mLamB \mLamCst)^{3/2}} \ ,\\
t_{1/2,\perp}=& \frac{\zeta s_- \sqrt{s_+}}{\sqrt{\mLamB \mLamCst}} \left(\frac{C_{T_1} \mLamCst }{\mLamB \mLamCst}+\frac{C_{T_2}\left(-\mLamB^2+\mLamCst^2-q^2\right)}{2 \mLamB^2 (\mLamB+\mLamCst) }-\frac{C_{T_3}\left(\mLamB^2-\mLamCst^2-q^2\right)}{2 (\mLamB+\mLamCst) \mLamB \mLamCst}\right)+\nn \\+&  \varepsilon_b\frac{\mLamCst \sqrt{s_+} (\mLamB-\mLamCst)}{(\mLamB+\mLamCst) \sqrt{\mLamB \mLamCst}}\left( 2 \zeta \bar\Lambda'  -\frac{\zeta \bar\Lambda \left(\mLamB^2+\mLamCst^2-q^2\right)}{\mLamB \mLamCst}\right)+\nn \\+&  \varepsilon_c \frac{\sqrt{s_+}\mLamCst(\mLamB-\mLamCst)}{(\mLamB+\mLamCst)\sqrt{\mLamB \mLamCst}}\left(2 \zeta \bar\Lambda - 4 \zeta_{\text{SL}} -\frac{\bar\Lambda'\zeta  \left(\mLamB^2+\mLamCst^2-q^2\right)}{\mLamB \mLamCst} \right) +\varepsilon_c\frac{\eta^{(c)}_\text{kin}  \mLamCst s_- \sqrt{s_+} }{(\mLamB \mLamCst)^{3/2}}\,,
\end{align}
 while for the pseudo-tensor form factors the matching gives
\begin{align}
t^5_{1/2,0}=& \frac{C_{T_1} \zeta  \mLamCst \sqrt{s_-} s_+}{(\mLamB \mLamCst)^{3/2}}+\varepsilon_b\frac{\mLamCst \sqrt{s_-}}{\sqrt{\mLamB \mLamCst}} \left(\frac{\zeta \bar\Lambda \left(\mLamB^2+\mLamCst^2-q^2\right)}{\mLamB \mLamCst}-2 \zeta \bar\Lambda'-4 \zeta_{\text{SL}} \right)+\nn \\+&\varepsilon_c\frac{\mLamCst \sqrt{s_-}}{\sqrt{\mLamB \mLamCst}} \left(2 \zeta \bar\Lambda  -4 \zeta_{\text{SL}} -\frac{\zeta \bar\Lambda'  \left(\mLamB^2+\mLamCst^2-q^2\right)}{\mLamB \mLamCst}\right)+\varepsilon_c\frac{\eta^{(c)}_\text{kin}  \mLamCst \sqrt{s_-} s_+}{(\mLamB \mLamCst)^{3/2}}\,,\\
t^5_{1/2,\perp}=& \frac{\zeta \sqrt{s_-} s_+}{\sqrt{\mLamB \mLamCst}}  \left(\frac{C_{T_1} \mLamCst }{\mLamB \mLamCst}-\frac{C_{T_2} s_-}{2 \mLamB^2 (\mLamB-\mLamCst) }-\frac{C_{T_3} s_-}{2 (\mLamB-\mLamCst) \mLamB \mLamCst}\right) +\nn \\+& \varepsilon_b\frac{\mLamCst \sqrt{s_-} (\mLamB+\mLamCst)}{ (\mLamB-\mLamCst) \sqrt{\mLamB \mLamCst}}\left(2\zeta\bar\Lambda'-\frac{\zeta \bar\Lambda \left(\mLamB^2+\mLamCst^2-q^2\right)}{\mLamB \mLamCst}\right)+\nn \\+&\varepsilon_c \frac{\mLamCst \sqrt{s_-} (\mLamB+\mLamCst)}{ (\mLamB-\mLamCst) \sqrt{\mLamB \mLamCst}}\left( 2 \zeta \bar\Lambda -4 \zeta_{\text{SL}}-\frac{\zeta \bar\Lambda' \left(\mLamB^2+\mLamCst^2-q^2\right)}{\mLamB \mLamCst}\right)+\varepsilon_c\frac{\eta^{(c)}_\text{kin}  \mLamCst \sqrt{s_-} s_+ }{(\mLamB \mLamCst)^{3/2}}\,.
\end{align}
For the $\LamCst[2625]$ final state, the tensor form factors are 
\begin{align}
    T_{1/2,0} =& \zeta\frac{\mLamCst\sqrt{s_+}s_-}{(\mLamB\mLamCst)^{3/2}}\left(C_{T_{1}}-C_{T_{2}}+C_{T_{3}}-\frac{s_+}{2\mLamB\mLamCst}C_{T_{4}}\right)+\nn \\
&+\varepsilon_b\frac{2\mLamCst\sqrt{s_+}}{\sqrt{\mLamB\mLamCst}}\left(-\zeta\bar\Lambda'-\zeta_{\text{SL}}+\zeta\frac{\mLamB^2+\mLamCst^2-q^2}{2\mLamB\mLamCst}\bar\Lambda\right)+\nn \\
+&\varepsilon_c\frac{2\mLamCst\sqrt{s_+}}{\sqrt{\mLamB\mLamCst}}\left(\zeta\bar\Lambda+\zeta_{\text{SL}}-\zeta\frac{\mLamB^2+\mLamCst^2-q^2}{2\mLamB\mLamCst}\bar\Lambda'\right)
+\varepsilon_c\frac{\mLamCst\sqrt{s_+}s_-}{(\mLamB\mLamCst)^{3/2}}\eta^{(c)}_{\text{kin}}
\,,\\
T_{1/2,\perp} =&\zeta\frac{\sqrt{s_+} s_-}{\mLamB\sqrt{\mLamB\mLamCst}}\left(C_{T_{1}}-\frac{\mLamB^2-\mLamCst^2+q^2}{2\mLamB(\mLamB+\mLamCst)}C_{T_{2}}-\frac{\mLamB^2-\mLamCst^2-q^2}{2\mLamCst(\mLamB+\mLamCst)}C_{T_{3}}\right)+\nn \\
&+\varepsilon_b\frac{2\mLamCst(\mLamB-\mLamCst)\sqrt{s_+}}{(\mLamB+\mLamCst)\sqrt{\mLamB\mLamCst}}\left(\zeta\bar\Lambda'-\zeta\frac{\mLamB^2+\mLamCst^2-q^2}{2\mLamB\mLamCst}\bar\Lambda\right)+\nn \\
&+\varepsilon_c\frac{2\mLamCst(\mLamB-\mLamCst)\sqrt{s_+}}{(\mLamB+\mLamCst)\sqrt{\mLamB\mLamCst}}\left(\zeta\bar\Lambda+\zeta_{\text{SL}}-\zeta\frac{\mLamB^2+\mLamCst^2-q^2}{2\mLamB\mLamCst}\bar\Lambda'\right)+\varepsilon_c\frac{\mLamCst \sqrt{s_+}s_-}{(\mLamB\mLamCst)^{3/2}}\eta^{(c)}_{\text{kin}}
\,, \\
 T_{3/2,\perp} =& -\varepsilon_b\frac{2(\mLamB-\mLamCst)\sqrt{s_+}}{\sqrt{\mLamB\mLamCst}}\zeta_{\text{SL}} \, ,
\end{align}
while for the pseudo-tensor form factor 
\begin{align}
T^5_{1/2,0} =&\frac{\zeta\mLamCst s_+\sqrt{s_-}}{(\mLamB\mLamCst)^{3/2}}C_{T_{1}}+\varepsilon_b\frac{2\mLamCst\sqrt{s_-}}{\sqrt{\mLamB\mLamCst}}\left(\zeta\frac{(\mLamB^2+\mLamCst^2-q^2)\bar\Lambda}{2\mLamB\mLamCst}+\zeta_{\text{SL}}-\zeta\bar\Lambda'\right)+\nn \\
&+\varepsilon_b\frac{2\mLamCst\sqrt{s_-}}{\sqrt{\mLamB\mLamCst}}\left(-\zeta\frac{(\mLamB^2+\mLamCst^2-q^2)\bar\Lambda'}{2\mLamB\mLamCst}+\zeta_{\text{SL}}+\zeta\bar\Lambda\right)+\varepsilon_c\frac{\mLamCst s_+\sqrt{s_-}}{(\mLamB\mLamCst)^{3/2}}\eta^{(c)}_{\text{kin}}\,,\\
T^5_{1/2,\perp} =& \zeta\frac{s_+\sqrt{s_-}}{\mLamB\sqrt{\mLamB\mLamCst}}\left(C_{T_{1}}-\frac{s_-}{2\mLamB(\mLamB-\mLamCst)}C_{T_{2}}-\frac{s_-}{2\mLamCst(\mLamB-\mLamCst)}\right)+\nn \\
&+\varepsilon_b\frac{2\mLamCst(\mLamB+\mLamCst)\sqrt{s_-}}{(\mLamB-\mLamCst)\sqrt{\mLamB\mLamCst}}\left(\zeta\bar\Lambda'-\zeta\frac{(\mLamB^2+\mLamCst^2-q^2)\bar\Lambda}{2\mLamB\mLamCst}\right)+\nn \\
&+\varepsilon_c\frac{2\mLamCst(\mLamB+\mLamCst)\sqrt{s_-}}{(\mLamB-\mLamCst)\sqrt{\mLamB\mLamCst}}\left(\zeta\bar\Lambda+\zeta_{\text{SL}}-\zeta\frac{(\mLamB^2+\mLamCst^2-q^2)\bar\Lambda'}{2\mLamB\mLamCst}\right)+\varepsilon_c\frac{\mLamCst s_+\sqrt{s_-}}{(\mLamB\mLamCst)^{3/2}}\eta^{(c)}_{\text{kin}}\,,\\
 T^5_{3/2,\perp} =&-\varepsilon_b\frac{2(\mLamB+\mLamCst)\sqrt{s_-}}{\sqrt{\mLamB\mLamCst}}\zeta_{\text{SL}}  \ .
\end{align}

\subsection{The chromomagnetic correction for tensor and pseudo-tensor form factors}
\label{sec:tensor_pseudotensor_form_factor_chromomagnetic}
The contribution of the chromomagnetic operator results in a shift to the form factors defined in \cref{sec:tensor_pseudotensor_form_factors_VC_limit}.
Concerning the $\LamCst[2595]$ final
state we find that the contributions of the chromomagnetic operator to the tensor form factors are:
\begin{align}
t^{mag}_{1/2,0}=&\frac{\mLamCst\,s_-\sqrt{s_+}}{(\mLamB \mLamCst)^{3/2}}(\varepsilon_b\eta_\text{mag}^{(b)}-\varepsilon_c\eta_\text{mag}^{(c)}) \,, \\
t^{mag}_{1/2,\perp}=&-\frac{\mLamCst\,s_-\sqrt{s_+}}{(\mLamB \mLamCst)^{3/2}}\varepsilon_c\,\eta_\text{mag}^{(c)} \,,
\end{align}
 while for the pseudo-tensor form factors the matching gives:
\begin{align}
t^{mag\,5}_{1/2,0}=&-\frac{\mLamCst\,s_+\sqrt{s_-}}{(\mLamB \mLamCst)^{3/2}}(\varepsilon_b\eta_\text{mag}^{(b)}+\varepsilon_c\eta_\text{mag}^{(c)})\,,\\
t^{mag\,5}_{1/2,\perp}=&  -\frac{\mLamCst\,s_+\sqrt{s_-}}{(\mLamB \mLamCst)^{3/2}}\varepsilon_c\eta_\text{mag}^{(c)}\,.
\end{align}
 For the $\LamCst[2625]$ final state, the contributions of the chromomagnetic operator to the tensor form factors are:
\begin{align}
    T^{mag}_{1/2,0} =&\frac{\mLamCst\,s_-\sqrt{s_+}}{2(\mLamB\mLamCst)^{3/2}}(-\varepsilon_b\,\eta_\text{mag}^{(b)}+\varepsilon_c\,\eta_\text{mag}^{(c)})\,,\\
T^{mag}_{1/2,\perp} = &\frac{\mLamCst\,s_-\sqrt{s_+}}{2(\mLamB\mLamCst)^{3/2}}\varepsilon_c\,\eta_\text{mag}^{(c)}\,, \\
 T^{mag}_{3/2,\perp} =&\frac{(\mLamB+\mLamCst)s_-\sqrt{s_+}}{2(\mLamB\mLamCst)^{3/2}}\varepsilon_b\,\eta_\text{mag}^{(b)}\,,
\end{align}
while for the pseudo-tensor form factors we obtain:
\begin{align}
T^{mag\,5}_{1/2,0} =&\frac{\mLamCst\,s_+\sqrt{s_-}}{2(\mLamB\mLamCst)^{3/2}}(\varepsilon_b\eta_\text{mag}^{(b)}+\varepsilon_c\eta_\text{mag}^{(c)})\,,\\
T^{mag\,5}_{1/2,\perp} =& \frac{\mLamCst\,s_+\sqrt{s_-}}{2(\mLamB\mLamCst)^{3/2}}\varepsilon_c\eta_\text{mag}^{(c)}\,,\\
 T^{mag\,5}_{3/2,\perp} =& \frac{(\mLamB-\mLamCst)\,s_+\sqrt{s_-}}{2(\mLamB\mLamCst)^{3/2}}\varepsilon_b\eta_\text{mag}^{(b)} \ .
\end{align}
 
The proportionality of the  chromomagnetic induced form factors with respect to $s_-$ guarantees that  the amplitudes vanish at zero recoil (e.g. $w=1$).

\vskip .2cm
Summarizing this section, we obtained non-local chromomagnetic and (pseudo)tensor form factor contributions that were absent in \cite{Boer:2018vpx} but  present in \cite{Papucci:2021pmj} expressed in a different  basis.  We explicitly checked that our results coincide with the ones of \cite{Papucci:2021pmj}.

\section{Next-to-next to leading order corrections}
\label{secondorder}
When delving into corrections at the next-to-next-to-leading order (NNLO) within the framework of HQET, it is necessary to turn our attention to all the $\mathcal{O}(1/m_{b,c}^{2}, 1/m_{b}m_{c})$ terms. This is important to ensure a comprehensive and accurate computation of the matrix elements that contributes to the $\Lambda_{b} \to \Lambda_{c}^{*}$ transition.

However, since the $\mathcal{O}(1/m_c^2)$ terms are numerically favoured with respect to $\mathcal{O}(1/m_b^2)$ and $\mathcal{O}(1/m_b m_c)$ ones, they are expected to give the major contribution to the amplitudes, as also underlined in \cite{Boer:2018vpx,Papucci:2021pmj}.

Furthermore, a comprehensive study of the complete second-order correction would entail a larger number of independent IW unknown parameters. Therefore, in order to extend the analysis beyond the next-to-leading order while keeping as few unconstrained parameters as possible, we limit our calculations of the form factors to the $\mathcal{O}(1/m_c^2)$ terms. 
In addition, we further simplify our study by employing both the Residual Chiral (RC) expansion and the Vanishing Chromomagnetic (VC) limit. These two  were introduced  in \cite{Bernlochner:2022ywh} in the context of the computation of the second order corrections to the mesonic $B\rightarrow D^{(*)}$ transitions.


We now proceed to parametrize the matrix elements at $\mathcal{O}(1/m_c^2)$ given in eq. \eqref{eqn:QCDmatchexp} which reads: 

\begin{equation}
\label{secondo11}
    \bra{\Lambda_c^{*}(k, \eta, s_c)} \Bar{c}^{v'}_{+}\Overleftarrow{\mJbar}^{\prime}_2\Pi_-' \Gamma \,b^{v}_{+} \ket{\Lambda_b(p, s_b)} = \sqrt{4}\Bar{u}_{\mu}\gamma_{\alpha}v'_{\beta}\Gamma u\, \psi^{\alpha \beta \mu}(v, v') \ ,
\end{equation}

where

\begin{equation}
    \psi^{\alpha\beta\mu}(v, v')=\psi_1(w)v^{\mu}(v^{\alpha}v'^{\beta}-v^{\beta}v'^{\alpha})+\psi_2(w)(g^{\alpha\mu}v^{\beta}-g^{\beta\mu}v^{\alpha})+\psi_3(w)(g^{\alpha\mu} v'^{\beta}-g^{\beta \mu}v'^{\alpha}) \ .
\end{equation}

To determine $ \mathcal{L}'_2 \circ \cbvp  \Gamma \bv$ we observe that the associated terms given in eq. \eqref{lagrcorr}, can be parametrized as the corresponding kinetic and chromomagnetic corrections at order $\mathcal{O}(1/m_c)$.
\begin{equation}
    \begin{aligned}
        &\bra{\Lambda_c^{*}(k, \eta, s_c)} \Bar{c}^{v'}_{+}\,v_\beta  D_\alpha G^{\ab}\,c^{v'}_{+}\circ\Bar{c}^{v'}_{+}\,\Gamma\,b^{v}_{+} \ket{\Lambda_b(p, s_b)}= \Bar{u}_{\mu}\Gamma\,u\, \lambda^{ \mu}(v, v') \ , \\
        & \bra{\Lambda_c^{*}(k, \eta, s_c)} - i\,  \Bar{c}^{v'}_{+} \,v_\alpha \sigma_{\beta\gamma} D^\gamma G^{\ab} c^{v'}_{+}\circ\Bar{c}^{v'}_{+} \, \Gamma\,b^{v}_{+} \ket{\Lambda_b(p, s_b)}= \eta(w) g^{\mu \alpha} v^\beta \Bar{u}_{\mu}\,i \sigma_{\alpha\beta}\Pi^{'}_{+} \Gamma\,u \ .
    \end{aligned}
\end{equation}
We then consider two times the insertion of the first order Lagrangian: $\mathcal{L}'_1 \circ \mathcal{L}'_1 \circ \cbvp \Gamma \, \bv \ $ and find
\begin{equation}
 \bra{\Lambda_c^{*}(k, \eta, s_c)} \Bar{c}^{v'}_{+}\,D^2\,c^{v'}_{+}\circ\Bar{c}^{v'}_{+}\,D^2\,c^{v'}_{+}\circ\Bar{c}^{v'}_{+}\,\Gamma\,b^{v}_{+} \ket{\Lambda_b(p, s_b)}=\Bar{u}_{\mu}\Gamma u\,\alpha^{\mu}(v, v') \ ,
\end{equation}

where

\begin{equation}
    \alpha^{\mu}(v, v')=\alpha(w)(v-v')^{\mu} \ .
\end{equation}

Then
\begin{equation}
    \bra{\Lambda_c^{*}(k, \eta, s_c)} \Bar{c}^{v'}_{+}\,\sigma_{\alpha\beta}G^{\alpha\beta}\,c^{v'}_{+}\circ\Bar{c}^{v'}_{+}
    \,\sigma_{\gamma\delta}G^{\gamma\delta}\,c^{v'}_{+}\circ\Bar{c}^{v'}_{+}\Gamma\,b^{v}_{+} \ket{\Lambda_b(p, s_b)} = \Bar{u}_{\mu} \sigma_{\alpha\beta}\Pi^{'}_{+}\sigma_{\gamma\delta}\Pi^{'}_{+}\Gamma\,u\, \alpha^{\prime \, \alpha\beta\gamma\delta\mu}(v, v') \ ,
\end{equation}

where 

\begin{equation}
\begin{aligned}
   &\alpha^{\prime \, \alpha\beta\gamma\delta\mu}(v, v')=\alpha_1^\prime(w)(g^{\alpha \gamma} g^{\beta \delta}-g^{\alpha \delta} g^{\beta \gamma})v^\mu+\alpha_2^\prime(w)(g^{\alpha \gamma} v^\beta v^\delta-g^{\beta \gamma} v^\alpha v^\delta-g^{\alpha \delta} v^\beta v^\gamma+g^{\beta \delta} v^\alpha v^\gamma)v^\mu + \\+&\alpha_3^\prime(w)(g^{\alpha \mu} g^{\beta \delta}v^\gamma-g^{\beta \mu} g^{\alpha \delta}v^\gamma +g^{\alpha \gamma} g^{\beta \mu}v^\delta -g^{\beta \gamma} g^{\alpha \mu}v^\delta+ g^{\mu \gamma} g^{\beta \delta}v^\alpha-g^{\beta \gamma} g^{\mu \delta}v^\alpha +g^{\alpha \gamma} g^{\mu \delta}v^\beta-g^{\mu \gamma} g^{\alpha \delta}v^\beta )\ .
    \end{aligned}
\end{equation}

We arrive at
\begin{equation}
     \bra{\Lambda_c^{*}(k, \eta, s_c)} \Bar{c}^{v'}_{+}\,D^2\,c^{v'}_{+}\circ\Bar{c}^{v'}_{+}\sigma_{\alpha\beta}G^{\alpha\beta}c^{v'}_{+}\circ\Bar{c}^{v'}_{+}\Gamma\,b^{v}_{+} \ket{\Lambda_b(p, s_b)} = \alpha^{\prime\prime}(w)\Bar{u}_{\mu} \sigma_{\alpha\beta}\Pi^{'}_{+}\Gamma\,u\,g^{\alpha\mu}v^{\beta}  \ .
\end{equation}
Turning our attention to  $ \mathcal{L}'_1 \circ \cbvp \Overleftarrow{\mJbar}^{\prime}_1\Pi_-' \Gamma \bv$ we see that the first term in $ \mathcal{L}'_1$  yields
\begin{equation}
    \bra{\Lambda_c^{*}(k, \eta, s_c)} \Bar{c}^{v'}_{+}\,D^2\,c^{v'}_{+}\circ\Bar{c}^{v'}_{+}\,\Overleftarrow{\mJbar}^{\prime}_1\,\Pi^{'}_{-}\,\Gamma\,b^{v}_{+} \ket{\Lambda_b(p, s_b)}= \Bar{u}_{\mu}\gamma_{\alpha}\,\Pi^{'}_{-}\Gamma\,u\, \beta^{\mu \alpha}(v, v') \ ,
\end{equation}

where

\begin{equation}
    \beta^{\mu \alpha}(v, v')=v^{\mu}(\beta_{1}(w)v^{\alpha}+\beta_{2}v'^{\alpha}) +\beta_{3}(w)\,g^{\mu\alpha}\ .
\end{equation}
The second term in $ \mathcal{L}'_1$ corresponds to 
\begin{equation}
    \bra{\Lambda_c^{*}(k, \eta, s_c)} \Bar{c}^{v'}_{+}\,\sigma_{\alpha\beta}G^{\alpha\beta}c^{v'}_{+}\circ\Bar{c}^{v'}_{+}\,\Overleftarrow{\mJbar}^{\prime}_1\,\Pi^{'}_{-}\Gamma\,b^{v}_{+} \ket{\Lambda_b(p, s_b)}= \Bar{u}_{\mu}\,\sigma_{\alpha\beta}\Pi^{'}_{+}\gamma_{\nu}\Pi^{'}_{-}\Gamma\,u\, \beta^{\prime\,\mu\alpha\beta\nu}(v, v') \ ,
\end{equation}
where 
\begin{equation}
    \beta^{\prime\,\mu\alpha\beta\nu}(v, v')= \beta_{1}^{\prime}(g^{\alpha \mu} g^{\beta \nu}-g^{\alpha \nu} g^{\beta \mu})+ \beta_{2}^{\prime}(g^{\alpha \mu} v^{\beta}-g^{\beta \mu} v^{\alpha})v^{\nu} +\beta_{3}^{\prime}(g^{\alpha \nu} v^{\beta}-g^{\beta \nu} v^{\alpha})v^{\mu}+ \beta_{4}^{\prime}(g^{\alpha \mu} v^{\beta}-g^{\beta \mu} v^{\alpha})v'^{\nu}\ .
\end{equation}
We observe that there are initially 17 independent IW functions. Nevertheless, as we consider the constraints stemming from the equation of motion, as well as the modified Ward identities, this number reduces to 15.\\
Moving forward, we now investigate the RC and VC limits,where the number of independent IW functions decreases to 3 and 2, respectively.

\subsection{Residual chiral limit at $\mathcal{O}(1/m_c^2,\theta^2)$}
The Residual Chiral expansion, initially introduced in \cite{Bernlochner:2022ywh}, aims at establishing an additional power counting besides the HQE one, identifying a subset of subdominant contributions that can be neglected. In fact, as detailed in \cite{Bernlochner:2022ywh}, both experimental data and theoretical predictions derived within specific quark-models suggest that there exists a hierarchy among the HQET matrix elements. This hierarchy depends on the number of insertions of the cross terms $\bar{Q}^v_+ \Dslash_\perp Q^v_-$ and $\bar{Q}^v_- \Dslash_\perp Q^v_+$, which contain the transverse derivative. Specifically, matrix elements with more insertions of the transverse derivative are observed to be suppressed when compared to those with fewer insertions. Since the cross terms break the accidental $U(1)_+ \times U(1)_-$ symmetry of the kinetic Lagrangian $\mathcal{L} = \bar{Q}^v_+ iv\cdot D Q^v_+ + \bar{Q}^v_- iv\cdot D Q^v_-$ in eq. \eqref{eqn:fullL} to a diagonal $U(1)$, we can systematically organize the perturbative HQ expansion by power counting the number of insertions of the cross operator, despite the absence of a small parameter for this symmetry breaking. This is achieved through the substitution $i \Dslash_\perp \to i \theta \Dslash_\perp$. 

Building upon the insights presented in \cite{Bernlochner:2022ywh}, the Residual Chiral expansion effectively introduces a power counting scheme where it is justified to retain  $\mathcal{O}(\theta^2)$ terms.
 Thus, at $\mathcal{O}(1/m_c^2,\theta^2)$ the expression eq. \eqref{eqn:QCDmatchexp} becomes:
\begin{align}
\label{eqn:RCexp}
	\frac{\langle \Hc | \cbar \,\Gamma\, b | \Hb \rangle}{\sqrt{m_{\Hc} m_{\Hb}}}  \simeq 
	&\big\langle \Hc^{v'} \big| \Bar{c}^{v'}_{+} \, \Gamma \, b^{v}_{+} \big| \Hb^v \big\rangle \nn +\\ 
	 + &\frac{1}{2m_c} \big\langle \Hc^{v'} \big|  \big(\Bar{c}^{v'}_{+}\Overleftarrow{\mJbar}^{\prime}_1+ \mathcal{L}'_1 \circ \Bar{c}^{v'}_{+} \big) \Gamma \, b^{v}_{+} \big| \Hb^v \big\rangle \nn  + \frac{1}{2m_b} \big\langle \Hc^{v'} \big| \Bar{c}^{v'}_{+}  \, \Gamma  \big(\Overrightarrow{\mJ}_1 b^{v}_{+} + b^{v}_{+} \circ \mathcal{L}_1 \big)  \big| \Hb^v \big\rangle  \nn +\\
	+ &\frac{1}{4m_c^2} \big\langle \Hc^{v'} \big| \big(\Bar{c}^{v'}_{+}\Overleftarrow{\mJbar}^{\prime}_2\Pi_-' + \mathcal{L}'_2 \circ \Bar{c}^{v'}_{+}) \Gamma \, b^{v}_{+} \big| \Hb^v \big\rangle  \ . 
\end{align}
The second order pure current correction can be written as:
\begin{equation}
\begin{aligned}
\label{secondo1}
    \bra{\Lambda_c^{*}(k, \eta, s_c)} \Bar{c}^{v'}_{+}\Overleftarrow{\mJbar}^{\prime}_2\Pi_-' \Gamma \,b^{v}_{+} \ket{\Lambda_b(p, s_b)} &= \bra{\Lambda_c^{*}(k, \eta, s_c)} \Bar{c}^{v'}_{+}\gamma_{\alpha}v'_{\beta}\,G^{\alpha\beta} \Gamma \,b^{v}_{+} \ket{\Lambda_b(p, s_b)} \\ &=\sqrt{4}\Bar{u}_{\mu}\gamma_{\alpha}v'_{\beta}\Gamma u\, \psi^{\alpha \beta \mu}(v, v') \ ,
\end{aligned}
\end{equation}
where the most general decomposition of the IW function $\psi^{\alpha\beta\mu}(v, v')$ is:
\begin{equation}
    \psi^{\alpha\beta\mu}(v, v')=\psi_1(w)v^{\mu}(v^{\alpha}v'^{\beta}-v^{\beta}v'^{\alpha})+\psi_2(w)(g^{\alpha\mu}v^{\beta}-g^{\beta\mu}v^{\alpha})+\psi_3(w)(g^{\alpha\mu} v'^{\beta}-g^{\beta \mu}v'^{\alpha}) \ .
\end{equation}

\begin{itemize}
    \item For the vector current $\Gamma=\gamma_\mu$ eq. \eqref{secondo1} gives:
    \begin{equation}
        4 \psi_1\Bar{u}_{\alpha} v^{\alpha}v_\mu u-2(1+w) \psi_1\Bar{u}_{\alpha} v^{\alpha}\gamma_\mu u+ 2(\psi_3+w\psi_2)\Bar{u}_{\alpha} \gamma^{\alpha}\gamma_\mu u \ ;
        \end{equation}
     \item For the axial current $\Gamma=\gamma_\mu \gamma_5$ eq. \eqref{secondo1} gives:
    \begin{equation}
        4 \psi_1\Bar{u}_{\alpha} v^{\alpha}v_\mu \gamma_5 u+2(1-w) \psi_1\Bar{u}_{\alpha} v^{\alpha}\gamma_\mu \gamma_5u+ 2(\psi_3+w\psi_2)\Bar{u}_{\alpha} \gamma^{\alpha}\gamma_\mu \gamma_5 u \ ;
    \end{equation}
       \item For the tensor current $\Gamma=i\sigma_{\mu \nu} q^{\nu}$ eq. \eqref{secondo1} gives:
    \begin{equation}
    \begin{aligned}
        &2 \psi_1(1+w)(\mLamB-\mLamCst)\Bar{u}_{\alpha} v^{\alpha}\gamma_\mu u+\psi_1[4\mLamCst-2(1+w)\mLamB] \Bar{u}_{\alpha} v^{\alpha}v_\mu u+\\+&2\psi_1(1-w)\mLamCst \Bar{u}_{\alpha} v^{\alpha}v'_\mu u+\\+& 2(\psi_3+w\psi_2)[(\mLamCst-\mLamB)\Bar{u}_{\alpha} \gamma^{\alpha}\gamma_\mu  u+\mLamB\Bar{u}_{\alpha} \gamma^{\alpha}v_\mu  u+\mLamCst\Bar{u}_{\alpha} \gamma^{\alpha}v'_\mu  u] \ ;
         \end{aligned}
    \end{equation}
     \item For the pseudo-tensor current $\Gamma=i\sigma_{\mu \nu} \gamma_5 q^{\nu}$ eq. \eqref{secondo1} gives:
    \begin{equation}
    \begin{aligned}
        &2 \psi_1(1-w)(\mLamB+\mLamCst)\Bar{u}_{\alpha} v^{\alpha}\gamma_\mu \gamma_5 u+\\+&\psi_1(4\mLamCst+2\mLamB(1-w))\Bar{u}_{\alpha} v^{\alpha}v_\mu \gamma_5 u-2\psi_1(1+w)\mLamCst\Bar{u}_{\alpha} v^{\alpha}v'_\mu \gamma_5 u+\\+& 2(\psi_3+w\psi_2)[(\mLamB+\mLamCst)\Bar{u}_{\alpha} \gamma^{\alpha}\gamma_\mu  \gamma_5 u+\mLamB\Bar{u}_{\alpha} \gamma^{\alpha}v_\mu  \gamma_5 u+\mLamCst\Bar{u}_{\alpha} \gamma^{\alpha}v'_\mu \gamma_5 u] \ .
         \end{aligned}
    \end{equation}
\end{itemize}

The matrix elements of the insertion of second order lagrangian $ \mathcal{L}'_2$ in the LO currents can easily be parametrized like the corresponding single  insertions of the first order lagrangian  \cite{Bernlochner:2022ywh}:
\begin{equation}
    \begin{aligned}
    \label{L2corr}
        &\bra{\Lambda_c^{*}(k, \eta, s_c)} \Bar{c}^{v'}_{+}\,v_\beta  D_\alpha G^{\ab}\,c^{v'}_{+}\circ\Bar{c}^{v'}_{+}\,\Gamma\,b^{v}_{+} \ket{\Lambda_b(p, s_b)}= \sqrt{4}\Bar{u}_{\mu}\Gamma\,u\, \lambda^{ \mu}(v, v') \ , \\
        & \bra{\Lambda_c^{*}(k, \eta, s_c)} - i\,  \Bar{c}^{v'}_{+} \,v_\alpha \sigma_{\beta\gamma} D^\gamma G^{\ab} c^{v'}_{+}\circ\Bar{c} \, \Gamma\,b^{v}_{+} \ket{\Lambda_b(p, s_b)}= \sqrt{4}\,\eta(w) g^{\mu \alpha} v^\beta \Bar{u}_{\mu}\,i \sigma_{\alpha\beta}\Pi^{'}_{+} \Gamma\,u \, .
    \end{aligned}
\end{equation}

At order $\mathcal{O}(1/m_c^2,\theta^2)$ they can be reabsorbed by redefining the IW functions $\eta^{(c)}_\text{kin}$ and $\eta^{(c)}_\text{mag}$ as follows:
\begin{align}
   & \eta^{(c)}_\text{kin}(w)+\frac{1}{2 m_c}\lambda(w)\to \eta^{(c)}_\text{kin} (w)\ , \\
 &\eta^{(c)}_\text{mag}(w)+\frac{1}{2 m_c}\eta(w)\to \eta^{(c)}_\text{mag}(w).   
\end{align}

\subsection{Form factors at $\mathcal{O}(1/m_c^2,\theta^2)$}

The NNLO contributions of the HQE for the matrix elements due to the current $\mJbar^{\prime}_{2}$ in the RC limit induces a shift in the form factors reported in \cref{sec:firstorder} which for the  $\LamCst[2595]$  vector current contribution  reads:  
\begin{align}
f^{\mJ^{\prime}_2}_{1/2,0}=&\varepsilon_c^2 \frac{\sqrt{s_+}(\mLamB-\mLamCst)}{(\mLamB+\mLamCst) \sqrt{\mLamB \mLamCst}}\left(3(\psi_3 +w \psi_2) -\frac{s_- s_+ \psi_1 }{4  (\mLamB \mLamCst)^{2}}\right)\,,\\
f^{\mJ^{\prime}_2}_{1/2,t}=&\varepsilon_c^2 \frac{\sqrt{s_-}(\mLamB+\mLamCst)}{(\mLamB-\mLamCst) \sqrt{\mLamB \mLamCst}}\left(3(\psi_3 +w \psi_2) -\frac{s_- s_+ \psi_1 }{4  (\mLamB \mLamCst)^{2}}\right)\,,\\
f^{\mJ^{\prime}_2}_{1/2,\perp}=&\varepsilon_c^2 \frac{\sqrt{s_+}}{\sqrt{\mLamB \mLamCst}}\left(3(\psi_3 +w \psi_2) -\frac{s_- s_+ \psi_1 }{4  (\mLamB \mLamCst)^{2}}\right) \,,
\end{align}
while for the axial-vector form factors we obtain:
\begin{align}
g^{\mJ^{\prime}_2}_{1/2,0}=&\varepsilon_c^2 \frac{\sqrt{s_-}(\mLamB+\mLamCst)}{(\mLamB-\mLamCst) \sqrt{\mLamB \mLamCst}}\left(3(\psi_3 +w \psi_2) -\frac{s_- s_+ \psi_1 }{4  (\mLamB \mLamCst)^{2}}\right)\,,\\
g^{\mJ^{\prime}_2}_{1/2,t}=&\varepsilon_c^2 \frac{\sqrt{s_+}(\mLamB-\mLamCst)}{(\mLamB+\mLamCst) \sqrt{\mLamB \mLamCst}}\left(3(\psi_3 +w \psi_2) -\frac{s_- s_+ \psi_1 }{4  (\mLamB \mLamCst)^{2}}\right)\,,\\
g^{\mJ^{\prime}_2}_{1/2,\perp}=&\varepsilon_c^2 \frac{\sqrt{s_-}}{\sqrt{\mLamB \mLamCst}}\left(3(\psi_3 +w \psi_2) -\frac{s_- s_+ \psi_1 }{4  (\mLamB \mLamCst)^{2}}\right) \,.
\end{align}

For the $\LamCst[2625]$ final state the vector current reads:
\begin{align}
 F^{\mJ^{\prime}_2}_{1/2,0} = &-\frac{s_+^{3/2}s_- (\mLamB-\mLamCst)}{4(\mLamB+\mLamCst)(\mLamB\mLamCst)^{5/2}}\varepsilon_c^2\psi_1\, , \\
    F^{\mJ^{\prime}_2}_{1/2,t} = & -\frac{s_+s_-^{3/2} (\mLamB+\mLamCst)}{4(\mLamB-\mLamCst)(\mLamB\mLamCst)^{5/2}}\varepsilon_c^2\psi_1 \ , \\
     F^{\mJ^{\prime}_2}_{1/2,\perp} =&  -\varepsilon_c^2\frac{s_+^{3/2}s_-}{4\,(\mLamB\mLamCst)^{5/2}}\psi_1\, , \\
 F^{\mJ^{\prime}_2}_{3/2,\perp} =& \,0 \,,
\end{align}
while for the axial-vector: 
\begin{align}
 G^{\mJ^{\prime}_2}_{1/2,0} =& -\frac{s_+s_-^{3/2}(\mLamB+\mLamCst)}{4(\mLamB-\mLamCst)(\mLamB\mLamCst)^{5/2}}\varepsilon_c^2\psi_1\, , \\
    G^{\mJ^{\prime}_2}_{1/2,t} =& -\frac{s_+^{3/2}s_-(\mLamB-\mLamCst)}{4(\mLamB+\mLamCst)(\mLamB\mLamCst)^{5/2}}\varepsilon_c^2\psi_1\, , \\
      G^{\mJ^{\prime}_2}_{1/2,\perp} =&-\frac{s_+ s_-^{3/2}}{4 (\mLamB\mLamCst)^{5/2}}\varepsilon_c^2\psi_1\, , \\
 G^{\mJ^{\prime}_2}_{3/2,\perp} =& \,0 \,.
\end{align}

The tensor contributions for  $\LamCst[2595]$ are:
\begin{align}
t^{\mJ^{\prime}_2}_{1/2,0}=&\varepsilon_c^2 \frac{\mLamCst \sqrt{s_+}}{\sqrt{\mLamB \mLamCst}}\left(6  (\psi_3 +w \psi_2)-\frac{s_- s_+ \psi_1}{2 (\mLamB \mLamCst)^{2}}\right)\,,\\
t^{\mJ^{\prime}_2}_{1/2,\perp}=&\varepsilon_c^2 \frac{\mLamCst \sqrt{s_+}(\mLamB-\mLamCst)}{(\mLamB+\mLamCst)\sqrt{\mLamB \mLamCst}}\left(6  (\psi_3 +w \psi_2)-\frac{s_- s_+ \psi_1}{2 (\mLamB \mLamCst)^{2}}\right) \,,
\end{align}
 while  the pseudo-tensor induced form factors read: 
\begin{align}
t^{5\,\mJ^{\prime}_2}_{1/2,0}=&\varepsilon_c^2 \frac{\mLamCst \sqrt{s_-}}{\sqrt{\mLamB \mLamCst}}\left(6  (\psi_3 +w \psi_2)-\frac{s_- s_+ \psi_1}{2 (\mLamB \mLamCst)^{2}}\right)\,,\\
t^{5\,\mJ^{\prime}_2}_{1/2,\perp}=&\varepsilon_c^2 \frac{\mLamCst \sqrt{s_+}(\mLamB+\mLamCst)}{(\mLamB-\mLamCst)\sqrt{\mLamB \mLamCst}}\left(6  (\psi_3 +w \psi_2)-\frac{s_- s_+ \psi_1}{2 (\mLamB \mLamCst)^{2}}\right) \,.
\end{align}
Similarly the tensor contributions for $\LamCst[2625]$  are
\begin{align}
    T^{\mJ^{\prime}_2}_{1/2,0} =&-\frac{\mLamCst\,s_+^{3/2}s_-}{2(\mLamB\mLamCst)^{5/2}}\varepsilon_c^2\psi_1 \,,\\
T^{\mJ^{\prime}_2}_{1/2,\perp} = & -\frac{\mLamCst(\mLamB-\mLamCst)s_+^{3/2}s_-}{2(\mLamB+\mLamCst)(\mLamB\mLamCst)^{5/2}}\varepsilon_c^2\psi_1\,, \\
 T^{\mJ^{\prime}_2}_{3/2,\perp} =& \,0 \,,
\end{align}
while the pseudo-tensor induced form factors are 
\begin{align}
T^{5\;\mJ^{\prime}_2}_{1/2,0} =&-\frac{\mLamCst\,s_+s_-^{3/2}}{2(\mLamB\mLamCst)^{5/2}}\varepsilon_c^2\psi_1\,,\\
T^{5\;\mJ^{\prime}_2}_{1/2,\perp} =&  -\frac{\mLamCst(\mLamB+\mLamCst)s_+s_-^{3/2}}{2(\mLamB-\mLamCst)(\mLamB\mLamCst)^{5/2}}\varepsilon_c^2 \psi_1\,,\\
 T^{5\;\mJ^{\prime}_2}_{3/2,\perp} =& \,0 \ .
\end{align}

\subsection{Vanishing chromomagnetic limit at $\mathcal{O}(1/m_c^2)$}
When taking the limit  $G_{\alpha \beta}\to0$ the number of matrix elements  drastically reduces with the first and second order lagrangian correction simplifying to \cite{Bernlochner:2022ywh}:
\begin{equation}
    \mathcal{L}_1  =   -\Qbar^v_+ D^2 Q^v_+ ;\quad \mathcal{L}_2=0 \ ,
\end{equation} 
and eq. \eqref{eqn:QCDmatchexp} becomes:
\begin{align}
\label{eqn:VClim}
	\frac{\langle \Hc | \cbar \,\Gamma\, b | \Hb \rangle}{\sqrt{m_{\Hc} m_{\Hb}}}  
	& \simeq \big\langle \Hc^{v'} \big| \Bar{c}^{v'}_{+} \, \Gamma \, b^{v}_{+} \big| \Hb^v \big\rangle  \nn +\\
	& + \frac{1}{2m_c} \big\langle \Hc^{v'} \big|  \big(\Bar{c}^{v'}_{+}\Overleftarrow{\mJbar}^{\prime}_1- \Bar{c}^{v'}_{+} D^2 c^{v'}_{+} \circ \Bar{c}^{v'}_{+} \big) \Gamma \, b^{v}_{+} \big| \Hb^v \big\rangle \nn  + \frac{1}{2m_b} \big\langle \Hc^{v'} \big| \Bar{c}^{v'}_{+}  \, \Gamma  \big(\Overrightarrow{\mJ}_1 b^{v}_{+} - b^{v}_{+} \circ \Bar{b}^{v}_{+} D^2 b^{v}_{+} \big)  \big| \Hb^v \big\rangle \nn +\\
	& + \frac{1}{4m_c^2} \big\langle \Hc^{v'} \big| \big(-\Bar{c}^{v'}_{+} D^2 c^{v'}_{+} \circ \Bar{c}^{v'}_{+}\Overleftarrow{\mJbar}^{\prime}_1\Pi_-'
		+ \frac{1}{2}\,\Bar{c}^{v'}_{+} D^2 c^{v'}_{+} \circ \Bar{c}^{v'}_{+} D^2 c^{v'}_{+} \circ \Bar{c}^{v'}_{+}\big) \Gamma \, b^{v}_{+} \big| \Hb^v \big\rangle \ .
\end{align}

The second order terms can be parametrized in the following way:

\begin{align}
\label{VCpar}
   & \bra{\Lambda_c^{*}(k, \eta, s_c)} -\Bar{c}^{v'}_{+}\,D^2\,c^{v'}_{+}\circ\Bar{c}^{v'}_{+}\,\Overleftarrow{\mJbar}^{\prime}_1\,\Pi^{'}_{-}\,\Gamma\,b^{v}_{+} \ket{\Lambda_b(p, s_b)}= \sqrt{4}\Bar{u}_{\mu}\gamma_{\alpha}\,\Pi^{'}_{-}\Gamma\,u\, \beta^{\mu \alpha}(v, v')\ , \\
   \label{eqn:double_kinetic_insertion}
    & \bra{\Lambda_c^{*}(k, \eta, s_c)} \Bar{c}^{v'}_{+}\,D^2\,c^{v'}_{+}\circ\Bar{c}^{v'}_{+}\,D^2\,c^{v'}_{+}\circ\Bar{c}^{v'}_{+}\,\Gamma\,b^{v}_{+} \ket{\Lambda_b(p, s_b)}=\sqrt{4}\Bar{u}_{\mu}\,\Gamma\,u\,\alpha^{\mu}(v, v') \ ,
\end{align}

where

\begin{align}
    &\beta^{\mu \alpha}(v, v')=v^{\mu}(\beta_{1}(w)v^{\alpha}+\beta_{2}(w)v'^{\alpha}) +\beta_{3}(w)\,g^{\mu\alpha}\ , \\
    \label{eqn:alpha_IsgurWise}
    &\alpha^{\mu}(v, v')=\alpha(w)(v-v')^{\mu} \ .
\end{align}

The equations of motion imply that $v_{\alpha}^{\prime}\beta^{\mu\alpha}=0$ leading to the following relations between the IW functions entering in the definition of $\beta^{\mu\alpha}$:
\begin{equation}
\label{eq:equation_motion_second_order_VC}
    \beta_{1}(w) w + \beta_{2}(w)=0 \ .
\end{equation}

Moreover things can be further simplified in equation eq. \eqref{VCpar} using that $\gamma_\alpha \Pi'_-= \Pi'_+ \gamma_\alpha -v'_\alpha$ leading to
\begin{align}
\label{betapar}
     \bra{\Lambda_c^{*}(k, \eta, s_c)} -\Bar{c}^{v'}_{+}\,D^2\,c^{v'}_{+}\circ\Bar{c}^{v'}_{+}\,\Overleftarrow{\mJbar}^{\prime}_1\,\Pi^{'}_{-}\,\Gamma\,b^{v}_{+} \ket{\Lambda_b(p, s_b)}=& \sqrt{4}\Bar{u}_{\mu}\gamma_{\alpha}\,\Gamma\,u\, \beta^{\mu \alpha}(v, v')-  \notag \\-&\sqrt{4}\Bar{u}_{\mu}\,\Gamma\,u\, v^\mu (\beta_1(w) w+\beta_{2}(w))\ .
\end{align}
We observe that $\sqrt{4}\Bar{u}_{\mu}\gamma_{\alpha}\,\Gamma\,u\, \beta^{\mu \alpha}(v, v')$  assumes the same shape as the first order pure current correction, while $\sqrt{4}\Bar{u}_{\mu}\,\Gamma\,u\, v^\mu (\beta_1(w) w+\beta_{2}(w))$ vanishes due to eq. \eqref{eq:equation_motion_second_order_VC}.

The matrix element in eq. \eqref{eqn:double_kinetic_insertion} has the same parametrisation of the LO order contribution to the matrix elements in HQE. Therefore, it shifts the form factors by a term proportional to the LO contribution  that can be reabsorbed into the definition of the Isgur-Wise function $\eta^{(c)}_\text{kin}$:

\begin{equation}
    \eta^{(c)}_\text{kin}(w)+\frac{1}{2 m_{c}}\alpha(w) \to \eta^{(c)}_\text{kin}(w) \ .
\end{equation}

\subsection{Form factors in the VC limit: including $\mathcal{O}(1/m_c^2)$ contributions}

For the  vector current contribution to   $\LamCst[2595]$   the shifts produced by the insertion of the first order pure current correction $\mJbar^{\prime}_{1}$ read:
\begin{align}
f^{\mathcal{L'}_1\circ\mJ^{\prime}_1}_{1/2,0}=&\varepsilon_c^2 \frac{\sqrt{s_+}(\mLamB-\mLamCst)}{(\mLamB+\mLamCst) \sqrt{\mLamB \mLamCst}}\left(3\beta_3 -\frac{s_- s_+ \beta_1 }{4  (\mLamB \mLamCst)^{2}}\right)\,,\\
f^{\mathcal{L'}_1\circ\mJ^{\prime}_1}_{1/2,t}=&\varepsilon_c^2 \frac{\sqrt{s_-}(\mLamB+\mLamCst)}{(\mLamB-\mLamCst) \sqrt{\mLamB \mLamCst}}\left(3\beta_3 -\frac{s_- s_+ \beta_1 }{4  (\mLamB \mLamCst)^{2}}\right)\,,\\
f^{\mathcal{L'}_1\circ\mJ^{\prime}_1}_{1/2,\perp}=&\varepsilon_c^2 \frac{\sqrt{s_+}}{\sqrt{\mLamB \mLamCst}}\left(3\beta_3 -\frac{s_- s_+ \beta_1 }{4  (\mLamB \mLamCst)^{2}}\right) \,,
\end{align}

while for the axial-vector form factors we obtain:
\begin{align}
g^{\mathcal{L'}_1\circ\mJ^{\prime}_1}_{1/2,0}=&\varepsilon_c^2 \frac{\sqrt{s_-}(\mLamB+\mLamCst)}{(\mLamB-\mLamCst) \sqrt{\mLamB \mLamCst}}\left(3\beta_3 -\frac{s_- s_+ \beta_1 }{4  (\mLamB \mLamCst)^{2}}\right)\,,\\
g^{\mathcal{L'}_1\circ\mJ^{\prime}_1}_{1/2,t}=&\varepsilon_c^2 \frac{\sqrt{s_+}(\mLamB-\mLamCst)}{(\mLamB+\mLamCst) \sqrt{\mLamB \mLamCst}}\left(3\beta_3 -\frac{s_- s_+ \beta_1 }{4  (\mLamB \mLamCst)^{2}}\right)\,,\\
g^{\mathcal{L'}_1\circ\mJ^{\prime}_1}_{1/2,\perp}=&\varepsilon_c^2 \frac{\sqrt{s_-}}{\sqrt{\mLamB \mLamCst}}\left(3\beta_3 -\frac{s_- s_+ \beta_1 }{4  (\mLamB \mLamCst)^{2}}\right) \,.
\end{align}

 Concerning the vector current contribution to  $\LamCst[2625]$ the shifts produced by the insertion of the second order pure current correction yield: 

\begin{align}
F^{\mathcal{L'}_1\circ\mJ^{\prime}_1}_{1/2,0}=&-\frac{(\mLamB-\mLamCst)s_+^{3/2}s_-}{4(\mLamB+\mLamCst)(\mLamB\mLamCst)^{5/2}}\varepsilon_c^2\beta_1\,,\\
F^{\mathcal{L'}_1\circ\mJ^{\prime}_1}_{1/2,t}=&-\frac{(\mLamB+\mLamCst)s_+s_-^{3/2}}{4(\mLamB-\mLamCst)(\mLamB\mLamCst)^{5/2}}\varepsilon_c^2\beta_1 \,,\\
F^{\mathcal{L'}_1\circ\mJ^{\prime}_1}_{1/2,\perp}=& -\frac{s_{+}^{3/2}s_{-}}{4(\mLamB\mLamCst)^{5/2}}\varepsilon_c^2\beta_1\,,\\
F^{\mathcal{L'}_1\circ\mJ^{\prime}_1}_{3/2,\perp}=&\,0 \ ,
\end{align}

while for the axial-vector form factors we obtain:

\begin{align}
G^{\mathcal{L'}_1\circ\mJ^{\prime}_1}_{1/2,0}=&-\frac{(\mLamB+\mLamCst)s_{+}s_{-}^{3/2}}{4(\mLamB-\mLamCst)(\mLamB\mLamCst)^{5/2}}\varepsilon_c^2\beta_1\,,\\
G^{\mathcal{L'}_1\circ\mJ^{\prime}_1}_{1/2,t}=&-\frac{(\mLamB-\mLamCst)s^{3/2}_{+}s_{-}}{4(\mLamB+\mLamCst)(\mLamB\mLamCst)^{5/2}}\varepsilon_c^2\beta_1\,,\\
G^{\mathcal{L'}_1\circ\mJ^{\prime}_1}_{1/2,\perp}=&  -\frac{s_{+}s_{-}^{3/2}}{4(\mLamB\mLamCst)^{5/2}}\varepsilon_c^2\beta_1 \,, \\
G^{\mathcal{L'}_1\circ\mJ^{\prime}_1}_{3/2,\perp}=&\,0 .
\end{align}

For the transition to $\LamCst[2595]$ mediated by the tensor current we have
\begin{align}
t^{\mJ^{\prime}_2}_{1/2,0}=&\varepsilon_c^2 \frac{\mLamCst \sqrt{s_+}}{\sqrt{\mLamB \mLamCst}}\left(6  \beta_3-\frac{s_- s_+ \beta_1}{2 (\mLamB \mLamCst)^{2}}\right)\,,\\
t^{\mJ^{\prime}_2}_{1/2,\perp}=&\varepsilon_c^2 \frac{\mLamCst \sqrt{s_+}(\mLamB-\mLamCst)}{(\mLamB+\mLamCst)\sqrt{\mLamB \mLamCst}}\left(6  \beta_3-\frac{s_- s_+ \beta_1}{2 (\mLamB \mLamCst)^{2}}\right) \,,
\end{align}
 while  the pseudo-tensor current we obtain: 
\begin{align}
t^{5\,\mJ^{\prime}_2}_{1/2,0}=&\varepsilon_c^2 \frac{\mLamCst \sqrt{s_-}}{\sqrt{\mLamB \mLamCst}}\left(6  \beta_3-\frac{s_- s_+ \beta_1}{2 (\mLamB \mLamCst)^{2}}\right)\,,\\
t^{5\,\mJ^{\prime}_2}_{1/2,\perp}=&\varepsilon_c^2 \frac{\mLamCst \sqrt{s_+}(\mLamB+\mLamCst)}{(\mLamB-\mLamCst)\sqrt{\mLamB \mLamCst}}\left(6 \beta_3-\frac{s_- s_+ \beta_1}{2 (\mLamB \mLamCst)^{2}}\right) \,.
\end{align}

Concerning the $\LamCst[2625]$ final state, we obtain the following shifts for the tensor form factors:
\begin{align}
T^{\mathcal{L'}_1\circ\mJ^{\prime}_1}_{1/2,0}=&-\frac{\mLamCst\,s_+^{3/2}s_-}{2(\mLamB\mLamCst)^{5/2}}\varepsilon_c^2\,\beta_1 \ ,\\
T^{\mathcal{L'}_1\circ\mJ^{\prime}_1}_{1/2,\perp}=&-\frac{\mLamCst(\mLamB-\mLamCst)s_+^{3/2}s_-}{2(\mLamB+\mLamCst)(\mLamB\mLamCst)^{5/2}}\varepsilon_c^2\,\beta_1 \ , \\
T^{\mathcal{L'}_1\circ\mJ^{\prime}_1}_{3/2,\perp}=&\,0\,,
\end{align}

while for the pseudo-tensor form factors the shifts read as:

\begin{align}
T^{5\,\mathcal{L'}_1\circ\mJ^{\prime}_1}_{1/2,0}=&-\frac{\mLamCst\,s_+s_-^{3/2}}{2(\mLamB\mLamCst)^{5/2}}\varepsilon_c^2\beta_1 \ ,\\
T^{5\,\mathcal{L'}_1\circ\mJ^{\prime}_1}_{1/2,\perp}=&-\frac{\mLamCst(\mLamB+\mLamCst)s_+s_-^{3/2}}{2(\mLamB-\mLamCst)(\mLamB\mLamCst)^{5/2}}\varepsilon_c^2\beta_1 \ ,\\
T^{5\,\mathcal{L'}_1\circ\mJ^{\prime}_1}_{3/2,\perp}=&\,0\,.
\end{align}
 Remarkably, we observe that the form factors within the VC limit can be derived from their counterparts in the RC expansion, by  making the following substitutions:
\begin{equation}
\label{betapsi}
    \psi_3+w \psi_2 \to \beta_3 \ , \; \psi_1 \to \beta_1 \ .
\end{equation} 

\section{Fit to the LQCD data}
\label{fitLQCD}
The form factors computed in the RC and VC limits depend on 7 and 5 unknown IW functions respectively. Although many models have been proposed in literature \cite{Isgur:1988gb,Scora:1995ty,Jenkins:1992zx,Jenkins:1992se,Wirbel:1985ji,Korner:1989qb,Pervin:2005ve,Gutsche:2012ze,Gutsche:2018nks,Becirevic:2020nmb}, it is not possible to determine the $q^2-$dependence of the IW functions from first principles in HQET  \cite{Isgur:1989vq,Isgur:1990yhj}. 

Owing to the limited availability of measurements for the observables associated with the $\Lambda_b \to \Lambda_c^*$ transitions, the fitting of the IW parameters to experimental data remains unattainable. In fact, the  measurements available to us pertain to the decay rate, which can be found in the reference \cite{LHCb:2017vhq}. Fortunately, we can use LQCD results to estimate these unknown terms following   e.g. Refs.  \cite{Papucci:2021pmj,Bordone:2021bop,Bernlochner:2018kxh}.

 For the transition $\Lambda_b\rightarrow \Lambda_c^{*}(2595)$ and $\Lambda_b\rightarrow \Lambda_c^{*}(2625)$ we employ the recent data of \cite{Meinel:2021mdj}, in which one finds updates of the early analysis presented in \cite{Meinel:2021rbm}, obtained by imposing the relations among the  form factors at zero-recoil \cite{Hiller:2013cza,Hiller:2021zth}.
The lattice results provide a continuum extrapolation of the QCD form factors $f_i$ in the low-recoil region assuming a linear dependence on the recoil parameter $w$:
\begin{equation}
\label{ffactorsexpansion}
    f_i=F_i+A_i(w-1) \ ,
\end{equation}
for vector, axial-vector, tensor and pseudo-tensor currents.  
 The best fit values of $F_i$ and $A_i$ and their covariance matrix are given in the ancillary files of \cite{Meinel:2021mdj}. 
Specifically, the LQCD results are valid only in the near zero-recoil regime ($w \lesssim 1.05$) and cannot be used to extrapolate the low $q^2-$dependence of the form factors.
We stress that the relations between the form factors adopted in the LQCD computations and the ones used in this work are given in appendix \ref{app:B}.

 We expand the IW functions to the first order in $(w-1)$ as follows:
\begin{equation}
\label{linearIW}
    \zeta\simeq\zeta^{(0)}+(w-1)\zeta^{(1)} \ .
\end{equation}
We use the same parametrisation for the subleading and the sub-subleading IW functions. Then we further expand the QCD form factors computed in the previous sections and collected in appendices \ref{ffsRC} and \ref{ffsVC} to the first order in $(w-1)$,   to be compared with the LQCD results. We follow the fitting procedure of \cite{Papucci:2021pmj} with the same input parameters summarized in \cref{tab:parameters}. 
\begin{table}[!htb!]
\begin{center}
\begin{tabular}{lc}
\toprule
\multicolumn{2}{c}{HQET parameters} \\
\midrule
$\alpha_s(\sqrt{m_b m_c})$ & $0.26$\\
$m_b$ & $4.78 \; \rm{GeV}$ \\
$m_c$ & $1.38 \;\rm{GeV}$ \\
$\varepsilon_b$ & $0.105 \; \rm{GeV}^{-1}$ \\
$\varepsilon_c$ & $0.36 \;\rm{GeV}^{-1}$ \\
$\mLamB$ & $5.61960 \;\rm{GeV}$ \\
$m_{\LamCst[2595]}$ & $2.59225 \;\rm{GeV}$ \\
$m_{\LamCst[2625]}$ & $2.62811 \;\rm{GeV}$ \\
$\bar\Lambda$ & $0.81\;\rm{GeV}$ \\
$\bar\Lambda'$ & $1.10\;\rm{GeV}$\\
\bottomrule
\end{tabular}
\end{center}
\caption{\em Values of the HQET input parameters from \cite{Papucci:2021pmj}.}
\label{tab:parameters}
\end{table}
\\
\\As noticed in \cite{Papucci:2021pmj}, neglecting the order $1/m_c^2$ in the HQE, we achieve a poor fit to the data with a  $\chi^2/d.o.f.\sim 23$. This suggests that higher-order corrections are needed to accommodate the LQCD results.
\\
We therefore start with examining the RC limit and notice that since the IW functions $\psi_2$ and $\psi_3$ always appear in the combination $\psi^{\prime}=\psi_3+w\,\psi_2$ we fit directly $\psi^{\prime}$.  After linearly expanding the IW functions as in eq. \eqref{linearIW} we further discover that the form factors are independent on $\psi^{(1)}_1$ leaving us with  13 unknown IW parameter to fit to the lattice data. The $\chi^2$ minimisation yields an excellent fit with  $\chi^2/d.o.f.\sim 0.89$. The best fit  and its uncertainties are summarized in \cref{tab:RC} and the correlation matrix is reported in \cref{tab:RCfull} in \cref{app:G}.
\begin{table}[!htb!]
\begin{center}
\begin{tabular}{ll}
\toprule
Parameter & Best fit \\
\midrule
$\zeta^{(0)}$ & $0.52\pm0.16$ \\
$\zeta^{(1)}$ & $-6.11\pm1.27$ \\
$\zeta_{SL}^{(0)}$ & $0.15\pm0.01$ \\
$\zeta_{SL}^{(1)}$ & $-0.38\pm0.11$ \\
$\eta_{kin,c}^{(0)}$ & $-0.26\pm0.44$ \\
$\eta_{kin,c}^{(1)}$ & $9.77\pm3.08$\\
$\eta_{mag,c}^{(0)}$ & $0.01\pm0.10$\\
$\eta_{mag,c}^{(1)}$ & $-0.15\pm1.44$ \\
$\eta_{mag,b}^{(0)}$ & $-0.08\pm0.06$ \\
$\eta_{mag,b}^{(1)}$ & $0.25\pm0.70$\\
$\psi_1^{(0)}$ &  $1.58\pm0.72$ \\
$\psi^{'(0)}$ &  $0.82\pm0.05$ \\
$\psi^{'(1)}$ &  $-1.13\pm0.52$ \\
\bottomrule
\end{tabular}
\end{center}
\caption{\em Best fit points and uncertainties for the IW parameters in the RC limit.}
\label{tab:RC}
\end{table}
\\
It is time to consider the independent VC limit and notice that for the linearized IW functions notice the form factors are $\beta^{(1)}_1$ independent. Therefore, we are left with 9 unknown IW parameters. Also in this case we find an excellent fit with $\chi^2/d.o.f.\sim 0.84$ as summarized in \cref{tab:VC}. The correlation matrix is reported in \cref{tab:VCfull} in \cref{app:G}.
\begin{table}[!htb!]
\begin{center}
\begin{tabular}{ll}
\toprule
Parameter & Best fit \\
\midrule
$\zeta^{(0)}$ & $0.52\pm0.16$ \\
$\zeta^{(1)}$ & $-5.97\pm1.24$  \\
$\zeta_{SL}^{(0)}$ & $0.15\pm0.01$ \\
$\zeta_{SL}^{(1)}$ & $-0.31\pm0.10$ \\
$\eta_{kin,c}^{(0)}$ &$-0.24\pm0.42$ \\
$\eta_{kin,c}^{(1)}$ & $9.48\pm3.06$ \\
$\beta_{1}^{(0)}$ & $1.58\pm0.71$ \\
$\beta_{3}^{(0)}$ &$0.81\pm0.05$\\
$\beta_{3}^{(1)}$ & $-0.96\pm0.50$\\
\bottomrule
\end{tabular}
\end{center}
\caption{\em Best fit points and uncertainties for the IW parameters in the VC limit.}
\label{tab:VC}
\end{table}\\

As a positive consistency check one observes that for the distinct  RC and VC limits reported in the tables \ref{tab:RC} and \ref{tab:VC}  the common IW form factors are compatible. 

In an effort to be as complete as possible we follow  the literature \cite{Bordone:2021bop,Bernlochner:2018kxh} and further  investigate the possibility to use lattice data for the (axial) vector currents as input to predict the (pseudo) tensor contributions  for both the RC and VC limits. Subsequently, we compare these results again with the lattice one. 
In this case, the $\chi^2$ minimisation yields $\chi^2/d.o.f.\sim 0.84$ for the RC limit and $\chi^2/d.o.f.\sim 0.72$ for the VC limit. The results of our fit are shown in \cref{tab:params} and the corresponding correlation matrices are shown in \cref{tab:RCred} and \cref{tab:VCred} in \cref{app:G}. In figure \ref{figurepred} we show the comparison between the LQCD  results of \cite{Meinel:2021mdj} (orange band) with the predictions for tensor and pseudo-tensor form factors derived from  fitting only the (axial) vector ones (\cref{tab:params}). The RC and VC limits are illustrated as the blue and green band respectively.  Our predictions for the form factors are compatible with the lattice results at $1-\sigma$ level.
The exception arises from the pseudo-tensor form factor $\tilde{h}_{\perp^\prime}^{(\frac{3}{2}^{-})} $ in the VC limit (see Fig. \ref{fig:subfig_h}). The larger than one sigma deviation occurs because in the VC limit the $1/m_c^2$ corrections are not entirely captured.  

Furthermore we observe  large uncertainties of the chromomagnetic IW functions in the RC limit, which are in fact compatible with zero. This  arises because the available LQCD data are confined to the low-recoil region, while the chromomagnetic corrections are expected to be significant in the large recoil region. As a result, it becomes meaningful to consider a scenario where we set the chromomagnetic IW functions to zero. It can be seen from eq. \eqref{betapsi} that this scenario corresponds exactly to the VC limit. Importantly, even in this case we obtain an excellent agreement with the lattice data.

For completeness, we have performed the same analysis above also using the baryon masses rather than the quark masses in the HQE. Here we observed that the overall qualitative picture remains unchanged with the second order coefficients getting substantially larger and $\eta_{kin,c}^{(1)}$ getting roughly 2 times  larger.

Our results clearly show that, both in the RC and VC limits,    $\mathcal{O}(1/m_c^2)$ corrections are essential to  match the lattice data, corroborating and quantifying the expectations of \cite{Papucci:2021pmj}.

\begin{table}[h]
\begin{subtable}{0.5\textwidth}
\centering
\begin{tabular}{cc}
\toprule
Parameter & Best fit \\
\midrule
$\zeta^{(0)}$ & $0.43\pm 0.50$  \\
$\zeta^{(1)}$ & $-6.59 \pm 1.62$ \\
$\zeta_{SL}^{(0)}$ & $0.16\pm 0.01$ \\
$\zeta_{SL}^{(1)}$& $-0.58\pm0.17$ \\
$\eta_{kin,c}^{(0)}$ &$ 0.01\pm1.30$ \\
$\eta_{kin,c}^{(1)}$ & $10.72\pm4.01$ \\
$\eta_{mag,c}^{(0)}$ & $-0.04\pm0.12$\\
$\eta_{mag,c}^{(1)}$  & $-0.07\pm1.61$ \\
$\eta_{mag,b}^{(0)}$  & $-0.01\pm 0.07$ \\
$\eta_{mag,b}^{(1)}$  & $-0.75\pm 0.93$ \\
$\psi_1^{(0)}$& $1.48\pm1.01$\\
$\psi^{'(0)}$ & $0.83\pm0.10$ \\
$\psi^{'(1)}$ & $-1.76\pm0.68$ \\
\bottomrule
\end{tabular}
\caption{}
\label{tab:params1}
\end{subtable}%
\begin{subtable}{0.5\textwidth}
\centering
\begin{tabular}{c c}
\toprule
Parameter & Best fit \\
\midrule
$\zeta^{(0)}$ & $0.52\pm 0.43$ \\
$\zeta^{(1)}$ & $-6.49\pm 1.55$ \\
$\zeta_{SL}^{(0)}$ & $0.16\pm0.01$ \\
$\zeta_{SL}^{(1)}$ & $-0.57\pm0.17$ \\
$\eta_{kin,c}^{(0)}$ & $-0.21\pm1.10$ \\
$\eta_{kin,c}^{(1)}$ & $10.56\pm 3.94$ \\
$\beta_{1}^{(0)}$ & $1.42\pm0.96$ \\
$\beta_{3}^{(0)}$ & $0.84\pm0.09$ \\
$\beta_{3}^{(1)}$ & $-1.69\pm0.64$ \\
\bottomrule
\end{tabular}
\caption{}
\label{tab:params2}
\end{subtable}
\caption{\em Best fit points and uncertainties for the IW parameters in the RC (a) and VC (b) limit.}
\label{tab:params}
\end{table}

\begin{figure}[h!]
\captionsetup[subfigure]{labelformat=empty}
    \centering
    \begin{subfigure}{0.48\textwidth}
        \includegraphics[width=\linewidth]{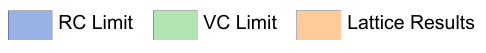}
    \end{subfigure}
    \hfill
    \\
    \begin{subfigure}{0.48\textwidth}
        \includegraphics[width=\linewidth]{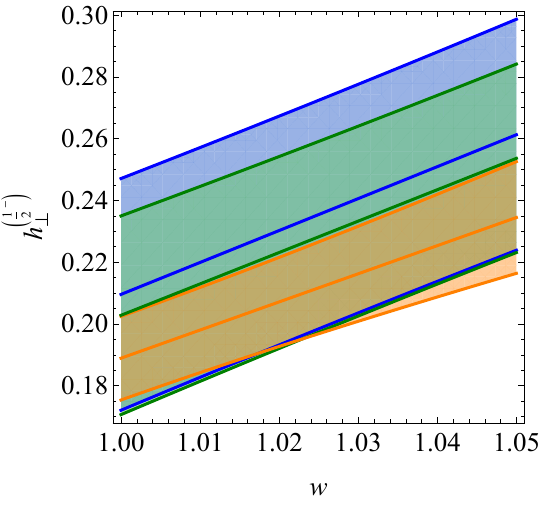}
        \subcaption{\em{(a)}}
    \label{fig:subfig_a}
    \end{subfigure}
    \hfill
    \begin{subfigure}{0.48\textwidth}
        \includegraphics[width=\linewidth]{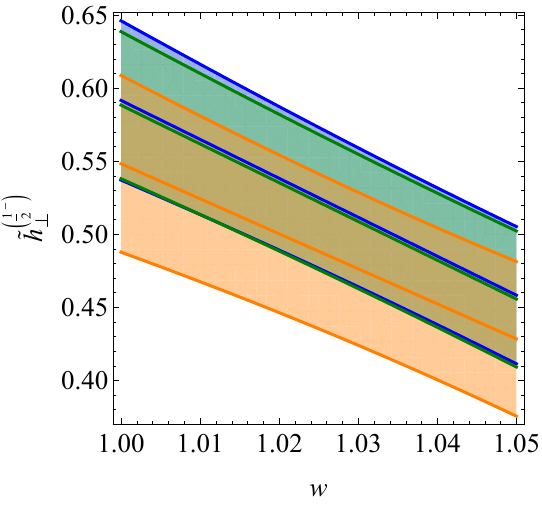}
        \subcaption{\em{(b)}}
    \label{fig:subfig_b}
    \end{subfigure}
\end{figure}
\newpage
\begin{figure}[h!]\ContinuedFloat
\captionsetup[subfigure]{labelformat=empty}
    \centering
    \begin{subfigure}{0.48\textwidth}
        \includegraphics[width=\linewidth]{legend.pdf}
    \end{subfigure}
    \hfill
    \\
    \begin{subfigure}{0.48\textwidth}
        \includegraphics[width=\linewidth]{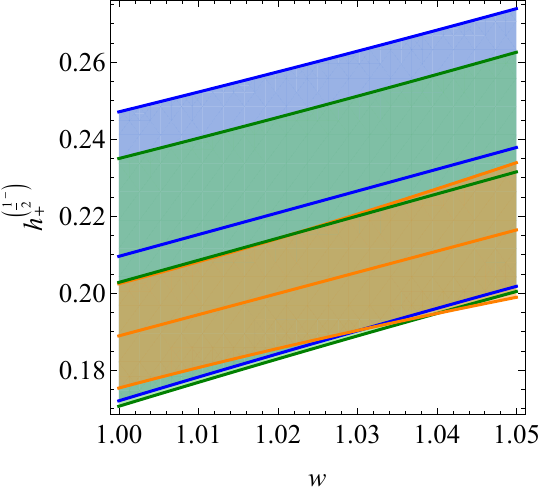}
        \subcaption{\em{(c)}}
    \label{fig:subfig_c}
    \end{subfigure}
    \hfill
    \begin{subfigure}{0.48\textwidth}
        \includegraphics[width=\linewidth]{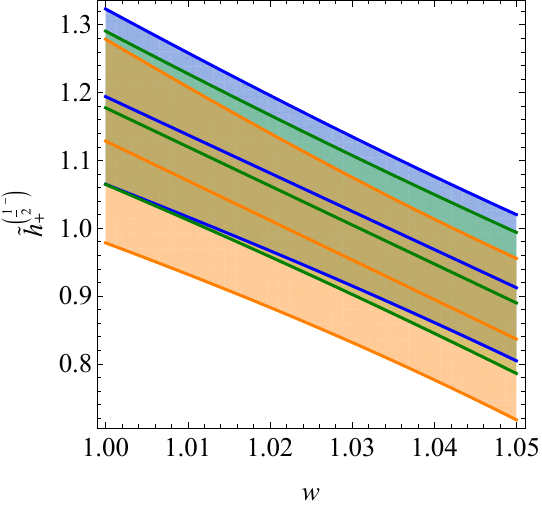}
        \subcaption{\em{(d)}}
    \label{fig:subfig_d}
    \end{subfigure}
    \\
    \vskip .5cm
    \begin{subfigure}{0.5\textwidth}
        \includegraphics[width=\linewidth]{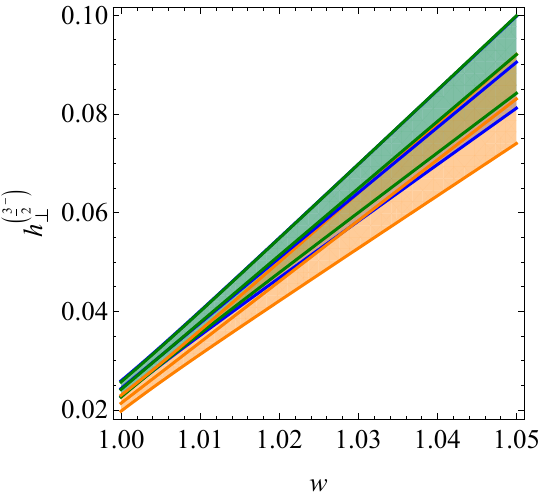}
        \subcaption{\em{(e)}}
    \label{fig:subfig_e}
    \end{subfigure}
    \hfill
    \begin{subfigure}{0.48\textwidth}
        \includegraphics[width=\linewidth]{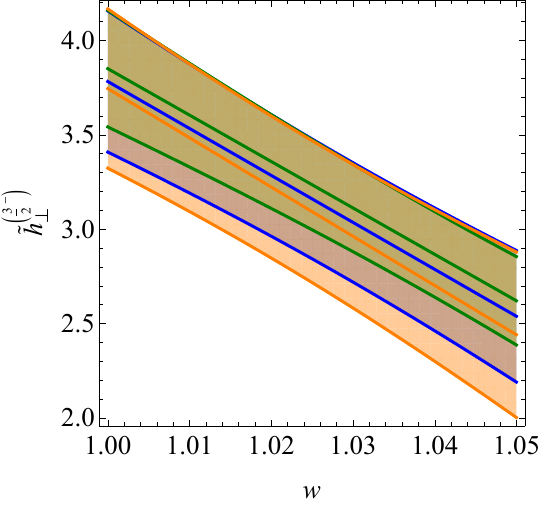}
        \subcaption{\em{(f)}}
    \label{fig:subfig_f}
    \end{subfigure}
    \end{figure}
    \begin{figure}[h!]\ContinuedFloat
    \captionsetup[subfigure]{labelformat=empty}
    \centering
    \begin{subfigure}{0.48\textwidth}
        \includegraphics[width=\linewidth]{legend.pdf}
    \end{subfigure}
    \hfill
    \\
    \begin{subfigure}{0.51\textwidth}
        \includegraphics[width=\linewidth]{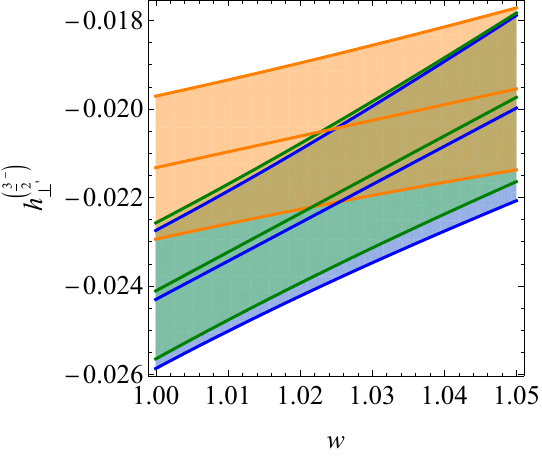}
        \subcaption{\em{(g)}}
    \label{fig:subfig_g}
    \end{subfigure}
    \hfill
    \begin{subfigure}{0.47\textwidth}
        \includegraphics[width=\linewidth]{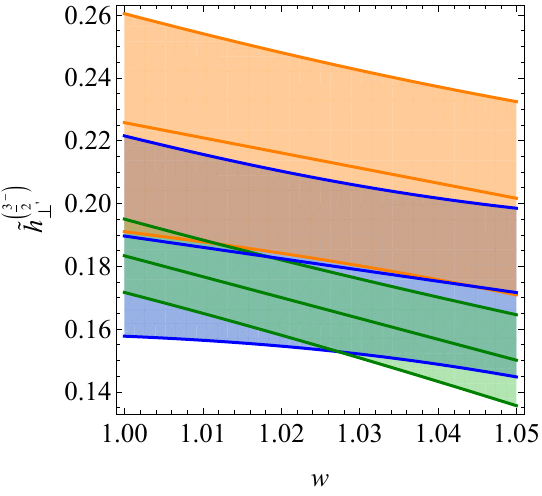}
        \subcaption{\em{(h)}}
    \label{fig:subfig_h}
    \end{subfigure}
    \\
    \vskip .5cm
    \begin{subfigure}{0.51\textwidth}
        \includegraphics[width=\linewidth]{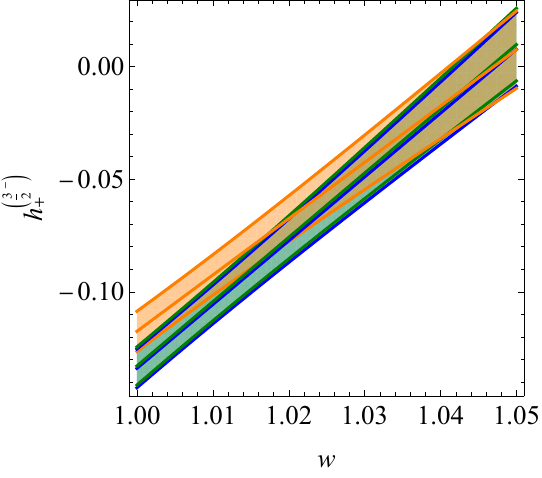}
        \subcaption{\em{(i)}}
    \label{fig:subfig_i}
    \end{subfigure}
    \hfill
    \begin{subfigure}{0.47\textwidth}
        \includegraphics[width=\linewidth]{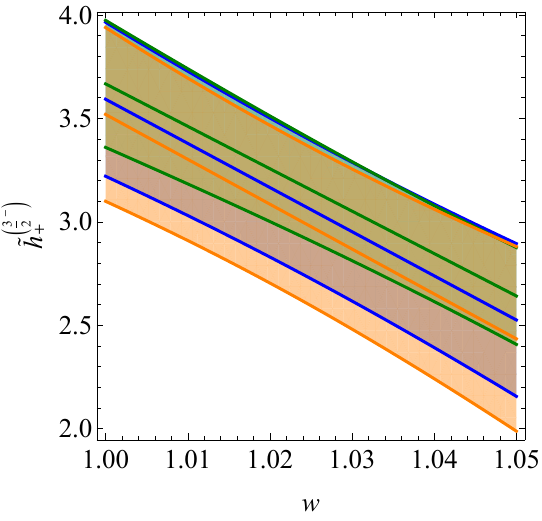}
        \subcaption{\em{(l)}}
    \label{fig:subfig_l}
    \end{subfigure}
    \caption{\em{The figures for each  sub-panels display the LQCD  results of \cite{Meinel:2021mdj} (depicted as the orange band) alongside the $1-\sigma$ level predictions for tensor and pseudo-tensor form factors coming from the independent RC (blue band) and VC (green band) limits in \cref{tab:params}.  }}
    \label{figurepred}
\end{figure}

\FloatBarrier
\section{Conclusions}
\label{conclusions}
We presented the set of form factors for the $\Lambda_b \to \Lambda_c^{*}$ transitions, accounting for the next-to-next-to-leading $\mathcal{O}(1/m_c^2)$ corrections in the RC and VC limits. We  determined the IW independent parameters via a fit to the lattice data from \cite{Meinel:2021mdj}. The latter compensate for the lack of experimental data for the $\Lambda_b \to \Lambda_c^{*} \ell \nu$ decay process. For the fits, we first enacted a $\chi^2$ minimization procedure using the entire lattice dataset and then  repeated the process using only the vector and axial form factors to generate predictions for the tensor and pseudo-tensor form factors. We discovered a notable agreement with the LQCD data, effectively resolving the previous tension underlined in \cite{Papucci:2021pmj}.

Our results demonstrates that  it is essential, to agree with the lattice data, to consider next-to-next-to leading order power corrections $\mathcal{O}(1/m_c^2)$ for these processes. Furthermore, barring potential lattice issues, these corrections are large and indicate a slow convergence of the series.  An independent test of both the effective approach and the lattice results will come from either more refined lattice simulations and/or future experiments.  We further stress that our findings  are  applicable only close to the zero-recoil region.  Future experimental measurements and theoretical efforts are needed to access the large-recoil regime which will allow us to fully describe  the physics of the  $\Lambda_b \to \Lambda_c^{*}$ decay processes.

\vskip 1cm
\noindent{\bf Acknowledgments}
\vskip .2cm
\noindent
We are indebted to Marzia Bordone for suggesting this project and for a fruitful initial collaboration.    We also thank Stefan Meinel, Gumaro Rendon, Michele Papucci, Dean~J.~Robinson, Michele Della Morte, Benjamin J\"ager and Antonio Rago for comments on the manuscript. 
The work of FS is partially supported by the Carlsberg Foundation, grant CF22-0922. 

\newpage

\appendix

\section{Rarita-Schwinger spinor}
\label{app:RSspinor}
 In the chiral representation,
a spinor $u$ with momentum 
\begin{equation}
    p^\mu = (p^0, |\vec{p}|\sin\theta\cos\phi, |\vec{p}|\sin\theta\sin\phi, |\vec{p}|\cos\theta), \quad  p^2 = m^2\ ,
\end{equation}
and helicity $h = \pm 1/2$ in the rest frame, can be written as \cite{Haber:1994pe, Boer:2018vpx}
\begin{align}
    u(p, h=+1/2) & = \frac{1}{\sqrt{2 (p^0 + m)}} \left[\begin{matrix}
    +(p^0 + m - |\vec{p}|) & \cos(\theta/2) & \\
    +(p^0 + m + |\vec{p}|) & \sin(\theta/2) & \exp(+i \phi)\\
    +(p^0 + m + |\vec{p}|) & \cos(\theta/2) & \\
    +(p^0 + m - |\vec{p}|) & \sin(\theta/2) & \exp(+i \phi)
    \end{matrix}\right] \ ,\\
    u(p, h=-1/2) & = \frac{1}{\sqrt{2 (p^0 + m)}} \left[\begin{matrix}
    -(p^0 + m - |\vec{p}|) & \sin(\theta/2) & \exp(-i\phi)\\
    +(p^0 + m + |\vec{p}|) & \cos(\theta/2) & \\
    -(p^0 + m + |\vec{p}|) & \sin(\theta/2) & \exp(-i\phi)\\
    +(p^0 + m - |\vec{p}|) & \cos(\theta/2) &
    \end{matrix}\right]\,.
\end{align}

We characterize a state with quantum numbers $J^P = 3/2^-$ using the spin-$3/2$ component $u_{(3/2)}^\alpha$ stemming from the projection of a general Rarita-Schwinger object $u^\alpha_\text{RS}(k, \eta) = \eta^\alpha u(k)$,
\begin{equation}
\begin{aligned}
    u_{(3/2)}^\alpha(k, \eta, s_c)
        & = \left[\eta^\alpha - \frac{1}{3}\left(\gamma^\alpha + \frac{k^\alpha}{\mLamCst}\right)\slashed{\eta}\right] u(k, s_c)\\
        & = \left[g^\alpha{}_\beta - \frac{1}{3}\left(\gamma^\alpha + \frac{k^\alpha}{\mLamCst}\right) \gamma_\beta\right]\, u_\text{RS}^\beta(k, \eta(\lambda), s_c)\\
        & \equiv \left[P_{3/2}\right]^{\alpha}{}_\beta\, u_\text{RS}^\beta(k, \eta(\lambda), s_c)\,.
\end{aligned}
\end{equation}
Here, $u(k, s_c)$ denotes the spin-$1/2^+$ spinor of four momentum $k$ and
rest-frame helicity $s_c = \pm 1/2$, as explicitly presented above. The vector $\eta$ is a polarisation vector with $J^P = 1^-$. 
Similarly, we can describe the state with spin-parity $J^{P}=1/2^-$, by means of the projection onto the spin-$1/2$ component, as follows:
\begin{align}
u^{\alpha}_{(1/2)}(k,\eta,s_c)
&=\frac{1}{3}\left[\gamma^{\alpha}+\frac{k^{\alpha}}{\mLamCst}\right]\slashed{\eta} \ u(k,s_c) \\
&=\frac{1}{3}\left[\gamma^{\alpha}+\frac{k^{\alpha}}{\mLamCst}\right]\gamma_{\beta} \ u_{\mathrm{RS}}^{\beta}(k,\eta(\lambda),s_c)  \\
       & \equiv \left[P_{1/2}\right]^{\alpha}{}_\beta\, u_\text{RS}^\beta(k, \eta(\lambda), s_c)\,.
       \end{align}
The Rarita-Schwinger fields satisfy the following useful identities
    \begin{align}
    k_\alpha u_{(3/2)}^\alpha(k, \eta, s_c) & =0=k_\alpha u_{(1/2)}^\alpha(k, \eta, s_c)   \,,\\
    \gamma_\alpha u_{(3/2)}^\alpha(k, \eta, s_c) & = 0\,,\\
    -i \sigma_{\alpha\beta} \, u_{(3/2)}^\alpha(k, \eta, s_c) & = u_{(3/2)}^\beta(k, \eta, s_c)\,.
\end{align}

\section{Form factor basis}
\label{app:ff-details}
Here, we present the spin structures $\Gamma^{\alpha\mu}_{J,i}$ that contribute to the transition $\Lambda_b\to\Lambda_c^{*}$. \\ 
For the vector and axial currents we follow the notation given in \cite{Boer:2018vpx}, while for the tensor and pseudo-tensor currents we use the parametrisation of Ref. \cite{Bordone:2021bop}.
For the vector current ($J=V$) mediating the transition to $\LamCst[2595]$ we find
\begin{equation}
\begin{aligned}
\label{eq:FF-12-basis-V}
    \Gamma_{V,(1/2,t)}^{\alpha\mu} & =  \frac{\sqrt{4 \mLamB \mLamCst}}{\sqrt{s_+}}\frac{2 \mLamCst}{\sqrt{s_ +s_-}} p^\alpha\,\frac{\mLamB - \mLamCst}{\sqrt{q^2}} \, \frac{q^\mu}{\sqrt{q^2}}\,,\\
\Gamma_{V,(1/2,0)}^{\alpha\mu}     & = \frac{\sqrt{4 \mLamB \mLamCst}}{\sqrt{s_-}} \frac{2 \mLamCst}{\sqrt{s_ +s_-}} p^\alpha\,\frac{\mLamB + \mLamCst}{s_+}\left[(p + k)^\mu - \frac{\mLamB^2 - \mLamCst^2}{q^2} q^\mu\right]\,,\\
\Gamma_{V,(1/2,\perp)}^{\alpha\mu} & = \frac{\sqrt{4 \mLamB \mLamCst}}{\sqrt{s_-}} \frac{2 \mLamCst}{\sqrt{s_ +s_-}} p^\alpha\,\left[\gamma^\mu - \frac{2 \mLamCst}{s_+} p^\mu - \frac{2 \mLamB}{s_+} k^\mu\right]\,,
\end{aligned}
\end{equation}
for the axial current ($J=A$) we have:
\begin{equation}
\begin{aligned}
\label{eq:FF-12-basis-A}
    \Gamma_{A,(1/2,t)}^{\alpha\mu} & =  \frac{\sqrt{4 \mLamB \mLamCst}}{\sqrt{s_-}}\frac{2 \mLamCst}{\sqrt{s_ +s_-}} p^\alpha\,\frac{\mLamB + \mLamCst}{\sqrt{q^2}} \, \frac{q^\mu}{\sqrt{q^2}}\,,\\
\Gamma_{A,(1/2,0)}^{\alpha\mu}     & =  \frac{\sqrt{4 \mLamB \mLamCst}}{\sqrt{s_+}}\frac{2 \mLamCst}{\sqrt{s_ +s_-}} p^\alpha\,\frac{\mLamB - \mLamCst}{s_-}\left[(p + k)^\mu - \frac{\mLamB^2 - \mLamCst^2}{q^2} q^\mu\right]\,,\\
\Gamma_{A,(1/2,\perp)}^{\alpha\mu} & =  \frac{\sqrt{4 \mLamB \mLamCst}}{\sqrt{s_+}}\frac{2 \mLamCst}{\sqrt{s_ +s_-}} p^\alpha\,\left[\gamma^\mu + \frac{2 \mLamCst}{s_-} p^\mu - \frac{2 \mLamB}{s_-} k^\mu\right]\,.
\end{aligned}
\end{equation}
For the tensor current ($J=T$) we find:
\begin{align}
\Gamma_{T,(1/2,0)}^{\alpha\mu}     & = \frac{\sqrt{4 \mLamB \mLamCst}}{\sqrt{s_+}}\, \frac{q^2}{s_+ s_-} p^\alpha\left[(p + k)^\mu - \frac{\mLamB^2 - \mLamCst^2}{q^2} q^\mu\right]\,,\\
\Gamma_{T,(1/2,\perp)}^{\alpha\mu} & = \frac{\sqrt{4 \mLamB \mLamCst}}{\sqrt{s_+}}\, \frac{\mLamB+\mLamCst}{s_-} p^\alpha\,\left[\gamma^\mu -2\frac{\mLamCst}{s_+} p^\mu  -2\frac{\mLamB}{s_+} k^\mu\right]\,,
\label{eq:FF-12-basis-T}
\end{align}

while for the pseudo-tensor current ($J=T5$) we obtain:

\begin{align}
\Gamma_{T5,(1/2,0)}^{\alpha\mu}     & = \frac{\sqrt{4 \mLamB \mLamCst}}{\sqrt{s_-}}\, \frac{q^2}{s_+ s_-} p^\alpha\left[(p + k)^\mu - \frac{\mLamB^2 - \mLamCst^2}{q^2} q^\mu\right]\,,\\
\Gamma_{T5,(1/2,\perp)}^{\alpha\mu} & = \frac{\sqrt{4 \mLamB \mLamCst}}{\sqrt{s_-}}\, \frac{\mLamB-\mLamCst}{s_+} p^\alpha\,\left[\gamma^\mu +2\frac{\mLamCst}{s_-} p^\mu  -2\frac{\mLamB}{s_-} k^\mu\right]\,.
\label{eq:FF-12-basis-T5}
\end{align}

When we consider the final state $\LamCst[2625]$, for the vector current ($J=V$) we get:
\begin{equation}
\begin{aligned}
\label{eq:FF-32-basis-V}
    \Gamma_{V,(1/2,t)}^{\alpha\mu} & = \frac{\sqrt{4 \mLamB \mLamCst}}{\sqrt{s_+}}\, \frac{2 \mLamCst}{\sqrt{s_ +s_-}} p^\alpha\,\frac{\mLamB - \mLamCst}{\sqrt{q^2}} \, \frac{q^\mu}{\sqrt{q^2}}\,,\\
\Gamma_{V,(1/2,0)}^{\alpha\mu}     & = \frac{\sqrt{4 \mLamB \mLamCst}}{\sqrt{s_-}}\, \frac{2 \mLamCst}{\sqrt{s_ +s_-}} p^\alpha\,\frac{\mLamB + \mLamCst}{s_+}\left[(p + k)^\mu - \frac{\mLamB^2 - \mLamCst^2}{q^2} q^\mu\right]\,,\\
\Gamma_{V,(1/2,\perp)}^{\alpha\mu} & = \frac{\sqrt{4 \mLamB \mLamCst}}{\sqrt{s_-}}\, \frac{2 \mLamCst}{\sqrt{s_ +s_-}} p^\alpha\,\left[\gamma^\mu - \frac{2 \mLamCst}{s_+} p^\mu - \frac{2 \mLamB}{s_+} k^\mu\right]\,,\\
\Gamma_{V,(3/2,\perp)}^{\alpha\mu} & = \frac{\sqrt{4 \mLamB \mLamCst}}{\sqrt{s_-}}\, \frac{-4 i \eps^{\alpha\mu p k}}{\sqrt{s_+ s_-}} \gamma_5 + \Gamma_{V,(1/2,\perp)}\,,
\end{aligned}
\end{equation}
while for the axial current ($J=A$) we use:
\begin{equation}
\begin{aligned}
\label{eq:FF-32-basis-A}
    \Gamma_{A,(1/2,t)}^{\alpha\mu} & = \frac{\sqrt{4 \mLamB \mLamCst}}{\sqrt{s_-}}\, \frac{2 \mLamCst}{\sqrt{s_ +s_-}} p^\alpha\,\frac{\mLamB + \mLamCst}{\sqrt{q^2}} \, \frac{q^\mu}{\sqrt{q^2}}\,,\\
\Gamma_{A,(1/2,0)}^{\alpha\mu}     & = \frac{\sqrt{4 \mLamB \mLamCst}}{\sqrt{s_+}}\, \frac{2 \mLamCst}{\sqrt{s_ +s_-}} p^\alpha\,\frac{\mLamB - \mLamCst}{s_-}\left[(p + k)^\mu - \frac{\mLamB^2 - \mLamCst^2}{q^2} q^\mu\right]\,,\\
\Gamma_{A,(1/2,\perp)}^{\alpha\mu} & = \frac{\sqrt{4 \mLamB \mLamCst}}{\sqrt{s_+}}\, \frac{2 \mLamCst}{\sqrt{s_ +s_-}} p^\alpha\,\left[\gamma^\mu + \frac{2 \mLamCst}{s_-} p^\mu - \frac{2 \mLamB}{s_-} k^\mu\right]\,,\\
\Gamma_{A,(3/2,\perp)}^{\alpha\mu} & = \frac{\sqrt{4 \mLamB \mLamCst}}{\sqrt{s_+}}\, \frac{-4 i \eps^{\alpha\mu p k}}{\sqrt{s_+ s_-}} \gamma_5 - \Gamma_{A,(1/2,\perp)}\,.
\end{aligned}
\end{equation}

For the tensor current ($J=T$) we find:

\begin{align}
\Gamma_{T,(1/2,0)}^{\alpha\mu}     & = \frac{\sqrt{4 \mLamB \mLamCst}}{\sqrt{s_+}}\, \frac{q^2}{s_+ s_-} p^\alpha\left[(p + k)^\mu - \frac{\mLamB^2 - \mLamCst^2}{q^2} q^\mu\right]\,,\\
\Gamma_{T,(1/2,\perp)}^{\alpha\mu} & = \frac{\sqrt{4 \mLamB \mLamCst}}{\sqrt{s_+}}\, \frac{\mLamB+\mLamCst}{s_-} p^\alpha\,\left[\gamma^\mu -2\frac{\mLamCst}{s_+} p^\mu  -2\frac{\mLamB}{s_+} k^\mu\right]\,,\\
\Gamma_{T,(3/2,\perp)}^{\alpha\mu} & = \frac{\sqrt{4 \mLamB \mLamCst}}{\sqrt{s_+}}\, \left[g^{\alpha\mu}+\frac{\mLamCst}{s_-}p^\alpha\left(\gamma^\mu-2\frac{1}{\mLamCst} k^\mu+ 2\frac{\mLamCst}{s_+}p^\mu +2\frac{\mLamB}{s_+}k^\mu\right)\right]\,,
\label{eq:FF-32-basis-T}
\end{align}

and for the pseudo-tensor current ($J=T5$) we obtain:

\begin{align}
\Gamma_{T5,(1/2,0)}^{\alpha\mu}     & = \frac{\sqrt{4 \mLamB \mLamCst}}{\sqrt{s_-}}\, \frac{q^2}{s_+ s_-} p^\alpha\left[(p + k)^\mu - \frac{\mLamB^2 - \mLamCst^2}{q^2} q^\mu\right]\,,\\
\Gamma_{T5,(1/2,\perp)}^{\alpha\mu} & = \frac{\sqrt{4 \mLamB \mLamCst}}{\sqrt{s_-}}\, \frac{\mLamB-\mLamCst}{s_+} p^\alpha\,\left[\gamma^\mu +2\frac{\mLamCst}{s_-} p^\mu  -2\frac{\mLamB}{s_-} k^\mu\right]\,,\\
\Gamma_{T5,(3/2,\perp)}^{\alpha\mu} & = \frac{\sqrt{4 \mLamB \mLamCst}}{\sqrt{s_-}}\, \left[g^{\alpha\mu}-\frac{\mLamCst}{s_+}p^\alpha\left(\gamma^\mu+2\frac{1}{\mLamCst} k^\mu-2\frac{\mLamCst}{s_-}p^\mu +2\frac{\mLamB}{s_-}k^\mu\right)\right]\,.
\label{eq:FF-32-basis-T5}
\end{align}
In the equations presented above, we have used the convention $\epsilon^{0123} = - \epsilon_{0123} = +1$ for the Levi-Civita tensor.

\subsection{Helicity amplitudes}
\label{app:helicity-amplitude-full-QCD}

By applying the principles of angular momentum composition,  the allowed on-shell amplitudes for the final state with a total angular momentum $J=1/2$ are \cite{Boer:2018vpx}:
\begin{equation}
\label{hel-amp12}
\begin{aligned}
    \mathcal{A}^{(1/2)}_\Gamma(+1/2, +1/2,  0) & \equiv -\sqrt{\frac{1}{3}} \mathcal{A}_\Gamma(+1/2, +1/2, 0, 0) + \sqrt{\frac{2}{3}} \mathcal{A}_\Gamma(+1/2, -1/2, +1, 0)\,,\\
    \mathcal{A}^{(1/2)}_\Gamma(+1/2, +1/2,  t) & \equiv -\sqrt{\frac{1}{3}} \mathcal{A}_\Gamma(+1/2, +1/2, 0, t) + \sqrt{\frac{2}{3}} \mathcal{A}_\Gamma(+1/2, -1/2, +1, t)\,,\\
    \mathcal{A}^{(1/2)}_\Gamma(+1/2, -1/2, -1) & \equiv \sqrt{\frac{1}{3}} \mathcal{A}_\Gamma(+1/2, -1/2, 0, -1) - \sqrt{\frac{2}{3}} \mathcal{A}_\Gamma(+1/2, +1/2, -1, -1)\,.
\end{aligned}
\end{equation}
On the other hand the complementary set of $J=3/2$ amplitudes reads as
\begin{equation}
\label{hel-amp32}
\begin{aligned}
    \mathcal{A}^{(3/2)}_\Gamma(+1/2, +3/2, +1) & \equiv \mathcal{A}_\Gamma(+1/2, +1/2, +1, +1)\,,\\
    \mathcal{A}^{(3/2)}_\Gamma(+1/2, +1/2,  0) & \equiv \sqrt{\frac{2}{3}} \mathcal{A}_\Gamma(+1/2, +1/2, 0, 0) + \sqrt{\frac{1}{3}} \mathcal{A}_\Gamma(+1/2, -1/2, +1, 0)\,,\\
    \mathcal{A}^{(3/2)}_\Gamma(+1/2, +1/2,  t) & \equiv \sqrt{\frac{2}{3}} \mathcal{A}_\Gamma(+1/2, +1/2, 0, t) + \sqrt{\frac{1}{3}} \mathcal{A}_\Gamma(+1/2, -1/2, +1, t)\,,\\
    \mathcal{A}^{(3/2)}_\Gamma(+1/2, -1/2, -1) & \equiv \sqrt{\frac{2}{3}} \mathcal{A}_\Gamma(+1/2, -1/2, 0,-1) + \sqrt{\frac{1}{3}} \mathcal{A}_\Gamma(+1/2, +1/2, -1,-1)\,.
\end{aligned}
\end{equation}
For the transitions to $J=1/2$ the set of amplitudes in eq. \eqref{hel-amp32} vanishes identically.
Similarly for transitions to $J=3/2$ the equations in eq. \eqref{hel-amp12} are zero.

We list the on-shell amplitudes for the transition $\Lambda_{b}\to\LamCst[2595]$.
For the vector current we find the following non-zero amplitudes:
\begin{align}
\label{eq:helamp12-vector-first}
    \mathcal{A}^{(1/2)}_V(+1/2, +1/2,  0) = \mathcal{A}^{(1/2)}_V(-1/2, -1/2,  0)
        & = -\sqrt{\frac{1}{3}} f_{1/2, 0} \frac{\mLamB +\mLamCst}{\sqrt{q^2}}  \sqrt{4\mLamB\mLamCst}\,,\\
    \mathcal{A}^{(1/2)}_V(+1/2, +1/2,  t) = \mathcal{A}^{(1/2)}_V(-1/2, -1/2,  t)
        & = -\sqrt{\frac{1}{3}} f_{1/2, t} \frac{\mLamB - \mLamCst}{\sqrt{q^2}}  \sqrt{4\mLamB\mLamCst}\,,\\
\label{eq:helamp12-vector-last}
    \mathcal{A}^{(1/2)}_V(+1/2, -1/2, -1) = \mathcal{A}^{(1/2)}_V(-1/2, +1/2, +1)
        & = -\sqrt{\frac{2}{3}} f_{1/2,\perp} \sqrt{4\mLamB\mLamCst} \, .
\end{align}
For the axialvector current we find similarly:
\begin{align}
\label{eq:helamp12-axialvector-first}
    \mathcal{A}^{(1/2)}_A(+1/2, +1/2,  0) = -\mathcal{A}^{(1/2)}_A(-1/2, -1/2,  0)
        & = -\sqrt{\frac{1}{3}} g_{1/2, 0} \frac{\mLamB - \mLamCst}{\sqrt{q^2}} \sqrt{4\mLamB\mLamCst}\,,\\
    \mathcal{A}^{(1/2)}_A(+1/2, +1/2,  t) = -\mathcal{A}^{(1/2)}_A(-1/2, -1/2,  t)
        & = -\sqrt{\frac{1}{3}} g_{1/2, t} \frac{\mLamB + \mLamCst}{\sqrt{q^2}}\sqrt{4\mLamB\mLamCst}\,,\\
\label{eq:helamp12-axialvector-last}
    \mathcal{A}^{(1/2)}_A(+1/2, -1/2, -1) = -\mathcal{A}^{(1/2)}_A(-1/2, +1/2, +1)
        & = +\sqrt{\frac{2}{3}} g_{1/2,\perp} \sqrt{4\mLamB\mLamCst}\,.
\end{align}

For the tensor current we find the following non-zero amplitude:
\begin{align}
\label{eq:helamp12-tensor-first}
    \mathcal{A}^{(1/2)}_T(+1/2, +1/2,  0) = \mathcal{A}^{(1/2)}_T(-1/2, -1/2,  0)
        & = -\sqrt{\frac{1}{3}} t_{1/2, 0} \frac{\mLamB \sqrt{q^2}}{\sqrt{\mLamB \mLamCst}} \,,\\ 
\label{eq:helamp12-tensor-last}
    \mathcal{A}^{(1/2)}_T(+1/2, -1/2, -1) = \mathcal{A}^{(1/2)}_T(-1/2, +1/2, +1)
        & = +\sqrt{\frac{2}{3}} t_{1/2,\perp} \frac{\mLamB (\mLamB+\mLamCst)}{\sqrt{\mLamB\mLamCst}}\,.
\end{align}

For the pseudo-tensor current we find similarly:
\begin{align}
\label{eq:helamp12-pseudotensor-first}
    \mathcal{A}^{(1/2)}_{T5}(+1/2, +1/2,  0) = -\mathcal{A}^{(1/2)}_{T5}(-1/2, -1/2,  0)
        & = -\sqrt{\frac{1}{3}} t^{5}_{1/2, 0} \frac{\mLamB \sqrt{q^2}}{\sqrt{\mLamB \mLamCst}} \,,\\ 
\label{eq:helamp12-pseudotensor-last}
    \mathcal{A}^{(1/2)}_{T5}(+1/2, -1/2, -1) = -\mathcal{A}^{(1/2)}_{T5}(-1/2, +1/2, +1)
        & = +\sqrt{\frac{2}{3}} t^{5}_{1/2,\perp} \frac{\mLamB (\mLamB-\mLamCst)}{\sqrt{\mLamB\mLamCst}}\,.
\end{align}
We list the on-shell amplitudes for the transition $\Lambda_{b}\to\LamCst[2625]$.

For the vector current we find the following non-zero amplitudes:
\begin{align}
\label{eq:helamp32-vector-first}
    \mathcal{A}^{(3/2)}_V(+1/2, +3/2, +1) = \mathcal{A}^{(3/2)}_V(-1/2, -3/2, -1)
        & = -2\,F_{3/2,\perp} \sqrt{4\,\mLamB\,\mLamCst}\,,\\
    \mathcal{A}^{(3/2)}_V(+1/2, +1/2,  0) = \mathcal{A}^{(3/2)}_V(-1/2, -1/2,  0)
        & = +\sqrt{\frac{2}{3}} F_{1/2, 0} \frac{\mLamB + \mLamCst}{\sqrt{q^2}} \sqrt{4\,\mLamB\,\mLamCst}\,,\\
    \mathcal{A}^{(3/2)}_V(+1/2, +1/2,  t) = \mathcal{A}^{(3/2)}_V(-1/2, -1/2,  t)
        & = +\sqrt{\frac{2}{3}} F_{1/2, t} \frac{\mLamB - \mLamCst}{\sqrt{q^2}} \sqrt{4\,\mLamB\,\mLamCst}\,,\\
\label{eq:helamp32-vector-last}
    \mathcal{A}^{(3/2)}_V(+1/2, -1/2, -1) = \mathcal{A}^{(3/2)}_V(-1/2, +1/2, +1)
        & = -\frac{2}{\sqrt{3}} F_{1/2,\perp} \sqrt{4\,\mLamB\,\mLamCst}\,.
\end{align}
For the axialvector current we find similarly:
\begin{align}
\label{eq:helamp32-axialvector-first}
    \mathcal{A}^{(3/2)}_A(+1/2, +3/2, +1) = -\mathcal{A}^{(3/2)}_A(-1/2, -3/2, -1)
        & = -2\,G_{3/2,\perp} \sqrt{4\,\mLamB\,\mLamCst}\,,\\
    \mathcal{A}^{(3/2)}_A(+1/2, +1/2,  0) = -\mathcal{A}^{(3/2)}_A(-1/2, -1/2,  0)
        & = +\sqrt{\frac{2}{3}} G_{1/2, 0} \frac{\mLamB - \mLamCst}{\sqrt{q^2}} \sqrt{4\,\mLamB\,\mLamCst}\,,\\
    \mathcal{A}^{(3/2)}_A(+1/2, +1/2,  t) = -\mathcal{A}^{(3/2)}_A(-1/2, -1/2,  t)
        & = +\sqrt{\frac{2}{3}} G_{1/2, t} \frac{\mLamB + \mLamCst}{\sqrt{q^2}} \sqrt{4\,\mLamB\,\mLamCst}\,,\\
\label{eq:helamp32-axialvector-last}
    \mathcal{A}^{(3/2)}_A(+1/2, -1/2, -1) = -\mathcal{A}^{(3/2)}_A(-1/2, +1/2, +1)
        & = +\frac{2}{\sqrt{3}} G_{1/2,\perp} \sqrt{4\,\mLamB\,\mLamCst}\,.
\end{align}

For the tensor current we find the following non-zero amplitudes:
\begin{align}
\label{eq:helamp32-tensor-first}
    \mathcal{A}^{(3/2)}_T(+1/2, +3/2, +1) = \mathcal{A}^{(3/2)}_T(-1/2, -3/2, -1)
        & = -2\,T_{3/2,\perp} \sqrt{\mLamB\,\mLamCst}\,,\\
    \mathcal{A}^{(3/2)}_T(+1/2, +1/2,  0) = \mathcal{A}^{(3/2)}_T(-1/2, -1/2,  0)
        & = -\sqrt{\frac{2}{3}} T_{1/2, 0} \frac{\mLamB \, \sqrt{q^2}}{ \sqrt{\mLamB\,\mLamCst}}\,,\\
\label{eq:helamp32-tensor-last}
    \mathcal{A}^{(3/2)}_T(+1/2, -1/2, -1) = \mathcal{A}^{(3/2)}_T(-1/2, +1/2, +1)
        & = \frac{2}{\sqrt{3}} T_{1/2,\perp} \frac{\mLamB\,(\mLamB+\mLamCst)}{\sqrt{\mLamB\,\mLamCst}}\, .
\end{align}

For the pseudo-tensor current we find similarly:
\begin{align}
\label{eq:helamp32-pseudotensor-first}
    \mathcal{A}^{(3/2)}_{T5}(+1/2, +3/2, +1) = -\mathcal{A}^{(3/2)}_{T5}(-1/2, -3/2, -1)
        & = 2\,T^{5}_{3/2,\perp} \sqrt{\mLamB\,\mLamCst}\,,\\
    \mathcal{A}^{(3/2)}_{T5}(+1/2, +1/2,  0) = -\mathcal{A}^{(3/2)}_{T5}(-1/2, -1/2,  0)
        & = \sqrt{\frac{2}{3}} T^{5}_{1/2, 0} \frac{\mLamB \, \sqrt{q^2}}{ \sqrt{\mLamB\,\mLamCst}}\,,\\
\label{eq:helamp32-pseudotensor-last}
    \mathcal{A}^{(3/2)}_{T5}(+1/2, -1/2, -1) = -\mathcal{A}^{(3/2)}_{T5}(-1/2, +1/2, +1)
        & = \frac{2}{\sqrt{3}} T^{5}_{1/2,\perp} \frac{\mLamB\,(\mLamB-\mLamCst)}{\sqrt{\mLamB\,\mLamCst}}\, .
\end{align}
We do not report the expressions of the on-shell amplitudes within the HQET framework. Nevertheless, one can readily reconstruct them by substituting the form factors provided in appendices \ref{ffsRC} and \ref{ffsVC} into the equations given above.
\subsection{Form factors at $\mathcal{O}(1/m_c^2)$ in the RC-limit}
\label{ffsRC}
Here we list the full set of form factors up to $\mathcal{O}(1/m_c^2, \theta^2)$.
For the final state $\LamCst[2595]$ the form factors are:
\begin{align}
   f_{1/2, 0} =&\frac{\zeta s_- \sqrt{s_+}}{2(\mLamB \mLamCst)^{3/2}}  \left(C_1+\frac{C_2 s_+}{2 \mLamB (\mLamB+\mLamCst)}+\frac{C_3 s_+}{2 \mLamCst (\mLamB+\mLamCst)}\right)+\nn \\+&\varepsilon_b \frac{ \sqrt{s_+} (\mLamB-\mLamCst)}{(\mLamB+\mLamCst)\sqrt{\mLamB \mLamCst}}  \left(\zeta \bar\Lambda'- 2 \zeta_{\text{SL}}-\frac{\zeta\bar\Lambda \left(\mLamB^2+\mLamCst^2-q^2\right)}{2 \mLamB \mLamCst} \right)+\nn \\+&\varepsilon_c \frac{ \sqrt{s_+} (\mLamB-\mLamCst)}{(\mLamB+\mLamCst)\sqrt{\mLamB \mLamCst}}  \left(\zeta \bar\Lambda- 2 \zeta_{\text{SL}}-\frac{\zeta\bar\Lambda' \left(\mLamB^2+\mLamCst^2-q^2\right)}{2 \mLamB \mLamCst} \right)+\nn \\+&\frac{s_-\sqrt{s_+}}{2(\mLamB\mLamCst)^{3/2}}(\varepsilon_c\eta^{(c)}_\text{kin} +\varepsilon_b \eta^{(b)}_{\text{mag}}-\varepsilon_c\eta^{(c)}_{\text{mag}}) +\nn \\+&\varepsilon_c^2 \frac{\sqrt{s_+}(\mLamB-\mLamCst)}{(\mLamB+\mLamCst) \sqrt{\mLamB \mLamCst}}\left(3(\psi_3 +w \psi_2) -\frac{s_- s_+ \psi_1 }{4  (\mLamB \mLamCst)^{2}}\right) \,, \\
 f_{1/2, t} =&\frac{\zeta s_+ \sqrt{s_-}}{2(\mLamB \mLamCst)^{3/2}}  \left(C_1+\frac{C_2 s_+}{2 \mLamB (\mLamB-\mLamCst)}+\frac{C_3 s_+}{2 \mLamCst (\mLamB-\mLamCst)}\right)+\nn \\+&\varepsilon_b \frac{ \sqrt{s_-} (\mLamB+\mLamCst)}{(\mLamB-\mLamCst)\sqrt{\mLamB \mLamCst}}  \left(\zeta \bar\Lambda'- 2 \zeta_{\text{SL}}-\frac{\zeta\bar\Lambda \left(\mLamB^2+\mLamCst^2-q^2\right)}{2 \mLamB \mLamCst} \right)+\nn \\+&\varepsilon_c \frac{ \sqrt{s_-} (\mLamB+\mLamCst)}{(\mLamB-\mLamCst)\sqrt{\mLamB \mLamCst}}  \left(\zeta \bar\Lambda- 2 \zeta_{\text{SL}}-\frac{\zeta\bar\Lambda' \left(\mLamB^2+\mLamCst^2-q^2\right)}{2 \mLamB \mLamCst} \right)+\nn \\+&\frac{s_+\sqrt{s_-}}{2(\mLamB\mLamCst)^{3/2}}(\varepsilon_c\eta^{(c)}_\text{kin}+\varepsilon_b \eta^{(b)}_{\text{mag}}-\varepsilon_c\eta^{(c)}_{\text{mag}}) +\nn \\+& \varepsilon_c^2 \frac{\sqrt{s_-}(\mLamB+\mLamCst)}{(\mLamB-\mLamCst) \sqrt{\mLamB \mLamCst}}\left(3(\psi_3 +w \psi_2) -\frac{s_- s_+ \psi_1 }{4  (\mLamB \mLamCst)^{2}}\right)\ ,  \\
 f_{1/2, \perp} =& \frac{C_1 \zeta  s_- \sqrt{s_+}}{2 (\mLamB \mLamCst)^{3/2}}+ \varepsilon_b \frac{\sqrt{s_+}}{\sqrt{\mLamB \mLamCst}} \left(\frac{\zeta \bar\Lambda  \left(\mLamB^2+\mLamCst^2-q^2\right)}{2 \mLamB \mLamCst}-\bar\Lambda'\zeta\right)+\nn \\+&\varepsilon_c \frac{\sqrt{s_+}}{\sqrt{\mLamB \mLamCst}} \left(\bar\Lambda\zeta-2\zeta_{\text{SL}}-\frac{\zeta \bar\Lambda'  \left(\mLamB^2+\mLamCst^2-q^2\right)}{2 \mLamB \mLamCst}\right)+\nn \\+&\frac{s_-\sqrt{s_+}}{2(\mLamB\mLamCst)^{3/2}}(\varepsilon_c\eta^{(c)}_\text{kin}-\varepsilon_c\eta^{(c)}_{\text{mag}})+\varepsilon_c^2 \frac{\sqrt{s_+}}{\sqrt{\mLamB \mLamCst}}\left(3(\psi_3 +w \psi_2) -\frac{s_- s_+ \psi_1 }{4  (\mLamB \mLamCst)^{2}}\right) \,, 
\end{align}
\newpage
\begin{align}
g_{1/2, 0} =& \frac{\zeta s_+ \sqrt{s_-}}{2(\mLamB \mLamCst)^{3/2}}  \left(C^5_1-\frac{C^5_2 s_-}{2 \mLamB (\mLamB-\mLamCst)}-\frac{C^5_3 s_-}{2 \mLamCst (\mLamB-\mLamCst)}\right)+\nn \\+&\varepsilon_b \frac{ \sqrt{s_-} (\mLamB+\mLamCst)}{(\mLamB-\mLamCst)\sqrt{\mLamB \mLamCst}}  \left(\zeta \bar\Lambda'+2 \zeta_{\text{SL}}-\frac{\zeta\bar\Lambda \left(\mLamB^2+\mLamCst^2-q^2\right)}{2 \mLamB \mLamCst} \right)+\nn \\+&\varepsilon_c \frac{ \sqrt{s_-} (\mLamB-\mLamCst)}{(\mLamB+\mLamCst)\sqrt{\mLamB \mLamCst}}  \left(\zeta \bar\Lambda- 2 \zeta_{\text{SL}}-\frac{\zeta\bar\Lambda' \left(\mLamB^2+\mLamCst^2-q^2\right)}{2 \mLamB \mLamCst} \right)+\nn \\+&\frac{s_+\sqrt{s_-}}{2(\mLamB\mLamCst)^{3/2}}(\varepsilon_c\eta^{(c)}_\text{kin}-\varepsilon_b \eta^{(b)}_{\text{mag}}-\varepsilon_c\eta^{(c)}_{\text{mag}}) +\nn \\+&\varepsilon_c^2 \frac{\sqrt{s_+}(\mLamB-\mLamCst)}{(\mLamB+\mLamCst) \sqrt{\mLamB \mLamCst}}\left(3(\psi_3 +w \psi_2) -\frac{s_- s_+ \psi_1 }{4  (\mLamB \mLamCst)^{2}}\right) \ , \\
    g_{1/2, t} =& \frac{\zeta s_- \sqrt{s_+}}{2(\mLamB \mLamCst)^{3/2}}  \left(C^5_1+\frac{C^5_2 s_+}{2 \mLamB (\mLamB+\mLamCst)}+\frac{C^5_3 s_+}{2 \mLamCst (\mLamB+\mLamCst)}\right)+\nn \\+&\varepsilon_b \frac{ \sqrt{s_+} (\mLamB+\mLamCst)}{(\mLamB-\mLamCst)\sqrt{\mLamB \mLamCst}}  \left(\zeta \bar\Lambda'+ 2 \zeta_{\text{SL}}-\frac{\zeta\bar\Lambda \left(\mLamB^2+\mLamCst^2-q^2\right)}{2 \mLamB \mLamCst} \right)+\nn \\+&\varepsilon_c \frac{ \sqrt{s_+} (\mLamB+\mLamCst)}{(\mLamB-\mLamCst)\sqrt{\mLamB \mLamCst}}  \left(\zeta \bar\Lambda- 2 \zeta_{\text{SL}}-\frac{\zeta\bar\Lambda' \left(\mLamB^2+\mLamCst^2-q^2\right)}{2 \mLamB \mLamCst} \right)+\nn \\+&\frac{s_-\sqrt{s_+}}{2(\mLamB\mLamCst)^{3/2}}(\varepsilon_c \eta^{(c)}_\text{kin}-\varepsilon_b \eta^{(b)}_{\text{mag}}-\varepsilon_c\eta^{(c)}_{\text{mag}}) +\nn \\+&\varepsilon_c^2 \frac{\sqrt{s_+}(\mLamB-\mLamCst)}{(\mLamB+\mLamCst) \sqrt{\mLamB \mLamCst}}\left(3(\psi_3 +w \psi_2) -\frac{s_- s_+ \psi_1 }{4  (\mLamB \mLamCst)^{2}}\right) \ , \\
    g_{1/2, \perp} =&\frac{C^5_1 \zeta  s_+ \sqrt{s_-}}{2 (\mLamB \mLamCst)^{3/2}}+ \varepsilon_b \frac{\sqrt{s_-}}{\sqrt{\mLamB \mLamCst}} \left(\frac{\zeta \bar\Lambda  \left(\mLamB^2+\mLamCst^2-q^2\right)}{2 \mLamB \mLamCst}- \zeta \bar\Lambda'\right)+\nn \\+&\varepsilon_c \frac{\sqrt{s_-}}{\sqrt{\mLamB \mLamCst}} \left(\zeta\bar\Lambda-2\zeta_{\text{SL}}-\frac{\zeta \bar\Lambda'  \left(\mLamB^2+\mLamCst^2-q^2\right)}{2 \mLamB \mLamCst}\right)+\nn \\+&\frac{s_+\sqrt{s_-}}{2(\mLamB\mLamCst)^{3/2}}(\varepsilon_c\eta^{(c)}_\text{kin}-\varepsilon_c\eta^{(c)}_{\text{mag}})+ \varepsilon_c^2 \frac{\sqrt{s_-}}{\sqrt{\mLamB \mLamCst}}\left(3(\psi_3 +w \psi_2)-\frac{s_- s_+ \psi_1 }{4  (\mLamB \mLamCst)^{2}}\right)  \ ,
\end{align}
\newpage
\begin{align}
t_{1/2, 0} =& \frac{\zeta  \mLamCst s_- \sqrt{s_+}}{(\mLamB \mLamCst)^{3/2}} \left(C_{T_1} -C_{T_2}+ C_{T_3}-\frac{C_{T_4} s_+}{2 \mLamB \mLamCst}\right)+\nn \\+&\varepsilon_b\frac{\mLamCst \sqrt{s_+}}{\sqrt{\mLamB \mLamCst}} \left(\frac{\zeta \bar\Lambda \left(\mLamB^2+\mLamCst^2-q^2\right)}{\mLamB \mLamCst}- 2 \zeta \bar\Lambda' + 4 \zeta_{\text{SL}} \right)+\nn \\+& \varepsilon_c\frac{\mLamCst \sqrt{s_+}}{\sqrt{\mLamB \mLamCst}} \left(2 \zeta \bar\Lambda - 4 \zeta_{\text{SL}}-\frac{\zeta \bar\Lambda' \left(\mLamB^2+\mLamCst^2-q^2\right)}{\mLamB \mLamCst} \right)+\nn \\+& \frac{\mLamCst\,s_-\sqrt{s_+}}{(\mLamB \mLamCst)^{3/2}}(\varepsilon_c\eta^{(c)}_\text{kin}+\varepsilon_b\eta_\text{mag}^{(b)}-\varepsilon_c\eta_\text{mag}^{(c)}) +\nn \\+& \varepsilon_c^2 \frac{\mLamCst \sqrt{s_+}}{\sqrt{\mLamB \mLamCst}}\left(6  (\psi_3 +w \psi_2)-\frac{s_- s_+ \psi_1}{2 (\mLamB \mLamCst)^{2}}\right)\ ,\\
t_{1/2, \perp} =& \frac{\zeta s_- \sqrt{s_+}}{\sqrt{\mLamB \mLamCst}} \left(\frac{C_{T_1} \mLamCst }{\mLamB \mLamCst}+\frac{C_{T_2}\left(-\mLamB^2+\mLamCst^2-q^2\right)}{2 \mLamB^2 (\mLamB+\mLamCst) }-\frac{C_{T_3}\left(\mLamB^2-\mLamCst^2-q^2\right)}{2 (\mLamB+\mLamCst) \mLamB \mLamCst}\right)+\nn \\+&  \varepsilon_b\frac{\mLamCst \sqrt{s_+} (\mLamB-\mLamCst)}{(\mLamB+\mLamCst) \sqrt{\mLamB \mLamCst}}\left( 2 \zeta \bar\Lambda'  -\frac{\zeta \bar\Lambda \left(\mLamB^2+\mLamCst^2-q^2\right)}{\mLamB \mLamCst}\right)+\nn \\+&  \varepsilon_c \frac{\sqrt{s_+}\mLamCst(\mLamB-\mLamCst)}{(\mLamB+\mLamCst)\sqrt{\mLamB \mLamCst}}\left(2 \zeta \bar\Lambda - 4 \zeta_{\text{SL}} -\frac{\bar\Lambda'\zeta  \left(\mLamB^2+\mLamCst^2-q^2\right)}{\mLamB \mLamCst} \right) +\nn \\+& \frac{\mLamCst\,s_-\sqrt{s_+}}{(\mLamB \mLamCst)^{3/2}}(\varepsilon_c\eta^{(c)}_\text{kin}-\varepsilon_c\,\eta_\text{mag}^{(c)}) +\nn \\+&\varepsilon_c^2 \frac{\mLamCst \sqrt{s_+}(\mLamB-\mLamCst)}{(\mLamB+\mLamCst)\sqrt{\mLamB \mLamCst}}\left(6  (\psi_3 +w \psi_2)-\frac{s_- s_+ \psi_1}{2 (\mLamB \mLamCst)^{2}}\right)\,,
\end{align}
\begin{align}
t^5_{1/2, 0} =& \frac{C_{T_1} \zeta  \mLamCst \sqrt{s_-} s_+}{(\mLamB \mLamCst)^{3/2}}+\varepsilon_b\frac{\mLamCst \sqrt{s_-}}{\sqrt{\mLamB \mLamCst}} \left(\frac{\zeta \bar\Lambda \left(\mLamB^2+\mLamCst^2-q^2\right)}{\mLamB \mLamCst}-2 \zeta \bar\Lambda'-4 \zeta_{\text{SL}} \right)+\nn \\+&\varepsilon_c\frac{\mLamCst \sqrt{s_-}}{\sqrt{\mLamB \mLamCst}} \left(2 \zeta \bar\Lambda  -4 \zeta_{\text{SL}} -\frac{\zeta \bar\Lambda'  \left(\mLamB^2+\mLamCst^2-q^2\right)}{\mLamB \mLamCst}\right)+\nn \\+& \frac{\mLamCst\,s_+\sqrt{s_-}}{(\mLamB \mLamCst)^{3/2}}(\varepsilon_c\eta^{(c)}_\text{kin}-\varepsilon_b\eta_\text{mag}^{(b)}-\varepsilon_c\eta_\text{mag}^{(c)})+\nn \\+&\varepsilon_c^2 \frac{\mLamCst \sqrt{s_-}}{\sqrt{\mLamB \mLamCst}}\left(6  (\psi_3 +w \psi_2)-\frac{s_- s_+ \psi_1}{2 (\mLamB \mLamCst)^{2}}\right)\,,\\
t^5_{1/2, \perp}=&\frac{\zeta \sqrt{s_-} s_+}{\sqrt{\mLamB \mLamCst}}  \left(\frac{C_{T_1} \mLamCst }{\mLamB \mLamCst}-\frac{C_{T_2} s_-}{2 \mLamB^2 (\mLamB-\mLamCst) }-\frac{C_{T_3} s_-}{2 (\mLamB-\mLamCst) \mLamB \mLamCst}\right) +\nn \\+& \varepsilon_b\frac{\mLamCst \sqrt{s_-} (\mLamB+\mLamCst)}{ (\mLamB-\mLamCst) \sqrt{\mLamB \mLamCst}}\left(2\zeta\bar\Lambda'-\frac{\zeta \bar\Lambda \left(\mLamB^2+\mLamCst^2-q^2\right)}{\mLamB \mLamCst}\right)+\nn \\+&\varepsilon_c \frac{\mLamCst \sqrt{s_-} (\mLamB+\mLamCst)}{ (\mLamB-\mLamCst) \sqrt{\mLamB \mLamCst}}\left( 2 \zeta \bar\Lambda -4 \zeta_{\text{SL}}-\frac{\zeta \bar\Lambda' \left(\mLamB^2+\mLamCst^2-q^2\right)}{\mLamB \mLamCst}\right)+\nn \\+& \frac{\mLamCst\,s_+\sqrt{s_-}}{(\mLamB \mLamCst)^{3/2}}(\varepsilon_c\eta^{(c)}_\text{kin}-\varepsilon_c\eta_\text{mag}^{(c)})+\nn \\+& \varepsilon_c^2 \frac{\mLamCst \sqrt{s_+}(\mLamB+\mLamCst)}{(\mLamB-\mLamCst)\sqrt{\mLamB \mLamCst}}\left(6  (\psi_3 +w \psi_2)-\frac{s_- s_+ \psi_1}{2 (\mLamB \mLamCst)^{2}}\right) \,.
\end{align}
\newpage
For the final state $\LamCst[2625]$ the form factors are:
\begin{align}
F_{1/2, 0} =&  \frac{\zeta s_- \sqrt{s_+}}{2(\mLamB \mLamCst)^{3/2}}  \left(C_1+\frac{C_2 s_+}{2 \mLamB (\mLamB+\mLamCst)}+\frac{C_3 s_+}{2 \mLamCst (\mLamB+\mLamCst)}\right)+\nn \\&+\varepsilon_b \frac{ \sqrt{s_+} (\mLamB-\mLamCst)}{(\mLamB+\mLamCst)\sqrt{\mLamB \mLamCst}}  \left(\zeta \bar\Lambda'+  \zeta_{\text{SL}}-\frac{\zeta\bar\Lambda \left(\mLamB^2+\mLamCst^2-q^2\right)}{2 \mLamB \mLamCst} \right)+\nn \\&+\varepsilon_c \frac{ \sqrt{s_+} (\mLamB-\mLamCst)}{(\mLamB+\mLamCst)\sqrt{\mLamB \mLamCst}}  \left(\zeta \bar\Lambda+ \zeta_{\text{SL}}-\frac{\zeta\bar\Lambda' \left(\mLamB^2+\mLamCst^2-q^2\right)}{2 \mLamB \mLamCst} \right)+ \nn \\
    &+\frac{\sqrt{s_+}s_-}{4(\mLamB\mLamCst)^{3/2}}(2\varepsilon_c\eta^{(c)}_{\text{kin}}-\varepsilon_b\eta^{(b)}_\text{mag}+\varepsilon_c\eta^{(c)}_\text{mag})-\nn \\
&-\varepsilon_c^2\frac{s_+^{3/2}s_- (\mLamB-\mLamCst)}{4(\mLamB+\mLamCst)(\mLamB\mLamCst)^{5/2}}\psi_1 \,,\\ 
F_{1/2, t} =& \frac{\zeta s_+ \sqrt{s_-}}{2(\mLamB \mLamCst)^{3/2}} \left( C_1+\frac{\mLamCst(\mLamB^2-\mLamCst^2+q^2)}{2(\mLamB-\mLamCst)\mLamB \mLamCst}C_2+\frac{\mLamB(\mLamB^2-\mLamCst^2-q^2)}{2(\mLamB-\mLamCst)\mLamB \mLamCst}C_3\right)+\nn \\
    &+\varepsilon_b \frac{(\mLamB+\mLamCst)\sqrt{s_-}}{(\mLamB-\mLamCst)\sqrt{\mLamB \mLamCst}}\left(\zeta \bar\Lambda'+\zeta_{\text{SL}}-\frac{\zeta \bar\Lambda(\mLamB^2+\mLamCst^2-q^2)}{2\mLamB\mLamCst}\right) +\nn \\
    &+\varepsilon_c \frac{(\mLamB+\mLamCst)\sqrt{s_-}}{(\mLamB-\mLamCst)\sqrt{\mLamB \mLamCst}}\left(\zeta\bar\Lambda+\zeta_{\text{SL}}-\frac{\zeta \bar\Lambda'(\mLamB^2+\mLamCst^2-q^2)}{2\mLamB\mLamCst}\right)+ \nn \\
    &+\frac{s_+\sqrt{s_-}}{4(\mLamB\mLamCst)^{3/2}}(2\varepsilon_c\eta^{(c)}_{\text{kin}}-\varepsilon_b\eta^{(b)}_\text{mag}+\varepsilon_c\eta^{(c)}_\text{mag})\,\notag
    -\\
&-\varepsilon_c^2\frac{s_+s_-^{3/2} (\mLamB+\mLamCst)}{4(\mLamB-\mLamCst)(\mLamB\mLamCst)^{5/2}}\psi_1  \,,\\ 
F_{1/2, \perp} =&  \frac{\zeta\,\sqrt{s_+}s_-}{2(\mLamB \mLamCst)^{3/2}}C_1+\varepsilon_b\zeta\frac{\sqrt{s_+}}{\sqrt{\mLamB\mLamCst}}\left(\frac{\bar\Lambda(\mLamB^2+\mLamCst^2-q^2)}{2\mLamB\mLamCst}-\bar\Lambda'\right)+ \nn \\
    &+\varepsilon_c \frac{\sqrt{s_+}}{\sqrt{\mLamB\mLamCst}}\left(\zeta\bar\Lambda+\zeta_{\text{SL}}-\zeta\frac{\bar\Lambda'(\mLamB^2+\mLamCst^2-q^2)}{2\mLamB\mLamCst}\right)+ \nn \\
    &+\frac{\sqrt{s_+}s_-}{4(\mLamB\mLamCst)^{3/2}}(2\varepsilon_c\eta^{(c)}_{\text{kin}}+\varepsilon_c\,\eta^{(c)}_\text{mag})-\nn \varepsilon_c^2\frac{s_+^{3/2}s_-}{4\,(\mLamB\mLamCst)^{5/2}}\psi_1\, ,
\\  
F_{3/2,\perp}=&-\varepsilon_b\frac{\sqrt{s_+}}{\sqrt{\mLamB\mLamCst}}\zeta_{\text{SL}}-\varepsilon_b\frac{ s_- \sqrt{ s_+}}{4(\mLamB\mLamCst)^{3/2}}\,\eta^{(b)}_\text{mag}  \, , \nn \\
\end{align}

\newpage

\begin{align}
    G_{1/2, 0} =&  \frac{\zeta s_+ \sqrt{s_-}}{2(\mLamB \mLamCst)^{3/2}}  \left(C^5_1-\frac{C^5_2 s_-}{2 \mLamB (\mLamB-\mLamCst)}-\frac{C^5_3 s_-}{2 \mLamCst (\mLamB-\mLamCst)}\right)+\nn \\+&\varepsilon_b \frac{ \sqrt{s_-} (\mLamB+\mLamCst)}{(\mLamB-\mLamCst)\sqrt{\mLamB \mLamCst}}  \left(\zeta \bar\Lambda'- \zeta_{\text{SL}}-\frac{\zeta\bar\Lambda \left(\mLamB^2+\mLamCst^2-q^2\right)}{2 \mLamB \mLamCst} \right)+\nn \\+&\varepsilon_c \frac{ \sqrt{s_-} (\mLamB+\mLamCst)}{(\mLamB-\mLamCst)\sqrt{\mLamB \mLamCst}}  \left(\zeta \bar\Lambda+ \zeta_{\text{SL}}-\frac{\zeta\bar\Lambda' \left(\mLamB^2+\mLamCst^2-q^2\right)}{2 \mLamB \mLamCst} \right)+ \nn \\
    &+\frac{s_+\sqrt{s_-}}{4(\mLamB\mLamCst)^{3/2}}(2\varepsilon_c\eta^{(c)}_{\text{kin}}+\varepsilon_b\eta^{(b)}_\text{mag}+\varepsilon_c\eta^{(c)}_\text{mag})
    -\nn \\
&-\varepsilon_c^2\frac{s_+s_-^{3/2}(\mLamB+\mLamCst)}{4(\mLamB-\mLamCst)(\mLamB\mLamCst)^{5/2}}\psi_1\,,\\
G_{1/2, t} =&  \frac{\zeta s_- \sqrt{s_+}}{2(\mLamB \mLamCst)^{3/2}} \left( C^5_1-\frac{(\mLamB^2-\mLamCst^2+q^2)}{2(\mLamB+\mLamCst)\mLamB }C^5_2-\frac{(\mLamB^2-\mLamCst^2-q^2)}{2(\mLamB+\mLamCst) \mLamCst}C^5_3\right)+\nn \\
    &+\varepsilon_b \frac{(\mLamB-\mLamCst)\sqrt{s_+}}{(\mLamB+\mLamCst)\sqrt{\mLamB \mLamCst}}\left(\zeta \bar\Lambda'-\zeta_{\text{SL}}-\frac{\zeta \bar\Lambda(\mLamB^2+\mLamCst^2-q^2)}{2\mLamB\mLamCst}\right) +\nn \\
    &+\varepsilon_c \frac{(\mLamB-\mLamCst)\sqrt{s_+}}{(\mLamB+\mLamCst)\sqrt{\mLamB \mLamCst}}\left(\zeta\bar\Lambda'+\zeta_{\text{SL}}-\frac{\zeta \bar\Lambda(\mLamB^2+\mLamCst^2-q^2)}{2\mLamB\mLamCst}\right)+ \nn \\
    &+\frac{\sqrt{s_+}s_-}{4(\mLamB\mLamCst)^{3/2}}(2\varepsilon_c\eta^{(c)}_{\text{kin}}+\varepsilon_b\eta^{(b)}_\text{mag}+\varepsilon_c\eta^{(c)}_\text{mag})\,\notag -\\
&-\varepsilon_c^2\frac{s_+^{3/2}s_-(\mLamB-\mLamCst)}{4(\mLamB+\mLamCst)(\mLamB\mLamCst)^{5/2}}\psi_1\,,\\
G_{1/2, \perp} =&  \frac{\zeta\,\sqrt{s_-}s_+}{2(\mLamB \mLamCst)^{3/2}}C^5_1+\varepsilon_b\zeta\frac{\sqrt{s_-}}{\sqrt{\mLamB\mLamCst}}\left(\frac{\bar\Lambda(\mLamB^2+\mLamCst^2-q^2)}{2\mLamB\mLamCst}-\bar\Lambda'\right)+ \nn \\
    &+\varepsilon_c \frac{\sqrt{s_-}}{\sqrt{\mLamB\mLamCst}}\left(\zeta\bar\Lambda+\zeta_{\text{SL}}-\zeta\frac{\bar\Lambda'(\mLamB^2+\mLamCst^2-q^2)}{2\mLamB\mLamCst}\right)+ \nn \\
    &+\frac{s_+\sqrt{s_-}}{4(\mLamB\mLamCst)^{3/2}}(2\varepsilon_c\eta^{(c)}_{\text{kin}}+\varepsilon_c\eta^{(c)}_\text{mag})\,\notag-\\
   &-\varepsilon_c^2\frac{s_+ s_-^{3/2}}{4 (\mLamB\mLamCst)^{5/2}}\psi_1\, , \\ 
G_{3/2, \perp} =& -\varepsilon_b\frac{\sqrt{s_-}}{\sqrt{\mLamB\mLamCst}}\zeta_{\text{SL}}-\varepsilon_b\frac{ s_+ \sqrt{ s_-}}{4(\mLamB\mLamCst)^{3/2}}\eta^{(b)}_\text{mag} \, \ , 
\end{align}

\newpage

\begin{align}
  T_{1/2, 0} =&  \zeta\frac{\mLamCst\sqrt{s_+}s_-}{(\mLamB\mLamCst)^{3/2}}\left(C_{T_{1}}-C_{T_{2}}+C_{T_{3}}-\frac{s_+}{2\mLamB\mLamCst}C_{T_{4}}\right)+\nn \\
&+\varepsilon_b\frac{2\mLamCst\sqrt{s_+}}{\sqrt{\mLamB\mLamCst}}\left(-\zeta\bar\Lambda'-\zeta_{\text{SL}}+\zeta\frac{\mLamB^2+\mLamCst^2-q^2}{2\mLamB\mLamCst}\bar\Lambda\right)+\nn \\
+&\varepsilon_c\frac{2\mLamCst\sqrt{s_+}}{\sqrt{\mLamB\mLamCst}}\left(\zeta\bar\Lambda+\zeta_{\text{SL}}-\zeta\frac{\mLamB^2+\mLamCst^2-q^2}{2\mLamB\mLamCst}\bar\Lambda'\right)
+\nn \\
&+\frac{\mLamCst\,\sqrt{s_+}s_-}{2(\mLamB\mLamCst)^{3/2}}(2\varepsilon_c\eta^{(c)}_{\text{kin}}-\varepsilon_b\,\eta_\text{mag}^{(b)}+\varepsilon_c\,\eta_\text{mag}^{(c)})-\varepsilon_c^2\frac{\mLamCst\,s_+^{3/2}s_-}{2(\mLamB\mLamCst)^{5/2}}\psi_1 \,,\\  
T_{1/2, \perp} =&\zeta\frac{\sqrt{s_+} s_-}{\mLamB\sqrt{\mLamB\mLamCst}}\left(C_{T_{1}}-\frac{\mLamB^2-\mLamCst^2+q^2}{2\mLamB(\mLamB+\mLamCst)}C_{T_{2}}-\frac{\mLamB^2-\mLamCst^2-q^2}{2\mLamCst(\mLamB+\mLamCst)}C_{T_{3}}\right)+\nn \\
&+\varepsilon_b\frac{2\mLamCst(\mLamB-\mLamCst)\sqrt{s_+}}{(\mLamB+\mLamCst)\sqrt{\mLamB\mLamCst}}\left(\zeta\bar\Lambda'-\zeta\frac{\mLamB^2+\mLamCst^2-q^2}{2\mLamB\mLamCst}\bar\Lambda\right)+\nn \\
&+\varepsilon_c\frac{2\mLamCst(\mLamB-\mLamCst)\sqrt{s_+}}{(\mLamB+\mLamCst)\sqrt{\mLamB\mLamCst}}\left(\zeta\bar\Lambda+\zeta_{\text{SL}}-\zeta\frac{\mLamB^2+\mLamCst^2-q^2}{2\mLamB\mLamCst}\bar\Lambda'\right)+\nn \\
&+\frac{\mLamCst \sqrt{s_+}s_-}{2(\mLamB\mLamCst)^{3/2}}(2\varepsilon_c\eta^{(c)}_{\text{kin}}+\varepsilon_c\,\eta_\text{mag}^{(c)})\,  -\varepsilon_c^2\frac{\mLamCst(\mLamB-\mLamCst)s_+^{3/2}s_-}{2(\mLamB+\mLamCst)(\mLamB\mLamCst)^{5/2}}\psi_1 \,,\\  
T_{3/2, \perp} =&-\varepsilon_b\frac{2(\mLamB-\mLamCst)\sqrt{s_+}}{\sqrt{\mLamB\mLamCst}}\zeta_{\text{SL}}
+\varepsilon_b\frac{(\mLamB+\mLamCst)\sqrt{s_+}s_-}{2(\mLamB\mLamCst)^{3/2}}\eta_\text{mag}^{(b)}\, ,
\end{align}
\newpage
\begin{align}
    T^5_{1/2, 0} =& \frac{\zeta\mLamCst s_+\sqrt{s_-}}{(\mLamB\mLamCst)^{3/2}}C_{T_{1}}+\varepsilon_b\frac{2\mLamCst\sqrt{s_-}}{\sqrt{\mLamB\mLamCst}}\left(\zeta\frac{(\mLamB^2+\mLamCst^2-q^2)\bar\Lambda}{2\mLamB\mLamCst}+\zeta_{\text{SL}}-\zeta\bar\Lambda'\right)+\nn \\
&+\varepsilon_b\frac{2\mLamCst\sqrt{s_-}}{\sqrt{\mLamB\mLamCst}}\left(-\zeta\frac{(\mLamB^2+\mLamCst^2-q^2)\bar\Lambda'}{2\mLamB\mLamCst}+\zeta_{\text{SL}}+\zeta\bar\Lambda\right)+\nn \\
&+\frac{\mLamCst s_+\sqrt{s_-}}{2(\mLamB\mLamCst)^{3/2}}(2\varepsilon_c\eta^{(c)}_{\text{kin}}+\varepsilon_b\eta_\text{mag}^{(b)}+\varepsilon_c\eta_\text{mag}^{(c)})\,\notag -\\
&-\varepsilon_c^2\frac{\mLamCst\,s_+s_-^{3/2}}{2(\mLamB\mLamCst)^{5/2}}\psi_1 \,,\\  
T^5_{1/2, \perp} =& \zeta\frac{s_+\sqrt{s_-}}{\mLamB\sqrt{\mLamB\mLamCst}}\left(C_{T_{1}}-\frac{s_-}{2\mLamB(\mLamB-\mLamCst)}C_{T_{2}}-\frac{s_-}{2\mLamCst(\mLamB-\mLamCst)}\right)+\nn \\
&+\varepsilon_b\frac{2\mLamCst(\mLamB+\mLamCst)\sqrt{s_-}}{(\mLamB-\mLamCst)\sqrt{\mLamB\mLamCst}}\left(\zeta\bar\Lambda'-\zeta\frac{(\mLamB^2+\mLamCst^2-q^2)\bar\Lambda}{2\mLamB\mLamCst}\right)+\nn \\
&+\varepsilon_c\frac{2\mLamCst(\mLamB+\mLamCst)\sqrt{s_-}}{(\mLamB-\mLamCst)\sqrt{\mLamB\mLamCst}}\left(\zeta\bar\Lambda+\zeta_{\text{SL}}-\zeta\frac{(\mLamB^2+\mLamCst^2-q^2)\bar\Lambda'}{2\mLamB\mLamCst}\right)+\nn \\
&+\frac{\mLamCst s_+\sqrt{s_-}}{2(\mLamB\mLamCst)^{3/2}}(2\varepsilon_c\eta^{(c)}_{\text{kin}}+\varepsilon_c\eta_\text{mag}^{(c)})-\varepsilon_c^2\frac{\mLamCst(\mLamB+\mLamCst)s_+s_-^{3/2}}{2(\mLamB-\mLamCst)(\mLamB\mLamCst)^{5/2}} \psi_1 \,,\\ 
T^5_{3/2, \perp} =&  -\varepsilon_b\frac{2(\mLamB+\mLamCst)\sqrt{s_-}}{\sqrt{\mLamB\mLamCst}}\zeta_{\text{SL}}
+\varepsilon_b\frac{(\mLamB-\mLamCst)\,s_+\sqrt{s_-}}{2(\mLamB\mLamCst)^{3/2}}\eta_\text{mag}^{(b)}  \ .
\end{align}
\subsection{Form factors at $\mathcal{O}(1/m_c^2)$ in the VC-limit}
\label{ffsVC}
Here we list the full set of form factors up to $\mathcal{O}(1/m_c^2)$ in the VC limit.
For the final state $\LamCst[2595]$ the form factors are:
\begin{align}
  f_{1/2,0}=&\frac{\zeta s_- \sqrt{s_+}}{2(\mLamB \mLamCst)^{3/2}}  \left(C_1+\frac{C_2 s_+}{2 \mLamB (\mLamB+\mLamCst)}+\frac{C_3 s_+}{2 \mLamCst (\mLamB+\mLamCst)}\right)+\nn \\+&\varepsilon_b \frac{ \sqrt{s_+} (\mLamB-\mLamCst)}{(\mLamB+\mLamCst)\sqrt{\mLamB \mLamCst}}  \left(\zeta \bar\Lambda'- 2 \zeta_{\text{SL}}-\frac{\zeta\bar\Lambda \left(\mLamB^2+\mLamCst^2-q^2\right)}{2 \mLamB \mLamCst} \right)+\nn \\+&\varepsilon_c \frac{ \sqrt{s_+} (\mLamB-\mLamCst)}{(\mLamB+\mLamCst)\sqrt{\mLamB \mLamCst}}  \left(\zeta \bar\Lambda- 2 \zeta_{\text{SL}}-\frac{\zeta\bar\Lambda' \left(\mLamB^2+\mLamCst^2-q^2\right)}{2 \mLamB \mLamCst} \right)+\nn \\+&\varepsilon_c\frac{\eta^{(c)}_\text{kin}  s_- \sqrt{s_+} }{2 (\mLamB \mLamCst)^{3/2}} +\varepsilon_c^2 \frac{\sqrt{s_+}(\mLamB-\mLamCst)}{(\mLamB+\mLamCst) \sqrt{\mLamB \mLamCst}}\left(3\beta_3 -\frac{s_- s_+ \beta_1 }{4  (\mLamB \mLamCst)^{2}}\right) \,, \\
f_{1/2,t}=&\frac{\zeta s_+ \sqrt{s_-}}{2(\mLamB \mLamCst)^{3/2}}  \left(C_1+\frac{C_2 s_+}{2 \mLamB (\mLamB-\mLamCst)}+\frac{C_3 s_+}{2 \mLamCst (\mLamB-\mLamCst)}\right)+\nn \\+&\varepsilon_b \frac{ \sqrt{s_-} (\mLamB+\mLamCst)}{(\mLamB-\mLamCst)\sqrt{\mLamB \mLamCst}}  \left(\zeta \bar\Lambda'- 2 \zeta_{\text{SL}}-\frac{\zeta\bar\Lambda \left(\mLamB^2+\mLamCst^2-q^2\right)}{2 \mLamB \mLamCst} \right)+\nn \\+&\varepsilon_c \frac{ \sqrt{s_-} (\mLamB+\mLamCst)}{(\mLamB-\mLamCst)\sqrt{\mLamB \mLamCst}}  \left(\zeta \bar\Lambda- 2 \zeta_{\text{SL}}-\frac{\zeta\bar\Lambda' \left(\mLamB^2+\mLamCst^2-q^2\right)}{2 \mLamB \mLamCst} \right)+\nn \\+&\varepsilon_c\frac{\eta^{(c)}_\text{kin}  s_+ \sqrt{s_-} }{2 (\mLamB \mLamCst)^{3/2}} + \varepsilon_c^2 \frac{\sqrt{s_-}(\mLamB+\mLamCst)}{(\mLamB-\mLamCst) \sqrt{\mLamB \mLamCst}}\left(3\beta_3 -\frac{s_- s_+ \beta_1 }{4  (\mLamB \mLamCst)^{2}}\right)\ ,  \\
 f_{1/2,\perp}=&\frac{C_1 \zeta  s_- \sqrt{s_+}}{2 (\mLamB \mLamCst)^{3/2}}+ \varepsilon_b \frac{\sqrt{s_+}}{\sqrt{\mLamB \mLamCst}} \left(\frac{\zeta \bar\Lambda  \left(\mLamB^2+\mLamCst^2-q^2\right)}{2 \mLamB \mLamCst}-\bar\Lambda'\zeta\right)+\nn \\+&\varepsilon_c \frac{\sqrt{s_+}}{\sqrt{\mLamB \mLamCst}} \left(\bar\Lambda\zeta-2\zeta_{\text{SL}}-\frac{\zeta \bar\Lambda'  \left(\mLamB^2+\mLamCst^2-q^2\right)}{2 \mLamB \mLamCst}\right)+\nn \\+&\varepsilon_c\frac{\eta^{(c)}_\text{kin}  s_- \sqrt{s_+} }{2 (\mLamB \mLamCst)^{3/2}} +\varepsilon_c^2 \frac{\sqrt{s_+}}{\sqrt{\mLamB \mLamCst}}\left(3\beta_3-\frac{s_- s_+ \beta_1 }{4  (\mLamB \mLamCst)^{2}}\right) \,,
\end{align}
\newpage
\begin{align}
 g_{1/2,0}=&\frac{\zeta s_+ \sqrt{s_-}}{2(\mLamB \mLamCst)^{3/2}}  \left(C^5_1-\frac{C^5_2 s_-}{2 \mLamB (\mLamB-\mLamCst)}-\frac{C^5_3 s_-}{2 \mLamCst (\mLamB-\mLamCst)}\right)+\nn \\+&\varepsilon_b \frac{ \sqrt{s_-} (\mLamB+\mLamCst)}{(\mLamB-\mLamCst)\sqrt{\mLamB \mLamCst}}  \left(\zeta \bar\Lambda'+2 \zeta_{\text{SL}}-\frac{\zeta\bar\Lambda \left(\mLamB^2+\mLamCst^2-q^2\right)}{2 \mLamB \mLamCst} \right)+\nn \\+&\varepsilon_c \frac{ \sqrt{s_-} (\mLamB-\mLamCst)}{(\mLamB+\mLamCst)\sqrt{\mLamB \mLamCst}}  \left(\zeta \bar\Lambda- 2 \zeta_{\text{SL}}-\frac{\zeta\bar\Lambda' \left(\mLamB^2+\mLamCst^2-q^2\right)}{2 \mLamB \mLamCst} \right)+\nn \\+&\varepsilon_c\frac{\eta^{(c)}_\text{kin}  s_+ \sqrt{s_-} }{2 (\mLamB \mLamCst)^{3/2}}+ \varepsilon_c^2 \frac{\sqrt{s_-}(\mLamB+\mLamCst)}{(\mLamB-\mLamCst) \sqrt{\mLamB \mLamCst}}\left(3\beta_3 -\frac{s_- s_+ \beta_1 }{4  (\mLamB \mLamCst)^{2}}\right)\ , \\
    g_{1/2,t}=&\frac{\zeta s_- \sqrt{s_+}}{2(\mLamB \mLamCst)^{3/2}}  \left(C^5_1+\frac{C^5_2 s_+}{2 \mLamB (\mLamB+\mLamCst)}+\frac{C^5_3 s_+}{2 \mLamCst (\mLamB+\mLamCst)}\right)+\nn \\+&\varepsilon_b \frac{ \sqrt{s_+} (\mLamB+\mLamCst)}{(\mLamB-\mLamCst)\sqrt{\mLamB \mLamCst}}  \left(\zeta \bar\Lambda'+ 2 \zeta_{\text{SL}}-\frac{\zeta\bar\Lambda \left(\mLamB^2+\mLamCst^2-q^2\right)}{2 \mLamB \mLamCst} \right)+\nn \\+&\varepsilon_c \frac{ \sqrt{s_+} (\mLamB+\mLamCst)}{(\mLamB-\mLamCst)\sqrt{\mLamB \mLamCst}}  \left(\zeta \bar\Lambda- 2 \zeta_{\text{SL}}-\frac{\zeta\bar\Lambda' \left(\mLamB^2+\mLamCst^2-q^2\right)}{2 \mLamB \mLamCst} \right)+\nn \\+&\varepsilon_c\frac{\eta^{(c)}_\text{kin}  s_- \sqrt{s_+} }{2 (\mLamB \mLamCst)^{3/2}}+\varepsilon_c^2 \frac{\sqrt{s_+}(\mLamB-\mLamCst)}{(\mLamB+\mLamCst) \sqrt{\mLamB \mLamCst}}\left(3\beta_3 -\frac{s_- s_+ \beta_1 }{4  (\mLamB \mLamCst)^{2}}\right)  \ , \\
    g_{1/2,\perp}=& \frac{C^5_1 \zeta  s_+ \sqrt{s_-}}{2 (\mLamB \mLamCst)^{3/2}}+ \varepsilon_b \frac{\sqrt{s_-}}{\sqrt{\mLamB \mLamCst}} \left(\frac{\zeta \bar\Lambda  \left(\mLamB^2+\mLamCst^2-q^2\right)}{2 \mLamB \mLamCst}- \zeta \bar\Lambda'\right)+\nn \\+&\varepsilon_c \frac{\sqrt{s_-}}{\sqrt{\mLamB \mLamCst}} \left(\zeta\bar\Lambda-2\zeta_{\text{SL}}-\frac{\zeta \bar\Lambda'  \left(\mLamB^2+\mLamCst^2-q^2\right)}{2 \mLamB \mLamCst}\right)+\nn \\+&\varepsilon_c\frac{\eta^{(c)}_\text{kin}  s_+ \sqrt{s_-} }{2 (\mLamB \mLamCst)^{3/2}} +\varepsilon_c^2 \frac{\sqrt{s_-}}{\sqrt{\mLamB \mLamCst}}\left(3\beta_3 -\frac{s_- s_+ \beta_1 }{4  (\mLamB \mLamCst)^{2}}\right)  \ ,
\end{align}
\newpage
\begin{align}
 t_{1/2,0}=&\frac{\zeta  \mLamCst s_- \sqrt{s_+}}{(\mLamB \mLamCst)^{3/2}} \left(C_{T_1} -C_{T_2}+ C_{T_3}-\frac{C_{T_4} s_+}{2 \mLamB \mLamCst}\right)+\nn \\+&\varepsilon_b\frac{\mLamCst \sqrt{s_+}}{\sqrt{\mLamB \mLamCst}} \left(\frac{\zeta \bar\Lambda \left(\mLamB^2+\mLamCst^2-q^2\right)}{\mLamB \mLamCst}- 2 \zeta \bar\Lambda' + 4 \zeta_{\text{SL}} \right)+\nn \\+& \varepsilon_c\frac{\mLamCst \sqrt{s_+}}{\sqrt{\mLamB \mLamCst}} \left(2 \zeta \bar\Lambda - 4 \zeta_{\text{SL}}-\frac{\zeta \bar\Lambda' \left(\mLamB^2+\mLamCst^2-q^2\right)}{\mLamB \mLamCst} \right)+\nn \\+& \varepsilon_c\frac{\eta^{(c)}_\text{kin}  \mLamCst s_- \sqrt{s_+} }{(\mLamB \mLamCst)^{3/2}} +\varepsilon_c^2 \frac{\mLamCst \sqrt{s_+}}{\sqrt{\mLamB \mLamCst}}\left(6  \beta_3-\frac{s_- s_+ \beta_1}{2 (\mLamB \mLamCst)^{2}}\right)\ ,\\
t_{1/2, \perp}=& \frac{\zeta s_- \sqrt{s_+}}{\sqrt{\mLamB \mLamCst}} \left(\frac{C_{T_1} \mLamCst }{\mLamB \mLamCst}+\frac{C_{T_2}\left(-\mLamB^2+\mLamCst^2-q^2\right)}{2 \mLamB^2 (\mLamB+\mLamCst) }-\frac{C_{T_3}\left(\mLamB^2-\mLamCst^2-q^2\right)}{2 (\mLamB+\mLamCst) \mLamB \mLamCst}\right)+\nn \\+&  \varepsilon_b\frac{\mLamCst \sqrt{s_+} (\mLamB-\mLamCst)}{(\mLamB+\mLamCst) \sqrt{\mLamB \mLamCst}}\left( 2 \zeta \bar\Lambda'  -\frac{\zeta \bar\Lambda \left(\mLamB^2+\mLamCst^2-q^2\right)}{\mLamB \mLamCst}\right)+\nn \\+&  \varepsilon_c \frac{\sqrt{s_+}\mLamCst(\mLamB-\mLamCst)}{(\mLamB+\mLamCst)\sqrt{\mLamB \mLamCst}}\left(2 \zeta \bar\Lambda - 4 \zeta_{\text{SL}} -\frac{\bar\Lambda'\zeta  \left(\mLamB^2+\mLamCst^2-q^2\right)}{\mLamB \mLamCst} \right) +\nn \\+& \varepsilon_c\frac{\eta^{(c)}_\text{kin}  \mLamCst s_- \sqrt{s_+} }{(\mLamB \mLamCst)^{3/2}}+\varepsilon_c^2 \frac{\mLamCst \sqrt{s_+}(\mLamB-\mLamCst)}{(\mLamB+\mLamCst)\sqrt{\mLamB \mLamCst}}\left(6  \beta_3-\frac{s_- s_+ \beta_1}{2 (\mLamB \mLamCst)^{2}}\right)\,,
\end{align}
\begin{align}
 t^5_{1/2, 0}=& \frac{C_{T_1} \zeta  \mLamCst \sqrt{s_-} s_+}{(\mLamB \mLamCst)^{3/2}}+\varepsilon_b\frac{\mLamCst \sqrt{s_-}}{\sqrt{\mLamB \mLamCst}} \left(\frac{\zeta \bar\Lambda \left(\mLamB^2+\mLamCst^2-q^2\right)}{\mLamB \mLamCst}-2 \zeta \bar\Lambda'-4 \zeta_{\text{SL}} \right)+\nn \\+&\varepsilon_c\frac{\mLamCst \sqrt{s_-}}{\sqrt{\mLamB \mLamCst}} \left(2 \zeta \bar\Lambda  -4 \zeta_{\text{SL}} -\frac{\zeta \bar\Lambda'  \left(\mLamB^2+\mLamCst^2-q^2\right)}{\mLamB \mLamCst}\right)+\nn \\+&\varepsilon_c\frac{\eta^{(c)}_\text{kin}  \mLamCst \sqrt{s_-} s_+}{(\mLamB \mLamCst)^{3/2}} +\varepsilon_c^2 \frac{\mLamCst \sqrt{s_-}}{\sqrt{\mLamB \mLamCst}}\left(6  \beta_3-\frac{s_- s_+ \beta_1}{2 (\mLamB \mLamCst)^{2}}\right) \,,\\
 t^5_{1/2, \perp}=&\frac{\zeta \sqrt{s_-} s_+}{\sqrt{\mLamB \mLamCst}}  \left(\frac{C_{T_1} \mLamCst }{\mLamB \mLamCst}-\frac{C_{T_2} s_-}{2 \mLamB^2 (\mLamB-\mLamCst) }-\frac{C_{T_3} s_-}{2 (\mLamB-\mLamCst) \mLamB \mLamCst}\right) +\nn \\+& \varepsilon_b\frac{\mLamCst \sqrt{s_-} (\mLamB+\mLamCst)}{ (\mLamB-\mLamCst) \sqrt{\mLamB \mLamCst}}\left(2\zeta\bar\Lambda'-\frac{\zeta \bar\Lambda \left(\mLamB^2+\mLamCst^2-q^2\right)}{\mLamB \mLamCst}\right)+\nn \\+&\varepsilon_c \frac{\mLamCst \sqrt{s_-} (\mLamB+\mLamCst)}{ (\mLamB-\mLamCst) \sqrt{\mLamB \mLamCst}}\left( 2 \zeta \bar\Lambda -4 \zeta_{\text{SL}}-\frac{\zeta \bar\Lambda' \left(\mLamB^2+\mLamCst^2-q^2\right)}{\mLamB \mLamCst}\right)+\nn \\+&\varepsilon_c\frac{\eta^{(c)}_\text{kin}  \mLamCst \sqrt{s_-} s_+ }{(\mLamB \mLamCst)^{3/2}}+\varepsilon_c^2 \frac{\mLamCst \sqrt{s_+}(\mLamB+\mLamCst)}{(\mLamB-\mLamCst)\sqrt{\mLamB \mLamCst}}\left(6 \beta_3-\frac{s_- s_+ \beta_1}{2 (\mLamB \mLamCst)^{2}}\right) \,.
\end{align}
\newpage
For the final state $\LamCst[2625]$ the form factors are:
\begin{align}
F_{1/2, 0} =&  \frac{\zeta s_- \sqrt{s_+}}{2(\mLamB \mLamCst)^{3/2}}  \left(C_1+\frac{C_2 s_+}{2 \mLamB (\mLamB+\mLamCst)}+\frac{C_3 s_+}{2 \mLamCst (\mLamB+\mLamCst)}\right)+\nn \\&+\varepsilon_b \frac{ \sqrt{s_+} (\mLamB-\mLamCst)}{(\mLamB+\mLamCst)\sqrt{\mLamB \mLamCst}}  \left(\zeta \bar\Lambda'+  \zeta_{\text{SL}}-\frac{\zeta\bar\Lambda \left(\mLamB^2+\mLamCst^2-q^2\right)}{2 \mLamB \mLamCst} \right)+\nn \\&+\varepsilon_c \frac{ \sqrt{s_+} (\mLamB-\mLamCst)}{(\mLamB+\mLamCst)\sqrt{\mLamB \mLamCst}}  \left(\zeta \bar\Lambda+ \zeta_{\text{SL}}-\frac{\zeta\bar\Lambda' \left(\mLamB^2+\mLamCst^2-q^2\right)}{2 \mLamB \mLamCst} \right)+ \nn \\
    &+\varepsilon_c\frac{\sqrt{s_+}s_-}{2(\mLamB\mLamCst)^{3/2}}\eta^{(c)}_{\text{kin}}-\varepsilon_c^2\frac{(\mLamB-\mLamCst)s_+^{3/2}s_-}{4(\mLamB+\mLamCst)(\mLamB\mLamCst)^{5/2}}\beta_1 \,,\\ 
F_{1/2, t} =& \frac{\zeta s_+ \sqrt{s_-}}{2(\mLamB \mLamCst)^{3/2}} \left( C_1+\frac{\mLamCst(\mLamB^2-\mLamCst^2+q^2)}{2(\mLamB-\mLamCst)\mLamB \mLamCst}C_2+\frac{\mLamB(\mLamB^2-\mLamCst^2-q^2)}{2(\mLamB-\mLamCst)\mLamB \mLamCst}C_3\right)+\nn \\
    &+\varepsilon_b \frac{(\mLamB+\mLamCst)\sqrt{s_-}}{(\mLamB-\mLamCst)\sqrt{\mLamB \mLamCst}}\left(\zeta \bar\Lambda'+\zeta_{\text{SL}}-\frac{\zeta \bar\Lambda(\mLamB^2+\mLamCst^2-q^2)}{2\mLamB\mLamCst}\right) +\nn \\
    &+\varepsilon_c \frac{(\mLamB+\mLamCst)\sqrt{s_-}}{(\mLamB-\mLamCst)\sqrt{\mLamB \mLamCst}}\left(\zeta\bar\Lambda+\zeta_{\text{SL}}-\frac{\zeta \bar\Lambda'(\mLamB^2+\mLamCst^2-q^2)}{2\mLamB\mLamCst}\right)+ \nn \\
    &+\varepsilon_c\frac{s_+\sqrt{s_-}}{2(\mLamB\mLamCst)^{3/2}}\eta^{(c)}_{\text{kin}}-\varepsilon_c^2\frac{(\mLamB+\mLamCst)s_+s_-^{3/2}}{4(\mLamB-\mLamCst)(\mLamB\mLamCst)^{5/2}}\beta_1 \,,\\ 
F_{1/2, \perp} =& \frac{\zeta\,\sqrt{s_+}s_-}{2(\mLamB \mLamCst)^{3/2}}C_1+\varepsilon_b\zeta\frac{\sqrt{s_+}}{\sqrt{\mLamB\mLamCst}}\left(\frac{\bar\Lambda(\mLamB^2+\mLamCst^2-q^2)}{2\mLamB\mLamCst}-\bar\Lambda'\right)+ \nn \\
    &+\varepsilon_c \frac{\sqrt{s_+}}{\sqrt{\mLamB\mLamCst}}\left(\zeta\bar\Lambda+\zeta_{\text{SL}}-\zeta\frac{\bar\Lambda'(\mLamB^2+\mLamCst^2-q^2)}{2\mLamB\mLamCst}\right)+ \nn \\
    &+\varepsilon_c\frac{\sqrt{s_+}s_-}{2(\mLamB\mLamCst)^{3/2}}\eta^{(c)}_{\text{kin}}  -\varepsilon_c^2\frac{s_{+}^{3/2}s_{-}}{4(\mLamB\mLamCst)^{5/2}}\beta_1\, ,\\ 
F_{3/2, \perp} =&   -\varepsilon_b\frac{\sqrt{s_+}}{\sqrt{\mLamB\mLamCst}}\zeta_{\text{SL}} \notag \, . \\
\end{align}

\newpage
\begin{align}
    G_{1/2, 0} =&   \frac{\zeta s_+ \sqrt{s_-}}{2(\mLamB \mLamCst)^{3/2}}  \left(C^5_1-\frac{C^5_2 s_-}{2 \mLamB (\mLamB-\mLamCst)}-\frac{C^5_3 s_-}{2 \mLamCst (\mLamB-\mLamCst)}\right)+\nn \\+&\varepsilon_b \frac{ \sqrt{s_-} (\mLamB+\mLamCst)}{(\mLamB-\mLamCst)\sqrt{\mLamB \mLamCst}}  \left(\zeta \bar\Lambda'- \zeta_{\text{SL}}-\frac{\zeta\bar\Lambda \left(\mLamB^2+\mLamCst^2-q^2\right)}{2 \mLamB \mLamCst} \right)+\nn \\+&\varepsilon_c \frac{ \sqrt{s_-} (\mLamB+\mLamCst)}{(\mLamB-\mLamCst)\sqrt{\mLamB \mLamCst}}  \left(\zeta \bar\Lambda+ \zeta_{\text{SL}}-\frac{\zeta\bar\Lambda' \left(\mLamB^2+\mLamCst^2-q^2\right)}{2 \mLamB \mLamCst} \right)+ \nn \\
    &+\varepsilon_c\frac{s_+\sqrt{s_-}}{2(\mLamB\mLamCst)^{3/2}}\eta^{(c)}_{\text{kin}}-\varepsilon_c^2\frac{(\mLamB+\mLamCst)s_{+}s_{-}^{3/2}}{4(\mLamB-\mLamCst)(\mLamB\mLamCst)^{5/2}}\beta_1
\,,\\
G_{1/2, t} =&   \frac{\zeta s_- \sqrt{s_+}}{2(\mLamB \mLamCst)^{3/2}} \left( C^5_1-\frac{(\mLamB^2-\mLamCst^2+q^2)}{2(\mLamB+\mLamCst)\mLamB }C^5_2-\frac{(\mLamB^2-\mLamCst^2-q^2)}{2(\mLamB+\mLamCst) \mLamCst}C^5_3\right)+\nn \\
    &+\varepsilon_b \frac{(\mLamB-\mLamCst)\sqrt{s_+}}{(\mLamB+\mLamCst)\sqrt{\mLamB \mLamCst}}\left(\zeta \bar\Lambda'-\zeta_{\text{SL}}-\frac{\zeta \bar\Lambda(\mLamB^2+\mLamCst^2-q^2)}{2\mLamB\mLamCst}\right) +\nn \\
    &+\varepsilon_c \frac{(\mLamB-\mLamCst)\sqrt{s_+}}{(\mLamB+\mLamCst)\sqrt{\mLamB \mLamCst}}\left(\zeta\bar\Lambda'+\zeta_{\text{SL}}-\frac{\zeta \bar\Lambda(\mLamB^2+\mLamCst^2-q^2)}{2\mLamB\mLamCst}\right)+ \nn \\
    &+\varepsilon_c\frac{\sqrt{s_+}s_-}{2(\mLamB\mLamCst)^{3/2}}\eta^{(c)}_{\text{kin}}-\varepsilon_c^2\frac{(\mLamB-\mLamCst)s^{3/2}_{+}s_{-}}{4(\mLamB+\mLamCst)(\mLamB\mLamCst)^{5/2}}\beta_1\,,\\
G_{1/2, \perp} =&  \frac{\zeta\,\sqrt{s_-}s_+}{2(\mLamB \mLamCst)^{3/2}}C^5_1+\varepsilon_b\zeta\frac{\sqrt{s_-}}{\sqrt{\mLamB\mLamCst}}\left(\frac{\bar\Lambda(\mLamB^2+\mLamCst^2-q^2)}{2\mLamB\mLamCst}-\bar\Lambda'\right)+ \nn \\
    &+\varepsilon_c \frac{\sqrt{s_-}}{\sqrt{\mLamB\mLamCst}}\left(\zeta\bar\Lambda+\zeta_{\text{SL}}-\zeta\frac{\bar\Lambda'(\mLamB^2+\mLamCst^2-q^2)}{2\mLamB\mLamCst}\right)+ \nn \\
    &+\varepsilon_c\frac{s_+\sqrt{s_-}}{2(\mLamB\mLamCst)^{3/2}}\eta^{(c)}_{\text{kin}} -\varepsilon_c^2\frac{s_{+}s_{-}^{3/2}}{4(\mLamB\mLamCst)^{5/2}}\beta_1 \,,\\
G_{3/2, \perp} =&  -\varepsilon_b\frac{\sqrt{s_-}}{\sqrt{\mLamB\mLamCst}}\zeta_{\text{SL}} \,\notag \\
\end{align}

\newpage
\begin{align}
  T_{1/2, 0} =&  \zeta\frac{\mLamCst\sqrt{s_+}s_-}{(\mLamB\mLamCst)^{3/2}}\left(C_{T_{1}}-C_{T_{2}}+C_{T_{3}}-\frac{s_+}{2\mLamB\mLamCst}C_{T_{4}}\right)+\nn \\
&+\varepsilon_b\frac{2\mLamCst\sqrt{s_+}}{\sqrt{\mLamB\mLamCst}}\left(-\zeta\bar\Lambda'-\zeta_{\text{SL}}+\zeta\frac{\mLamB^2+\mLamCst^2-q^2}{2\mLamB\mLamCst}\bar\Lambda\right)+\nn \\
+&\varepsilon_c\frac{2\mLamCst\sqrt{s_+}}{\sqrt{\mLamB\mLamCst}}\left(\zeta\bar\Lambda+\zeta_{\text{SL}}-\zeta\frac{\mLamB^2+\mLamCst^2-q^2}{2\mLamB\mLamCst}\bar\Lambda'\right)
+\varepsilon_c\frac{\mLamCst\sqrt{s_+}s_-}{(\mLamB\mLamCst)^{3/2}}\eta^{(c)}_{\text{kin}}- \nn \\ 
&-\varepsilon_c^2\frac{\mLamCst\,s_+^{3/2}s_-}{2(\mLamB\mLamCst)^{5/2}}\,\beta_1 \,,\\  
T_{1/2, \perp} =& \zeta\frac{\sqrt{s_+} s_-}{\mLamB\sqrt{\mLamB\mLamCst}}\left(C_{T_{1}}-\frac{\mLamB^2-\mLamCst^2+q^2}{2\mLamB(\mLamB+\mLamCst)}C_{T_{2}}-\frac{\mLamB^2-\mLamCst^2-q^2}{2\mLamCst(\mLamB+\mLamCst)}C_{T_{3}}\right)+\nn \\
&+\varepsilon_b\frac{2\mLamCst(\mLamB-\mLamCst)\sqrt{s_+}}{(\mLamB+\mLamCst)\sqrt{\mLamB\mLamCst}}\left(\zeta\bar\Lambda'-\zeta\frac{\mLamB^2+\mLamCst^2-q^2}{2\mLamB\mLamCst}\bar\Lambda\right)+\nn \\
&+\varepsilon_c\frac{2\mLamCst(\mLamB-\mLamCst)\sqrt{s_+}}{(\mLamB+\mLamCst)\sqrt{\mLamB\mLamCst}}\left(\zeta\bar\Lambda+\zeta_{\text{SL}}-\zeta\frac{\mLamB^2+\mLamCst^2-q^2}{2\mLamB\mLamCst}\bar\Lambda'\right)+\nn \\
&+\varepsilon_c\frac{\mLamCst \sqrt{s_+}s_-}{(\mLamB\mLamCst)^{3/2}}\eta^{(c)}_{\text{kin}}-\varepsilon_c^2\frac{\mLamCst(\mLamB-\mLamCst)s_+^{3/2}s_-}{2(\mLamB+\mLamCst)(\mLamB\mLamCst)^{5/2}}\,\beta_1 \,,\\  
T_{3/2, \perp} =& -\varepsilon_b\frac{2(\mLamB-\mLamCst)\sqrt{s_+}}{\sqrt{\mLamB\mLamCst}}\zeta_{\text{SL}}\,\notag \ ,\\  
    T^5_{1/2, 0} =&  \frac{\zeta\mLamCst s_+\sqrt{s_-}}{(\mLamB\mLamCst)^{3/2}}C_{T_{1}}+\varepsilon_b\frac{2\mLamCst\sqrt{s_-}}{\sqrt{\mLamB\mLamCst}}\left(\zeta\frac{(\mLamB^2+\mLamCst^2-q^2)\bar\Lambda}{2\mLamB\mLamCst}+\zeta_{\text{SL}}-\zeta\bar\Lambda'\right)+\nn \\
&+\varepsilon_b\frac{2\mLamCst\sqrt{s_-}}{\sqrt{\mLamB\mLamCst}}\left(-\zeta\frac{(\mLamB^2+\mLamCst^2-q^2)\bar\Lambda'}{2\mLamB\mLamCst}+\zeta_{\text{SL}}+\zeta\bar\Lambda\right)+\nn \\
&+\varepsilon_c\frac{\mLamCst s_+\sqrt{s_-}}{(\mLamB\mLamCst)^{3/2}}\eta^{(c)}_{\text{kin}}-\varepsilon_c^2\frac{\mLamCst\,s_+s_-^{3/2}}{2(\mLamB\mLamCst)^{5/2}}\beta_1\,,\\  
T^5_{1/2, \perp} =&  \zeta\frac{s_+\sqrt{s_-}}{\mLamB\sqrt{\mLamB\mLamCst}}\left(C_{T_{1}}-\frac{s_-}{2\mLamB(\mLamB-\mLamCst)}C_{T_{2}}-\frac{s_-}{2\mLamCst(\mLamB-\mLamCst)}\right)+\nn \\
&+\varepsilon_b\frac{2\mLamCst(\mLamB+\mLamCst)\sqrt{s_-}}{(\mLamB-\mLamCst)\sqrt{\mLamB\mLamCst}}\left(\zeta\bar\Lambda'-\zeta\frac{(\mLamB^2+\mLamCst^2-q^2)\bar\Lambda}{2\mLamB\mLamCst}\right)+\nn \\
&+\varepsilon_c\frac{2\mLamCst(\mLamB+\mLamCst)\sqrt{s_-}}{(\mLamB-\mLamCst)\sqrt{\mLamB\mLamCst}}\left(\zeta\bar\Lambda+\zeta_{\text{SL}}-\zeta\frac{(\mLamB^2+\mLamCst^2-q^2)\bar\Lambda'}{2\mLamB\mLamCst}\right)+\nn \\
&+\varepsilon_c\frac{\mLamCst s_+\sqrt{s_-}}{(\mLamB\mLamCst)^{3/2}}\eta^{(c)}_{\text{kin}}-\varepsilon_c^2\frac{\mLamCst(\mLamB+\mLamCst)s_+s_-^{3/2}}{2(\mLamB-\mLamCst)(\mLamB\mLamCst)^{5/2}}\beta_1 \,,\\ 
T^5_{3/2, \perp} =&  -\varepsilon_b\frac{2(\mLamB+\mLamCst)\sqrt{s_-}}{\sqrt{\mLamB\mLamCst}}\zeta_{\text{SL}} \notag\, \ .
\end{align}

\section{Polarisation vectors}
\label{app:kin}
In this section, we present the definitions of the polarisation vectors that have been employed in the determiation of the on-shell amplitude, according to the procedure outlined in Ref. \cite{Boer:2018vpx}.
We have selected the $z$ axis to align with the flight direction of the $\Lambda_c^{*}$. Consequently, within the rest frame of the $\Lambda_b$ ($\Lambda_b-RF$), we find:
\begin{align}
    p^\mu\big|_\text{$\Lambda_b$-RF} & = (m_{\Lambda_b}, 0, 0, 0)\,,\\
    q^\mu\big|_\text{$\Lambda_b$-RF} & = (q^0, 0, 0, -|\vec{q}\,|)\,,\\
    k^\mu\big|_\text{$\Lambda_b$-RF} & = (m_{\Lambda_b} - q^0, 0, 0, +|\vec{q}\,|)\,.
\end{align}
Expressed in term of the invariant $q^2$, we obtain
\begin{align}
        q^0\big|_{\Lambda_b-RF} & = \frac{\mLamB^2 - \mLamCst^2 + q^2}{2 \mLamB}\,, &
|\vec{q}\,|\big|_{\Lambda_b-RF} & = \frac{\sqrt{\lambda(\mLamB^2, \mLamCst^2, q^2)}}{2 \mLamB}\,,
\end{align}
where $\lambda$ is the K\"all\'en function 
\begin{equation}
\lambda(a,b,c)= a^2+b^2+c^2-2ab-2bc-2ac \ .    
\end{equation}
The definition of the Rarita-Schwinger spinor involves a spin-$1$ polarisation vector denoted as $\eta(m)$. Following Refs.  \cite{Haber:1994pe,Boer:2018vpx} we have
\begin{align}
    \eta(\pm)|_{\Lambda_b-RF} & = (0, \mp 1, - i, 0) / \sqrt{2}\,,\\
    \eta(0)|_{\Lambda_b-RF}   & = (|\vec{q}\,|, 0, 0, \mLamB - q^0) / \mLamCst\,.
\end{align}
For the exchange of the virtual $W$ we use the following polarisation vectors \cite{Haber:1994pe, Boer:2018vpx}:
\begin{align}
    \eps^\mu(t)\big|_{\Lambda_b-RF}   & = (q^0, 0, 0, -|\vec{q}\,|) / \sqrt{q^2} = q^\mu / \sqrt{q^2}\,,\\
    \eps^\mu(\pm)\big|_{\Lambda_b-RF}  & = (0, \pm 1, - i, 0) / \sqrt{2}\,,\\
    \eps^\mu(0)\big|_{\Lambda_b-RF}   & = (+|\vec{q}\,|, 0, 0, -q_0) / \sqrt{q^2}\,.
\end{align}
\section{Relations with lattice form factors}
\label{app:B}
Our definitions of form factors are related to the lattice one in Refs.~\cite{Meinel:2021mdj,Meinel:2021rbm} by the following relations:
\begin{equation}
\begin{aligned}
F_{1/2,t} =&\, \frac{1}{2}\sqrt{\frac{s_{-}}{4\mLamB \mLamCst}} f_0^{(\frac{3}{2}^{-})}\,, & F_{1/2,0} =&\, \frac{1}{2}\sqrt{\frac{s_{+}}{4\mLamB \mLamCst}} f_+^{(\frac{3}{2}^{-})}\,, \\
F_{1/2,\perp} =&\, \frac{1}{2}\sqrt{\frac{s_{+}}{4\mLamB \mLamCst}} f_\perp^{(\frac{3}{2}^{-})}\,, & F_{3/2,\perp} =&\, -\frac{1}{2}\sqrt{\frac{s_{+}}{4\mLamB \mLamCst}} f_{\perp^\prime}^{(\frac{3}{2}^{-})}\,, \\
G_{1/2,t} =&\, \frac{1}{2} \sqrt{\frac{s_{+}}{4\mLamB\mLamCst}} g_0^{(\frac{3}{2}^{-})}\,, & G_{1/2,0} =&\, \frac{1}{2} \sqrt{\frac{s_{-}}{4\mLamB\mLamCst}} g_+^{(\frac{3}{2}^{-})}\,, \\
G_{1/2,\perp} =&\, \frac{1}{2} \sqrt{\frac{s_{-}}{4\mLamB\mLamCst}} g_\perp^{(\frac{3}{2}^{-})}\,, & G_{3/2,\perp} =&\, \frac{1}{2} \sqrt{\frac{s_{-}}{4\mLamB\mLamCst}} g_{\perp^\prime}^{(\frac{3}{2}^{-})}\,, \\
f_{1/2,t} =&\, \frac{1}{2}\sqrt{\frac{3 s_{-}}{\mLamB \mLamCst}}\frac{\mLamB+\mLamCst}{\mLamB-\mLamCst}  f_0^{(\frac{1}{2}^{-})}\,, & f_{1/2,0} =&\, \frac{1}{2}\sqrt{\frac{3 s_{+}}{\mLamB \mLamCst}} \frac{\mLamB-\mLamCst}{\mLamB+\mLamCst} f_+^{(\frac{1}{2}^{-})}\,, \\
f_{1/2,\perp} =&\, \frac{1}{2}\sqrt{\frac{3 s_{+}}{\mLamB \mLamCst}} f_\perp^{(\frac{1}{2}^{-})}\,, &  \\
g_{1/2,t} =&\, \frac{1}{2} \sqrt{\frac{3 s_{+}}{\mLamB\mLamCst}} \frac{\mLamB-\mLamCst}{\mLamB+\mLamCst}  g_0^{(\frac{1}{2}^{-})}\,, & g_{1/2,0} =&\, \frac{1}{2} \sqrt{\frac{3 s_{-}}{\mLamB\mLamCst}}\frac{\mLamB+\mLamCst}{\mLamB-\mLamCst}  g_+^{(\frac{1}{2}^{-})}\,, \\
g_{1/2,\perp} =&\, \frac{1}{2} \sqrt{\frac{3 s_{-}}{\mLamB\mLamCst}} g_\perp^{(\frac{1}{2}^{-})}\,, &  \\
T_{1/2,0} =&\,  s_{+}^{1/2} \sqrt{\frac{\mLamCst}{4\mLamB}}h_{+}^{(\frac{3}{2}^{-})}\,,  \\
T_{1/2,\perp} =&\,  s_{+}^{1/2} \sqrt{\frac{\mLamCst}{4\mLamB}} h_{\perp}^{(\frac{3}{2}^{-})}\,,  & T_{3/2,\perp} =&\, \sqrt{\frac{s_{+}}{4\mLamB\mLamCst}}(\mLamB+\mLamCst) h_{\perp^\prime}^{(\frac{3}{2}^{-})}\,,\\
T^5_{1/2,0} =&\,  s_{-}^{1/2} \sqrt{\frac{\mLamCst}{4\mLamB}}\tilde{h}_{+}^{(\frac{3}{2}^{-})}\,,  \\
T^5_{1/2,\perp} =&\,  s_{-}^{1/2} \sqrt{\frac{\mLamCst}{4\mLamB}} \tilde{h}_{\perp}^{(\frac{3}{2}^{-})}\,,  & T^5_{3/2,\perp} =&\,- \sqrt{\frac{s_{-}}{4\mLamB\mLamCst}}(\mLamB-\mLamCst)\tilde{h}_{\perp^\prime}^{(\frac{3}{2}^{-})} \ , \\
t_{1/2,0}=& \sqrt{\frac{3s_{+}\mLamCst}{\mLamB}}h_{+}^{(\frac{1}{2}^{-})} \ ,
&t_{1/2,\perp} = & \sqrt{\frac{3s_{+}\mLamCst}{\mLamB}}\frac{\mLamB-\mLamCst}{\mLamB+\mLamCst} h_{\perp}^{(\frac{1}{2}^{-})} \ , \\
t^{5}_{1/2,0}=& \sqrt{\frac{3s_{-}\mLamCst}{\mLamB}}\Tilde{h}_{+}^{(\frac{1}{2}^{-})} \ ,
&t^{5}_{1/2,\perp} = & \sqrt{\frac{3s_{-}\mLamCst}{\mLamB}}\frac{\mLamB+\mLamCst}{\mLamB-\mLamCst} \Tilde{h}_{\perp}^{(\frac{1}{2}^{-})} \ .
\end{aligned}
\end{equation}
\section{Correlation matrices from LQCD fit}
\label{app:G}
In this section, we provide the correlation matrices that correspond to the fitting results within the RC and VC limits. Our analysis produces the outcomes presented in \cref{tab:RCfull} and \cref{tab:VCfull} when considering the complete lattice dataset. When focusing solely on the vector and axial lattice form factors, we observe the results outlined in \cref{tab:RCred} and \cref{tab:VCred}.
\begin{table}[h!]
    \centering
    \scriptsize
    \begin{tabular}{ccccccccccccc}
    \toprule
 1 & 0.211 & 0.195 & 0.011 & -0.972 & -0.214 & 0.204 & 0.075 & -0.025 & 0.004 & -0.298 & 0.825 & 0.103 \\
  0.211 & 1 & -0.055 & 0.201 & -0.316 & -0.891 & -0.022 & 0.174 & 0.104 & -0.097 & -0.907 & 0.132 & 0.047 \\
  0.195 & -0.055 & 1 & -0.072 & -0.176 & 0.064 & 0.097 & -0.005 & -0.141 & 0.012 & -0.030 & 0.540 & -0.054 \\
  0.011 & 0.201 & -0.072 & 1 & 0.005 & -0.141 & 0.052 & 0.175 & 0.428 & -0.270 & 0.004 & -0.023 & 0.629 \\
 -0.972 & -0.316 & -0.176 & 0.005 & 1 & 0.256 & -0.176 & -0.085 & 0.022 & 0.003 & 0.423 & -0.793 & -0.058 \\
 -0.214 & -0.891 & 0.064 & -0.141 & 0.256 & 1 & 0.008 & -0.032 & -0.080 & 0.067 & 0.752 & -0.133 & -0.038 \\
  0.204 & -0.022 & 0.097 & 0.052 & -0.176 & 0.008 & 1 & -0.525 & -0.032 & 0.033 & 0.109 & 0.154 & 0.173 \\
  0.075 & 0.174 & -0.005 & 0.175 & -0.085 & -0.032 & -0.525 & 1 & 0.073 & -0.076 & -0.209 & 0.068 & -0.027 \\
 -0.025 & 0.104 & -0.141 & 0.428 & 0.022 & -0.080 & -0.032 & 0.073 & 1 & -0.592 & -0.020 & -0.077 & 0.224 \\
  0.004 & -0.097 & 0.012 & -0.270 & 0.003 & 0.067 & 0.033 & -0.076 & -0.592 & 1 & 0.052 & 0.009 & -0.118 \\
 -0.298 & -0.907 & -0.030 & 0.004 & 0.423 & 0.752 & 0.109 & -0.209 & -0.020 & 0.052 & 1 & -0.232 & 0.172 \\
  0.825 & 0.132 & 0.540 & -0.023 & -0.793 & -0.133 & 0.154 & 0.068 & -0.077 & 0.009 & -0.232 & 1 & 0.060 \\
  0.103 & 0.047 & -0.054 & 0.629 & -0.058 & -0.038 & 0.173 & -0.027 & 0.224 & -0.118 & 0.172 & 0.060 & 1 \\
    \bottomrule
  \end{tabular}
  \caption{\em Correlation matrix for the IW parameters in the RC limit, when using the full lattice dataset.}
\label{tab:RCfull}
\end{table}

\begin{table}[h!]
    \centering
   \scriptsize
    \begin{tabular}{ccccccccccccc}
    \toprule
 1 & 0.483 & 0.014 & 0.084 & -0.997 & -0.464 & 0.423 & 0.136 & -0.007 & -0.019 & -0.639 & 0.942 & 0.247 \\
0.483 & 1 & -0.066 & 0.196 & -0.515 & -0.924 & 0.137 & 0.183 & 0.068 & -0.075 & -0.899 & 0.436 & 0.173 \\
0.014 & -0.066 & 1 & -0.086 & -0.013 & 0.074 & 0.046 & -0.016 & -0.142 & 0.029 & -0.008 & 0.268 & -0.080 \\
0.084 & 0.196 & -0.086 & 1 & -0.077 & -0.160 & 0.142 & 0.143 & 0.129 & -0.123 & 0.045 & 0.057 & 0.757 \\
-0.997 & -0.515 & -0.013 & -0.077 & 1 & 0.478 & -0.413 & -0.139 & 0.006 & 0.020 & 0.674 & -0.938 & -0.232 \\
-0.464 & -0.924 & 0.074 & -0.160 & 0.478 & 1 & -0.139 & -0.048 & -0.056 & 0.062 & 0.796 & -0.417 & -0.163 \\
0.423 & 0.137 & 0.046 & 0.142 & -0.413 & -0.139 & 1 & -0.389 & -0.045 & 0.028 & -0.114 & 0.395 & 0.290 \\
0.136 & 0.183 & -0.016 & 0.143 & -0.139 & -0.048 & -0.389 & 1 & 0.074 & -0.069 & -0.212 & 0.132 & 0.034 \\
-0.007 & 0.068 & -0.142 & 0.129 & 0.006 & -0.056 & -0.045 & 0.074 & 1 & -0.679 & -0.032 & -0.047 & 0.081 \\
-0.019 & -0.075 & 0.029 & -0.123 & 0.020 & 0.062 & 0.028 & -0.069 & -0.679 & 1 & 0.045 & -0.005 & -0.076 \\
-0.639 & -0.899 & -0.008 & 0.045 & 0.674 & 0.796 & -0.114 & -0.212 & -0.032 & 0.045 & 1 & -0.601 & 0.041 \\
0.942 & 0.436 & 0.268 & 0.057 & -0.937 & -0.417 & 0.395 & 0.131 & -0.046 & -0.005 & -0.601 & 1 & 0.214 \\
0.247 & 0.173 & -0.080 & 0.757 & -0.232 & -0.163 & 0.290 & 0.034 & 0.081 & -0.076 & 0.041 & 0.214 & 1 \\
\bottomrule
\end{tabular}
\caption{\em Correlation matrix for the IW parameters in the RC limit, when using only the vector and axial lattice form factors.}
\label{tab:RCred}
\end{table}

\begin{table}[h!]
    \centering
    \begin{tabular}{ccccccccccccc}
    \toprule
1 & 0.190 & 0.170 & -0.044 & -0.970 & -0.222 & -0.302 & 0.815 & 0.065 \\
0.190 & 1 & -0.056 & 0.138 & -0.301 & -0.902 & -0.910 & 0.114 & 0.018 \\
0.170 & -0.056 & 1 & -0.037 & -0.151 & 0.058 & -0.030 & 0.527 & -0.044 \\
-0.044 & 0.138 & -0.037 & 1 & 0.061 & -0.115 & 0.051 & -0.045 & 0.607 \\
-0.970 & -0.301 & -0.151 & 0.061 & 1 & 0.263 & 0.431 & -0.780 & -0.021 \\
-0.222 & -0.902 & 0.058 & -0.115 & 0.263 & 1 & 0.764 & -0.139 & -0.021 \\
-0.302 & -0.910 & -0.030 & 0.051 & 0.431 & 0.764 & 1 & -0.229 & 0.179 \\
0.815 & 0.114 & 0.527 & -0.045 & -0.780 & -0.139 & -0.229 & 1 & 0.043 \\
0.065 & 0.018 & -0.044 & 0.607 & -0.021 & -0.021 & 0.179 & 0.043 & 1 \\
\bottomrule
\end{tabular}
\caption{\em Correlation matrix for the IW parameters in the VC limit, when using the full lattice dataset.}
\label{tab:VCfull}
\end{table}

\begin{table}[h!]
    \centering
    \begin{tabular}{ccccccccccccc}
    \toprule
1 & 0.423 & -0.011 & -0.056 & -0.995 & -0.445 & -0.620 & 0.923 & 0.091 \\
0.423 & 1 & -0.072 & 0.123 & -0.461 & -0.930 & -0.890 & 0.369 & 0.096 \\
-0.011 & -0.072 & 1 & -0.085 & 0.012 & 0.078 & -0.003 & 0.281 & -0.091 \\
-0.056 & 0.123 & -0.085 & 1 & 0.063 & -0.116 & 0.137 & -0.075 & 0.745 \\
-0.995 & -0.461 & 0.012 & 0.063 & 1 & 0.461 & 0.661 & -0.917 & -0.075 \\
-0.445 & -0.930 & 0.078 & -0.116 & 0.461 & 1 & 0.799 & -0.389 & -0.109 \\
-0.620 & -0.890 & -0.003 & 0.137 & 0.661 & 0.799 & 1 & -0.571 & 0.136 \\
0.923 & 0.369 & 0.281 & -0.075 & -0.917 & -0.389 & -0.571 & 1 & 0.065 \\
0.091 & 0.096 & -0.091 & 0.745 & -0.075 & -0.109 & 0.136 & 0.065 & 1 \\
\bottomrule
\end{tabular}
\caption{\em Correlation matrix for the IW parameters in the RC limit, when using only the vector and axial lattice form factors.}
\label{tab:VCred}
\end{table}

\end{document}